\documentclass[aps,twocolumn,amsmath,amsmath,amssymb,superscriptaddress,longbibliography]{revtex4-1}

\usepackage{graphicx}
\usepackage{dcolumn}
\usepackage{bm}
\usepackage[colorlinks,linkcolor=blue,citecolor=blue,urlcolor=blue]{hyperref}
\usepackage{float}
\usepackage[dvipsnames]{xcolor}
\usepackage[utf8]{inputenc} 
\usepackage[T1]{fontenc}
\usepackage[normalem]{ulem}
\usepackage{amsmath}
\usepackage{lineno}

\begin{document}

\title{Electron-phonon-dominated charge-density-wave fluctuations in TiSe$_2$ accessed by ultrafast nonequilibrium dynamics}

\author{Sotirios Fragkos}
\affiliation{Universit\'e de Bordeaux - CNRS - CEA, CELIA, UMR5107, F33405 Talence, France}

\author{Hibiki Orio}
\email{hibiki.orio@uni-wuerzburg.de}
\affiliation{Experimentelle Physik VII and Würzburg-Dresden Cluster of Excellence ctd.qmat, Universität Würzburg, D-97074, Würzburg, Germany}

\author{Nina Girotto Erhardt}
\affiliation{Centre for Advanced Laser Techniques, Institute of Physics, 10000 Zagreb, Croatia}
\affiliation{European Theoretical Spectroscopy Facility, Institute of Condensed Matter and Nanosciences, Université catholique de Louvain, Louvain-la-Neuve, Belgium}

\author{Akib Jabed}
\affiliation{Universit\'e de Bordeaux - CNRS - CEA, CELIA, UMR5107, F33405 Talence, France}
\affiliation{Institut National de la Recherche Scientifique – Énergie Matériaux Télécommunications Varennes, QC J3X 1S2, Canada}

\author{Sarath Sasi}
\affiliation{New Technologies-Research Center, University of West Bohemia, 30614, Pilsen, Czech Republic}

\author{Quentin Courtade}
\affiliation{Universit\'e de Bordeaux - CNRS - CEA, CELIA, UMR5107, F33405 Talence, France}

\author{Muthu P. T. Masilamani}
\affiliation{Experimentelle Physik VII and Würzburg-Dresden Cluster of Excellence ctd.qmat, Universität Würzburg, D-97074, Würzburg, Germany}

\author{Maximilian Ünzelmann}
\affiliation{Experimentelle Physik VII and Würzburg-Dresden Cluster of Excellence ctd.qmat, Universität Würzburg, D-97074, Würzburg, Germany}

\author{Florian Diekmann}
\affiliation{Ruprecht Haensel Laboratory, Deutsches Elektronen-Synchrotron DESY, D-22607, Hamburg, Germany}
\affiliation{Institute of Experimental and Applied Physics, Kiel University, D-24098, Kiel, Germany}

\author{Baptiste Hildebrand}
\affiliation{University of Fribourg and Fribourg Centre for Nanomaterials, Chemin du Musée 3, CH-1700 Fribourg, Switzerland}

\author{Dominique Descamps}
\affiliation{Universit\'e de Bordeaux - CNRS - CEA, CELIA, UMR5107, F33405 Talence, France}

\author{Stéphane Petit}
\affiliation{Universit\'e de Bordeaux - CNRS - CEA, CELIA, UMR5107, F33405 Talence, France}

\author{Fabio Boschini}
\affiliation{Advanced Laser Light Source, Institut National de la Recherche Scientifique, Varennes QC J3X 1S2 Canada}

\author{Ján Minár}
\affiliation{New Technologies-Research Center, University of West Bohemia, 30614, Pilsen, Czech Republic}

\author{Yann Mairesse}
\affiliation{Universit\'e de Bordeaux - CNRS - CEA, CELIA, UMR5107, F33405 Talence, France}

\author{Friedrich Reinert}
\affiliation{Experimentelle Physik VII and Würzburg-Dresden Cluster of Excellence ctd.qmat, Universität Würzburg, D-97074, Würzburg, Germany}

\author{Kai Rossnagel}
\affiliation{Ruprecht Haensel Laboratory, Deutsches Elektronen-Synchrotron DESY, D-22607, Hamburg, Germany}
\affiliation{Institute of Experimental and Applied Physics, Kiel University, D-24098, Kiel, Germany}

\author{Dino Novko}
\email{dino.novko@gmail.com}
\affiliation{Centre for Advanced Laser Techniques, Institute of Physics, 10000 Zagreb, Croatia}

\author{Samuel Beaulieu}
\affiliation{Universit\'e de Bordeaux - CNRS - CEA, CELIA, UMR5107, F33405 Talence, France}

\author{Jakub Schusser}
\email{schusser@ntc.zcu.cz}
\affiliation{Experimentelle Physik VII and Würzburg-Dresden Cluster of Excellence ctd.qmat, Universität Würzburg, D-97074, Würzburg, Germany}
\affiliation{New Technologies-Research Center, University of West Bohemia, 30614, Pilsen, Czech Republic}

\begin{abstract}
The complex phase diagram of 1T-TiSe$_2$ consists of a charge density wave (CDW) below 200\,K, and CDW fluctuations of still unknown origin at higher temperatures. Here, we use time-resolved extreme ultraviolet momentum microscopy and density functional perturbation theory to uncover the formation mechanism of CDW fluctuations and their spectral features at 295\,K. We investigated the transient dynamics of fluctuations upon nonresonant ultrafast photoexcitation, and directly correlate it with the CDW soft-phonon hardening. Surprisingly, our results show that the coherent amplitude mode modulating ultrafast CDW recovery persists above $T_{\rm CDW}$, and reveal that CDW fluctuations are dominated by the electron-phonon interaction rather than excitonic correlations as commonly believed. Our findings on these microscopic CDW fluctuations clarify the complex interplay between electronic and lattice degrees of freedom at elevated temperatures and, therefore, could be useful in understanding the nature of the CDW phase transition in 1T-TiSe$_2$ and similar quantum materials.

\end{abstract}

\maketitle
\section*{Introduction}
Understanding the interplay between electronic and lattice degrees of freedom is essential for comprehending phase transitions, superconductivity, transport properties, and the underlying mechanism of complex quantum phases. Disentangling the origin of the charge density wave (CDW) phase in 1T-TiSe$_2$, with $T_{\rm CDW}=200$\,K and a $2\times2\times2$ lattice reconstruction~\cite{di1976electronic,woo1076superlattice}, provides the right platform to do so. Despite extensive experimental and theoretical investigations, the driving mechanism of the CDW phase remains a debated question\,\cite{Rossnagel_2011}. One perspective attributes the origin to electron-hole coupling\,\cite{jerome1967excitonic}, as in the excitonic insulator scenario~\cite{di1976electronic,kidd2002electron,cercellier2007evidence,monney2009spontaneous,monney2010temperature,monney2012electron,kogar2017sign}, while alternative argues that the dominant mechanism is electron-phonon coupling, represented by the band-type Jahn-Teller effect~\cite{hughes1977structural,yoshida1980epc,whangbo1992analogies,rossnagel2002charge,weber2011epc}. Recently, a series of ultrafast time-resolved studies made an effort to disentangle the two contributions\,\cite{porer2014nonthermal,monney2016revealing,karam2018strong,hedayat2019exc,Lian2020,otto2021epc,cheng2022light,heinrich2023elec}, and reach a consensus that electron-hole and electron-lattice interactions are intertwined in driving the CDW transition~\cite{van2010exciton}. The ultrafast femtosecond dynamics (i.e., melting and recovery) of the order parameter (e.g., the CDW gap)\,\cite{rohwer2011collapse,hellmann2012time} was used as an argument for the former, while the laser-induced coherent phonon motion of the CDW amplitude mode\,\cite{porer2014nonthermal,monney2016revealing,hedayat2019exc,duan2021optical} for the latter interaction.

\begin{figure*}[!t]
\begin{center}
\includegraphics[width=18cm]{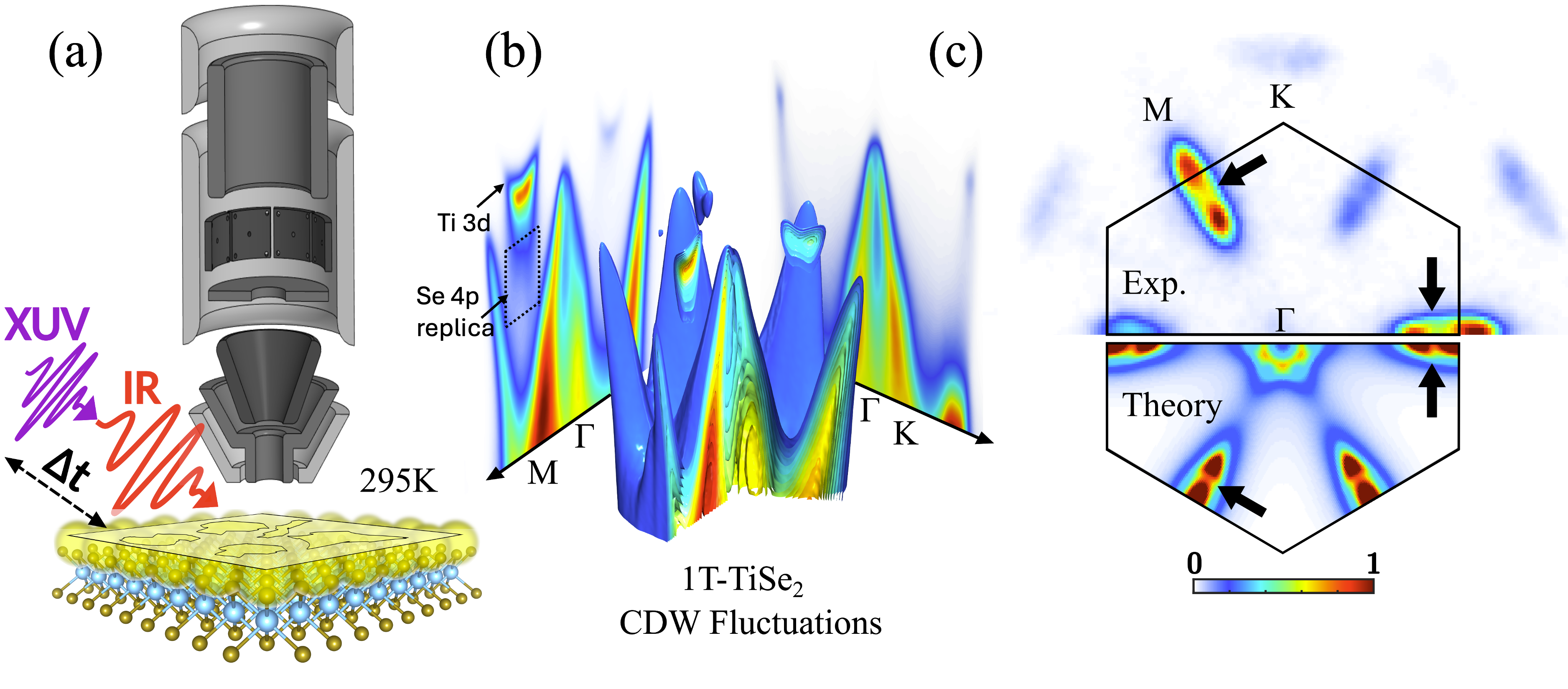}
\caption{\textbf{Experimental and theoretical methodologies for investigating charge-density-wave (CDW) fluctuations in 1T-TiSe$_2$.} \textbf{(a)} Polarization-tunable infrared pump (1030~nm, 135~fs, 0.95 mJ/cm$^2$) and extreme ultraviolet (XUV) probe pulses (21.6~eV) are focused at an incidence angle of 65$^{\circ}$ on a 1T-TiSe$_2$ sample at room temperature (295~K) in the interaction chamber of a time-of-flight momentum microscope, capable of collecting photoelectrons from the entire 2$\pi$ solid angle in an energy range of several eV. \textbf{(b)} Three-dimensional photoemission intensity $I(E_\mathrm{B}, k_x, k_y)$ and associated energy-momentum cut along M-$\Gamma$-M' and  K-$\Gamma$-K', acquired using s-polarized IR pump pulse (1.2~eV, 135~fs, $\sim$0.95~mJ/cm$^2$) and while continuously rotating the linear polarization axis angle of the XUV probe beam (21.6~eV) at the pump-probe temporal overlap. The photoemission intensity is integrated for all XUV polarization axis angles. \textbf{(c)} Comparison between experimental and theoretical constant energy contours (CECs), near the Fermi level. The measured loss of spectral weight within M pockets (black arrows) is well captured by theoretical calculations, which include dynamical electron-phonon interaction for temperatures above the CDW transition temperature $T_{\rm CDW}$. The colorbar indicates the intensity (experiment)/spectral weight (theory). $\Delta$t indicates time delay between the pump and the probe.}
\label{Fig1}
\end{center}
\end{figure*}

Several studies have reported that the CDW phase persists above $T_{\rm CDW}$ due to the fluctuation of the order parameter related to enigmatic CDW fluctuations~\cite{woo1976lattice,miyahara1995sts,holt2001x,kidd2002electron,chen2016hidden,cercellier2007evidence,monney2010temperature,monney2016revealing}, where the presence of finite correlation lengths leads to phase incoherent CDW domains across the sample. Despite recent reports discussing the CDW fluctuations in 1T-TiSe$_2$ at such elevated temperatures~\cite{mizukoshi2023ultrafast, zhang2022second, cheng2022light,guo2025xrd} with some reporting CDW fluctuation signatures in static angle-resolved photoemission spectroscopy (ARPES) \cite{Jaouen2019}, none of these directly mapped the nonequilibrium light-induced dynamics of these fluctuations in energy and momentum space. Applying the latter technique to explore this exotic phase could provide fertile ground to scrutinize the interplay and robustness of electron-hole and electron-phonon interactions at elevated temperatures. Indeed, a study using time-resolved ARPES (tr-ARPES) demonstrated sub-100\,fs melting of the fluctuating phase, which was used as an evidence for the exciton-driven CDW fluctuations\,\cite{monney2016revealing}.
Contrary to these claims, a recent time-dependent density functional theory study showed that exciton order is melted above $T_{\rm CDW}$\,\cite{chao2019hampers}.
Moreover, the coherent oscillation of the CDW amplitude mode was observed under ambient conditions using time-resolved optical spectroscopy\,\cite{mizukoshi2023ultrafast}, suggesting a lattice involvement in the dynamics of CDW fluctuations. As evidenced by these opposing claims, the CDW fluctuating phase in 1T-TiSe$_2$ is still an elusive puzzle, while it would be highly valuable to understand its origin considering its recently discovered connection to unconvetional superconductivity\,\cite{joe2014walls,Jaouen2019,roemer2024lifshitz}, as well as its resemblance to the fluctuations preceding ordered phases (e.g., pseudogap) in other transition metal dichalcogenides (TMDCs)\,\cite{borisenko2008pg} and strongly correlated materials, such as cuprates\,\cite{arpaia2019dyn} and kagome metals\,\cite{zhong2024,liu2025fluct}. Unraveling nonequilibrium properties of CDW fluctuations in TiSe$_2$ would also be highly useful for comprehending recently discovered light-induced hidden phases in TiSe$_2$\,\cite{duan2023,huber2024ultrafast}.

In this work, we resolve the debate on the origin of CDW fluctuations in 1T-TiSe$_2$ by using time-resolved extreme ultraviolet (XUV) momentum microscopy, which allows ultrafast light-induced dynamics directly in energy-momentum space. The band structure mapping reveals that the backfolded Se 4p band associated with the CDW phase remains detectable even at room temperature (RT), in agreement with previous reports\,\cite{kidd2002electron,cercellier2007evidence,Jaouen2019}. We also measured the partial melting of these CDW fluctuations on an ultrafast timescale, followed by a rapid recovery ($\sim 700$\,fs). This ultrafast melting signal of CDW fluctuations is offset from the maximum of the hot electron temperature by around $100-200$\,fs, and overlaps with the population dynamics of excited Ti 3d states. Together, this demonstrates that CDW fluctuations are able to withstand a strong laser excitation and are not directly linked to the hot electron temperature, suggesting an involvement of the phonon mechanism.
Interestingly, we also observe that the recovery of the CDW fluctuations is modulated by coherent CDW amplitude phonon modes. We challenge the belief that the fluctuating phase is ruled solely by the excitonic correlations\,\cite{monney2016revealing} by reporting the periodic modulations of the backfolded band spectral weight at RT together with density-functional-theory (DFT) based calculations, indicating not only the preserved electron-phonon coupling mechanism at elevated temperatures but also confirming this to be the main contributor in the hybrid mechanism. More specifically, our calculations incorporating the dynamical electron-phonon scattering confirm that the strongly-coupled CDW-related soft phonon is relevant for the observed equilibrium and transient spectral features of CDW fluctuations. Namely, strong interaction between the soft (yet stable) phonon and Se 4p states leads to the fluctuating replica, while light-induced excitations harden the phonon and screen this interaction, consequently quenching the electron-phonon-driven CDW fluctuations. 
More generally, our study poses a series of strong arguments favoring the electron-phonon mechanism of CDW fluctuation formation at elevated temperature, which also puts a strong emphasis on the electron-phonon scenario in the debate on the origin of the regular CDW phase. Besides the fundamental importance, correct interpretation of this mechanism could unleash the full potential of unconventional CDW-related features, such as  CDW chirality\,\cite{ishioka2010chiral}, tunable nonlinear optical response\,\cite{zhang2022second}, optical gyrotropy\,\cite{xu2020gyro}, and superconductivity\,\cite{morosan2006superconductivity}

\begin{figure}[t]
\begin{center}
\includegraphics[width=8cm]{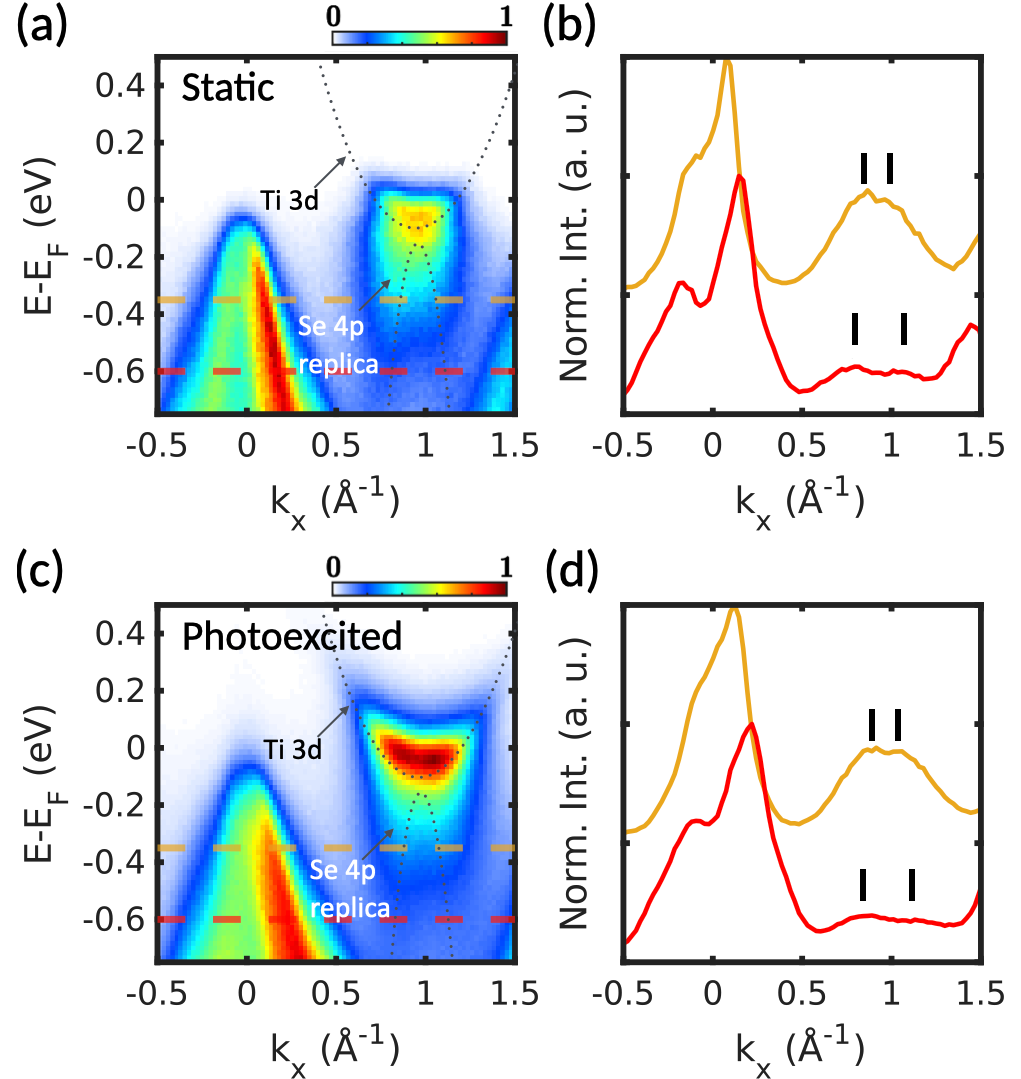}
\caption{\textbf{Equilibrium and excited state band structure in 1T-TiSe$_2$ at Room Temperature.} The first row \textbf{(a,b)} presents the static band structure before infrared pump excitation. In contrast, the second row \textbf{(c,d)} describes the excited band structure at the pump-probe pulse time overlap at time zero. \textbf{(a, c)} Energy-momentum cut along the M-$\Gamma$-M' direction before and at the excitation by the pump pulse. The intensity is integrated for continuously varied linear polarization between horizontal and vertical. Dashed markers indicate where the momentum distribution curves (MDCs) in \textbf{(b, d)} were extracted. The lines in \textbf{(b, d)} indicate the position of Se 4p replica, and \textbf{(a,c)} are overlaid with a sketch of the normal phase band structure. The colorbar captures ARPES intensity.}
\label{Fig2}
\end{center}
\end{figure}

\section*{Results}

\textbf{Equilibrium and Nonequilibrium Electronic Band Structure Mapping.}
The experiments were performed at the Centre Lasers Intenses et Applications (CELIA) in Bordeaux, France. The time-resolved momentum microscopy setup featuring a polarization-tunable infrared (1.2~eV, sub-150$\mu m$) pump and XUV probe (21.6~eV, sub-50$\mu m$) pulses [Fig.~\ref{Fig1}(a)] allows for mapping of the occupied and unoccupied states (up to 1.2~eV above the Fermi level) over the full photoemission horizon [Fig.~\ref{Fig1}(b)]. More details about the experimental setup can be found elsewhere~\cite{Fragkos2025-ob, tkach24, tkach24-2} and in the Methods section. 

To perform photoexcited band mapping, we sit at the pump-probe temporal overlap (s-polarized pump pulse) and measure the photoemission intensity while continuously modulating the XUV linear polarization axis angle  (going from s- to p-polarization through all intermediate linear polarization axis angles). The three-dimensional photoemission intensity $I(E_B, k_x, k_y)$ shown in Fig.~\ref{Fig1}(b) is obtained by summing over all XUV polarization states, allowing for minimized spectral weight suppression caused by photoemission matrix-element effects~\cite{Moser17, Schusser2024, PhysRevLett.129.246404, Fragkos2025-ob}.
The low-energy electronic band structure of 1T-TiSe$_2$ is characterized by a Se 4p valence band located at the $\Gamma$(A) point and Ti 3d conduction band pocket at the M(L) point. As shown in Fig.~\ref{Fig1}(b), one clear indication of the CDW phase in static ARPES is the backfolded Se 4p band at the M point\,\cite{kidd2002electron,cercellier2007evidence}. Figure \ref{Fig1}(c) shows a comparison between experimental and theoretical constant energy contours (CEC), near the Fermi level. The measured loss of spectral weight within M pockets (black arrows) at RT is another fingerprint of the CDW fluctuations, and it is well captured by theoretical calculations, which include dynamical electron-phonon interaction for temperatures above $T_{\rm CDW}$. Detailed analysis of the theoretical calculations will be given later. These scattering hot spots are reminiscent of the pseudogap openings and loss of spectral weight at $T>T_{\rm CDW}$ and certain momenta relevant for the CDW in other TMDCs, such as 2H-TaSe$_2$\,\cite{borisenko2008pg} and 2H-NbSe$_2$\,\cite{borisenko2009pg}.

To gain further insight into the signatures of RT-CDW fluctuations, we perform an in-depth analysis of the spectral features around the M(L) pocket for unexcited [Figs.~\ref{Fig2} (a)-(b)] and photoexcited 1T-TiSe$_2$ (at pump-probe temporal overlap) [Figs.~\ref{Fig2} (c)-(d)]. The photoemission intensity along the $\Gamma$(A) - M(L) cut  [Fig.~\ref{Fig2}(a)] reveal spectral features associated with RT-CDW fluctuations. First, we note the presence of the well-established backfolded Se 4p valence band, typically used as the ARPES evidence of the emergence of CDW order, exhibited by a faint yet discernible down-dispersive band feature at the M(L) point below the Ti 3d band in Fig.~\ref{Fig2}(a). This feature is more apparent in the momentum distribution curves (MDC) around $k_x \approx 1.0 $ \text{\AA}$^{-1}$, as indicated by the black bars in Fig.~\ref{Fig2}(b). The distance between the two peaks increases with higher binding energies, demonstrating the hole-like Se 4p band character. The backfolded Se 4p feature is also observed at pump-probe overlap in both the photoemission intensity image and MDC cut [Figs.~\ref{Fig2}(c) and (d)], indicating that CDW fluctuations are not completely melted and persist upon photoexcitation at this relatively high fluence (0.95\,mJ/cm$^2$). Additionally, a first-principles theoretical study of the electronic response demonstrated that the excitonic mode is quenched upon heating above $T_{\rm CDW}$\,\cite{chao2019hampers}, while ultrafast pump-probe optical spectroscopy and tr-ARPES studies of 1T-TiSe$_2$ in the CDW regime ($T<T_{\rm CDW}$) argued that the excitonic correlations should be destroyed above the critical fluence of 0.04\,mJ/cm$^2$\,\cite{porer2014nonthermal,huber2022fluct,cheng2022light}. This robustness of the fluctuating replica band at RT and high fluence conditions suggests that the CDW fluctuations are not solely of excitonic origin.

\begin{figure*}[t]
\begin{center}
\includegraphics[width=17cm]{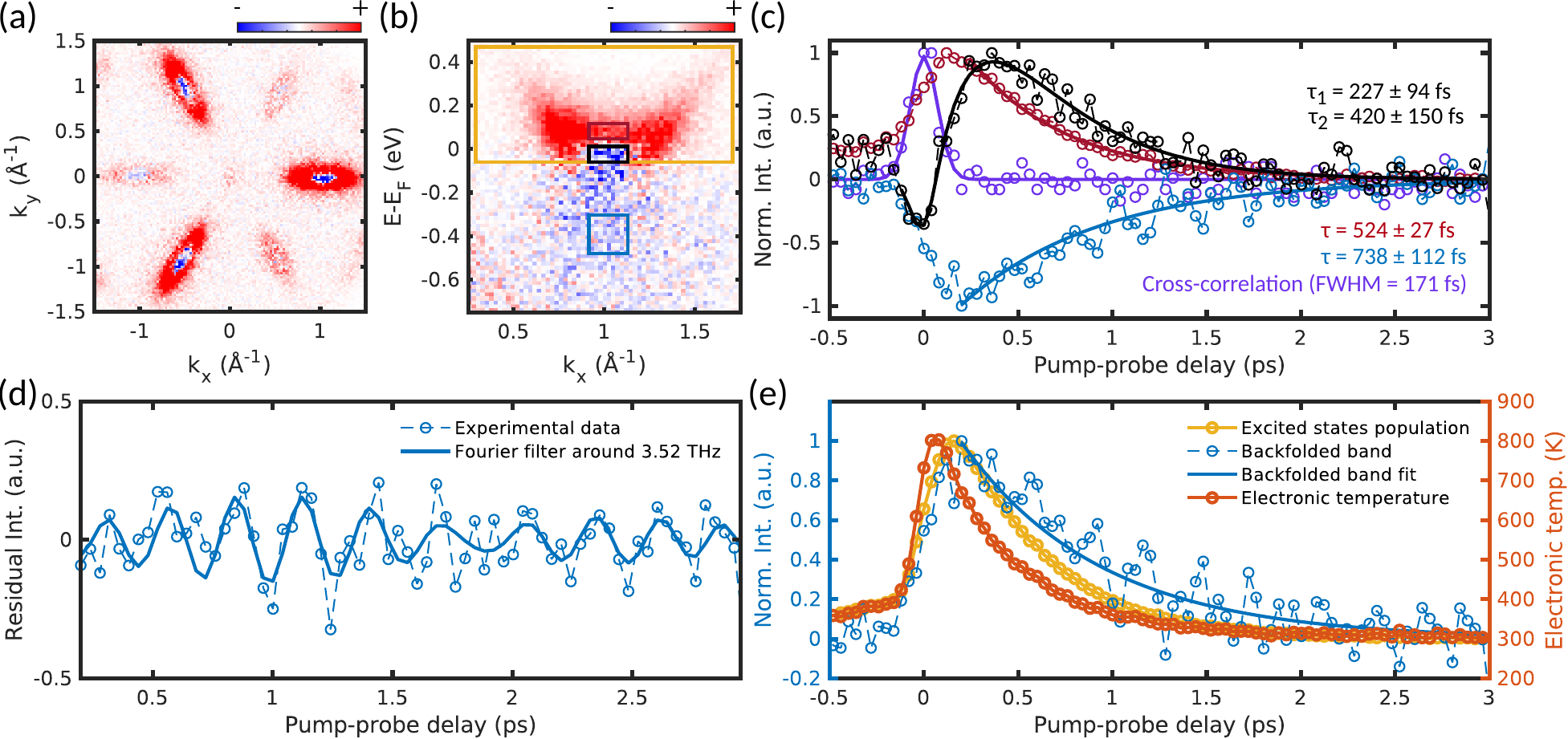}
\caption{\textbf{Ultrafast melting and recovery of room temperature charge-density-wave (RT-CDW) fluctuations in 1T-TiSe$_2$.}  \textbf{(a)-(b)} Differential constant energy contour (integrated over -23~meV$< \mathrm{E-E_F} <$12~meV) and energy-momentum cut near the M pocket, obtained by subtracting photoemission intensity before (integration between -880~fs and -720~fs) and during pump-probe temporal overlap (integration between -80~fs and 80~fs). \textbf{(c)} Time-dependent photoemission intensity for different regions of interest (ROIs) along with corresponding fits. Purple dots represent the infrared (IR)/extreme ultraviolet (XUV) cross-correlation extracted from the time-dependent laser-assisted photoemission signal (ROI not shown), with an FWHM duration of 171~fs (Gaussian fit, purple line). Red, black, and blue dots denote time-resolved photoemission intensity from the respective ROIs shown in (b), while the associated lines indicate fits. \textbf{(d)} Blue dots and dashed line represent the residual intensity obtained by subtracting the double exponential fit from the photoemission intensity in the blue ROI in (b) (CDW backfolded band), while the thick blue line represents the Fourier filtered data (bandpass filter around 3.52~THz) indicating coherent oscillation of the CDW amplitude mode. \textbf{(e)} The electronic temperature as a function of pump-probe delay (orange) extracted by Fermi-Dirac distribution fitting, alongside with the backfolded band intensity (blue) and the excited state population dynamics (yellow). The colorbar captures differential ARPES intensity. Panel (c) shows exponential decay fits: single-exponential decays characterized by a time constant $\tau$ for the red and blue curves, and a bi-exponential decay with two time constants ($\tau_1$ and $\tau_2$) for the black curve.}
\label{Fig3}
\end{center}
\end{figure*}

\textbf{Ultrafast Melting and Recovery of RT-CDW Fluctuations.}
Now that we have established the equilibrium spectral signatures associated with RT-CDW fluctuations, we turn our attention to their ultrafast melting and recovery upon photoexcitation to gain detailed insight into the microscopic origin underlying the formation of this exotic phase\,\cite{boschini2024}.  The tr-ARPES enables systematic temporal separation of fundamental electronic and structural processes, allowing the identification of the dominant interactions\,\cite{hellmann2012time}. 

In this spirit, we record the three-dimensional photoemission intensity as a function of pump-probe delay (between -0.5~ps to 3~ps), yielding a four-dimensional dataset $I(E_\mathrm{B}, k_x, k_y, \Delta t)$. From this multidimensional dataset, one can hunt for subtle signatures of RT-CDW fluctuations melting and recovery along energy, momentum, and time axes. We start by looking at the differential (pre-time-zero signal subtracted) CEC near the Fermi level ($E_B$ = -0.07~eV), at a time delay of $\Delta t=0$ [Figs.~\ref{Fig3} (a) and (b)]. These CECs reveal the population of the conduction Ti 3d band (hollow elliptical positive differential signal) and suppression of the CDW backfolded band spectral weight (central negative differential signal) upon photoexcitation, around M(L) pockets. At this binding energy, after 400~fs, the negative differential signal associated with the melting of CDW order vanishes, as the filling of the bottom of the conduction band becomes dominant. This behavior is also evidenced in the differential energy-momentum region around the L(M) point, extracted for the same pump-probe delays. 

To get more insight into the concomitant dynamics of hot electrons and CDW melting/recovery, we analyze the time-resolved photoemission intensities [Fig.~\ref{Fig3}(c)] in selected regions-of-interest (ROIs), shown as colored boxes in Fig.~\ref{Fig3}(b). The purple dots represent the IR/XUV cross-correlation extracted from the time-dependent laser-assisted photoemission signal (ROI not shown, cross-correlation shown in Supplementary Figure 6), with an FWHM duration of 171~fs (Gaussian fit, yellow line). The dark red curve represents hot electron populations in Ti 3d conduction band, showing a fast light-induced rise followed by a single exponential decay behavior. The timescale of this relaxation process is strongly dependent on the excess energy of these hot electrons, as shown in the Supplementary Figure 1 in Supplementary Information (SI). As discussed in the context of the analysis of the CECs, the black ROI is spectrally congested, with some overlapping contribution of both the bottom of the Ti 3d conduction band and the top of the Se 4p CDW backfolded band. The dynamics in this ROI feature a fast decay (experimental-resolution-limited) associated with the melting of CDW fluctuations, followed by a refilling and subsequent decay associated with the intraband hot electron dynamics.

To track the CDW fluctuation melting and recovery dynamics in a background-free manner, i.e., in an energy-momentum area where there is no overlap with the spectral feature from the normal phase, we look at the time-resolved photoemission intensity in the blue ROI, which emerges solely due to CDW-fluctuation-driven backfolded bands. The dynamics in this ROI is characterized by an ultrafast melting on a timescale limited by our temporal resolution ($\sim$170~fs), followed by a recovery on a sub-picosecond timescale, i.e., $\tau \sim 700$\,fs. On top of the sub-picosecond recovery dynamics, some (low contrast) periodic modulations of the photoemission intensity can be observed with a long lifetime. To isolate the contribution from these periodic oscillations, we subtracted the single exponential fitted function from the experimental signal, yielding the residual intensity shown in Fig.~\ref{Fig3}(d). By performing Fourier analysis, we found that the CDW fluctuations recovery dynamics is periodically modulated by a mode around 3.5~THz. This mode was previously observed in time-resolved photoemission measurements in 1T-TiSe$_2$ below $T_{\rm CDW}$\,\cite{porer2014nonthermal,monney2016revealing,hedayat2019exc,duan2021optical} and has been identified as the $A_{1g}$ CDW amplitude phonon mode corresponding to displacements along the CDW periodic lattice distortions. This represents the first time-resolved photoemission observation of coherent amplitude phonon mode surviving in the CDW-fluctuations regime at a temperature strongly exceeding the transition temperature and characterized by an unusually long lifetime. 

Another interesting feature in the time-resolved photoemission signal is evident in panel \ref{Fig3}(e), where we compare the dynamics of the backfolded Se 4p band with the time evolution of the electronic temperature (extracted from the Fermi-Dirac distribution fitting) and of the total excited state population taken from the yellow ROI in Fig.\,\ref{Fig3}(b), representing the excitation and thermalization of Ti 3d states in a wider energy and momentum region. Note that the increase in electronic temperature and excited-state population observed at negative time delays is attributed to a weak prepulse extending up to a few hundred femtoseconds in the temporal profile of the infrared pump pulses. More importantly, the minimum of the backfolded band intensity is offset from the maximum of the electronic temperature, while it approximately follows the early dynamics of the excited state population of Ti 3d bands. This comparison suggests a mechanism beyond the excitonic insulator scenario, considering that the excitonic order is expected to be melted when the electron temperature reaches its maximum.  Namely, exciton, if it exists, would be formed from the holes in Se 4p and electrons in Ti 3d bands, and, therefore, its quenching should be correlated with the total photo-excited carrier density, which can be described with the effective temperature. Furthermore, it is expected that the suppression of the excitonic mode is very fast and that it should correspond to the fastest signal, however, the RT-CDW signal suppression is offset from $\Delta t=0$ by approximately $140$\,fs.

The concomitant observation of partial melting under relatively high-fluence condition (0.95 mJ/cm$^2$), the finite delay between the peak of transient electronic temperature and quenching of the backfolded Se 4p band, as well as coherent CDW amplitude phonon mode modulating the recovery dynamics, supports the idea of the dominant phonon contribution in the assumed hybrid exciton and electron-phonon mechanisms at such elevated temperatures. 
The idea of dominant phonon contribution is further supported by a recent study~\cite{buchberger2025} which shows that while the band gap is sensitive to substrate screening, the temperature-dependent evolution of its electronic structure remains unaffected, indicating that excitons are not a primary mechanism that drives the CDW transition in TiSe$_2$.

\begin{figure*}[t]
\begin{center}
\includegraphics[width=17cm]{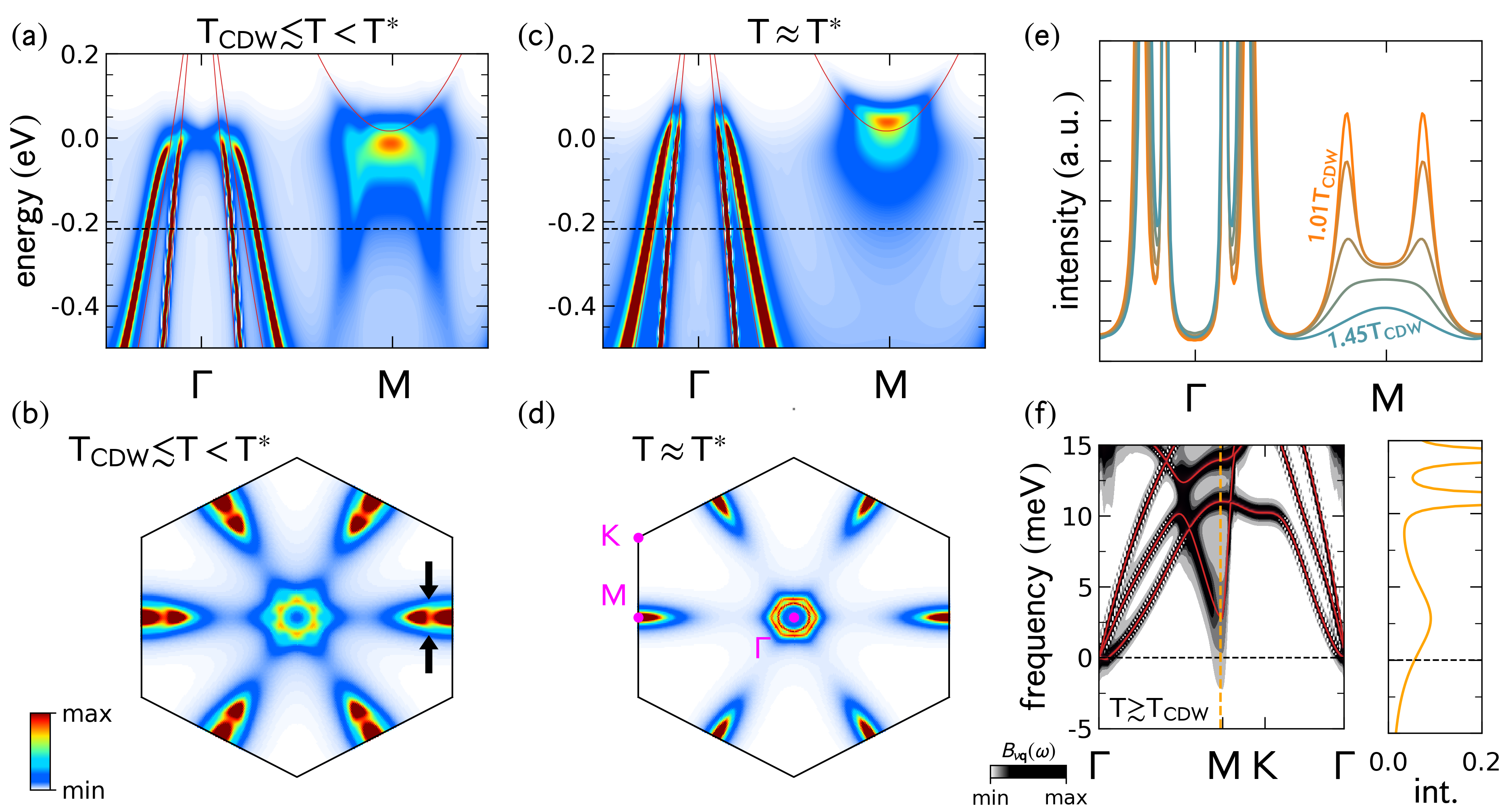}
\caption{\textbf{Signatures of charge-density-wave (CDW) fluctuations from dynamical electron-phonon scattering.} \textbf{(a)} Electron spectral function of 1T-TiSe$_2$ along $\Gamma$-M high symmetry direction with dynamical electron-phonon interaction included for temperature just above the CDW transition temperature $T_{\rm CDW}$ (i.e., $T=1150$\,K). The theoretical value for the transition temperature as obtained with harmonic density functional perturbation theory (DFPT) is $T_{\rm CDW}=1105$\,K. \textbf{(b)} The corresponding constant energy contour, where black arrows show the loss of spectral weight within the M pocket. \textbf{(c)-(d)} Same as \textbf{(a)-(b)} but close to the temperature $T^{\ast}$, i.e. when CDW fluctuations are diminished. Our theoretical calculations give $T^{\ast}\approx 1600$\,K. Red lines represent the bare bands without dynamical electron-phonon coupling included. \textbf{(e)} MDC cuts for several temperatures above the $T_{\rm CDW}$ for energy well below the Ti-3d conduction band [see dashed line in panels \textbf{(a)} and \textbf{(b)}]. \textbf{(f)} Phonon spectral function $B_{\nu\mathbf{q}}(\omega)$ along the $\Gamma$-M-K-$\Gamma$ high-symmetry path, with dynamical electron-phonon coupling included. The temperature is 50\,K above $T_{\rm CDW}$ (i.e., $T=1150$\,K). The red curve shows the phonon dispersions without including spectral broadening due to electron-phonon scattering. The right panel shows the intensity profile of the phonon spectral function at the M point (orange dashed line in the left panel). The colorbar represents spectral weight.}
\label{Fig4}
\end{center}
\end{figure*}

\section*{Discussion}

To further elaborate on the physical mechanism of CDW-fluctuation melting, two possible suppression pathways can be considered: (i) incoherent suppression, in which coherence is abruptly destroyed by hot electrons on a fast timescale ($10$--$100$~fs), and (ii) coherent suppression, mediated by the excitation of the amplitude mode. As previously noted, the comparison of timescales for the maximum suppression of the backfolded band in Fig.~\ref{Fig3}(e) reveals a near coincidence with the peak in the excited-state population above the Fermi level. Both are delayed relative to the peak in electronic temperature by approximately $140$~fs, which closely corresponds to half the period of the amplitude mode oscillation. While it is likely that both coherent and incoherent mechanisms contribute to the suppression, this characteristic time delay—matching half an amplitude mode cycle—indicates a dominant role for coherent suppression.

Further insight can be gained by comparing the time-resolved dynamics of the backfolded band with the timescale of signal suppression at the M point in RT ultrafast electron diffraction\,\cite{otto2021epc}. It was shown that the diffuse signal at M point is quenched within the first 130\,fs (limited by the instrument response function), and subsequently recovered within the time scale of 700\,fs, which remarkably follows the melting time scales of backfolded Se 4p band observed here.
The latter quench of the M point diffuse intensity was attributed to hardening of the M acoustic phonon mode by the photo-excited carriers, following the same time dynamics. This directly relates the quenching of the CDW fluctuating band with the hardening of the CDW-related M acoustic phonon.

Another recent ultrafast low-energy electron diffuse scattering study investigated the melting process of charge order in TiSe$_2$ and estimated roughly
30:70 ratio of excitonic versus electron-phonon contributions to the total
lattice distortion in the low-temperature CDW phase\,\cite{kurtz2024nontherm}. These findings highlight that both excitonic effects and electron-phonon coupling play significant roles in stabilizing the CDW order.
While short-range CDW fluctuations are a dynamical phenomenon—fluctuating over both time and space—the mechanisms that govern their stabilization and ultrafast dynamics are similar to those in conventional low-temperature CDW phases. These short-range CDW fluctuations and their coexistence with some normal (non-CDW) domains can explain the low contrast of the coherent oscillations associated with the CDW amplitude phonon mode. Indeed, in low-temperature standard CDW phases, high-contrast modulations were measured by tr-ARPES\,\cite{monney2016revealing,hedayat2019exc,duan2021optical}.

\textbf{Theoretical Analysis of Dynamical Electron-Phonon Interactions.}
To further disentangle the influence of the electron-phonon coupling on the bare electronic structure above the transition temperature, we turn our attention to the theoretical framework that includes electron self-energy corrections necessary for the inclusion of electron-phonon interaction. Notably, no term responsible for the electron-hole coupling was used here. Yet, the spectral function strikingly matches the experimentally measured features that we attribute to the CDW-fluctuating phase.
In Fig.\,\ref{Fig4}, we show the impact of dynamical electron-phonon scattering on the electron and phonon properties of 1T-TiSe$_2$ when the system is just above the structural instability $T>T_{\rm CDW}$. Temperatures used in the theoretical simulations are scaled with respect to the $T_{\rm CDW}$ value from density functional perturbation theory (DFPT), which is 1105\,K. The DFPT $T_{\rm CDW}$ is the electronic temperature at which the acoustic mode at the M point becomes soft in the harmonic Born-Oppenheimer approximation. Panels (a)-(d) show the electron spectral functions along the $\Gamma$-M high symmetry direction and CECs (at Fermi energy) in the first Brillouin zone. The results for a temperature above but close to the CDW instability $T_{\rm CDW}$ are depicted in Fig.\,\ref{Fig4} (a) and (b), while the other two panels are for $T=T^{\ast}\gg T_{\rm CDW}$, when the CDW fluctuating replica are gone and system enters into the normal phase. The electron-phonon interaction is included into the spectral function by using the dynamical Fan-Migdal electron self-energy $\Sigma_{n\mathbf{k}}(\omega)$\,\cite{giustino2017epc,lee2023epw} based on inputs derived within the density functional perturbation theory\,\cite{baroni2001dfpt,giannozzi2017qe}. Even though we use the original unit cell with no CDW periodic lattice distortions, we observe strong renormalizations of electron bands for $T\gtrsim T_{\rm CDW}$ very similar to the characteristic modifications observed in the CDW state, i.e., (i) finite spectral weight below the Ti 3d conduction band at the M point, coming from replicas of Se 4p states at the $\Gamma$ point, (ii) the flower-shaped contour at the $\Gamma$ point close to the Fermi level coming from the replicas of the three non-equivalent Ti 3d M pockets, and (iii) the loss of spectral weight at the particular momentum of the M pocket close to the Fermi level [see Fig.~\ref{Fig1} (c)]. These signatures come from the electron-phonon fluctuations and very much resemble the CDW fluctuations observed in our ARPES data at RT. With the rise of the temperature, these electron-phonon fluctuations are reduced [see Fig.\,\ref{Fig4}(e) and Supplementary Figure 2 in SI], and the electronic band structure becomes closer to the bare DFT result [red curves in Fig.\,\ref{Fig4}(c)]. Note that the bimodal intensity profile in our simulations disappears continuously, while it is completely vanished above $T^{\ast}$=1600\,K, which is 500\,K above the theoretical $T_{\rm CDW}$. We observe that these features are related to the soft acoustic phonon mode at the M point [see Fig.\ref{Fig4}(f)] and are strongest when the energy of this mode is close to zero, while the electron-phonon coupling is the strongest (see Supplementary Figure 2 in SI). When the temperature is elevated, the M-point acoustic phonon hardens, which quenches the electron-phonon scatterings between Se 4p and Ti 3d states. This further supports the correlation discussed above between the melting and recovery dynamics of the CDW fluctuating Se 4p band and hardening of the M phonon mode\,\cite{otto2021epc}.
Similar microscopic explanation, which includes an active soft bosonic mode, was used many years ago to explain the fluctuating (or so-called pseudogap) region above the CDW phase\,\cite{sadovskii2001pg} in the quasi-1D Peierls systems\,\cite{lee1973fluct,mckenzie1995pg}, as well as in the high-$T_c$ cuprates\,\cite{pines1997hightc}, and more recently in metallic 2H-NbSe$_2$ and 2H-TaSe$_2$\,\cite{kuchinskii2012tmds}. Further, spectral weight redistribution around the Fermi level due to electron-phonon-related fluctuations for $T\gtrsim T_{\rm CDW}$ was once discussed for 1T-TiSe$_2$\,\cite{yoshiyama1986tise2}, but it was never thoroughly appreciated and later connected to the replicas and other photoemission signatures in 1T-TiSe$_2$.
Our microscopic interpretation is also in line with the observation that the CDW fluctuations reach a maximum at $T_{\rm CDW}$, as observed with X-ray thermal diffuse scattering\,\cite{holt2001x}. The temperature dependence of the phonon scattering lifetime characterized by hardening of the soft phonon mode with increasing temperature and a concomitant reduction of the electron scattering rates (i.e., $\mathrm{Im}\,\Sigma$), consistent with transport measurements\,\cite{velebit2016scatt}, is shown in Supplementary Figures 2 and 3. It shows the proportionality between the larger CDW gap and the longer lifetime of the amplitude mode. This is most likely caused by less quasiparticle damping and suggests that even in the short-range ordered CDW regions, there is a finite CDW gap that suppresses the damping of the coherent (and re-hardened) amplitude mode oscillation. Notably, the figures also show that artificial introduction of small periodic lattice distortions leads to an increase in the amplitude mode lifetime and better lifetime agreement with the experiment. However, the dynamical electron-phonon scattering alone redistributes the density of states enough to account for a pseudogap formation, as shown in Supplementary Figure 3 in SI. We therefore conclude that while static CDW gaps vanish above the transition temperature, the dynamic pseudogap persists. This affects both the electronic scattering and phonon suppression rates.

Figure \ref{Fig4}(f) additionally shows the phonon spectral function of 1T-TiSe$_2$ in the fluctuation regime with included phonon-electron coupling, where part of the M phonon spectral weight extends up to negative frequencies. This indicates that in the fluctuation region where $T_{\rm CDW}<T<T^{\ast}$, the atomic motions fluctuate between normal and CDW states. Note that equilibrium electron diffraction measurements\,\cite{cheng2022light} as well as X-ray diffuse scattering\,\cite{woo1976lattice,guo2025xrd} observed both L and M diffusive peaks at RT, suggesting the presence of the fluctuating periodic lattice distortions above $T_{\rm CDW}$.
In Supplementary Figure 4 we demonstrate how the dynamical electron-phonon scattering introduces very similar modification to the electronic structure as the symmetry-breaking periodic lattice distortions in the CDW state.
The system could, therefore, support the excitation of the CDW amplitude mode at $T>T_{\rm CDW}$, as it is observed experimentally following optical photoexcitation [Fig.~\ref{Fig3}(f)]. These theoretical results demonstrate that the features (i)-(iii) as well as the CDW amplitude mode observed in the CDW fluctuation regime could be explained by the dynamical electron-phonon effects (also supported by our scanning tunneling microscopy measurements - see Supplementary Figure 5 in SI), which opposes the common belief that the fluctuation regime is dominated by the electron-hole correlations\,\cite{monney2012electron,monney2015elcorr}. The latter mechanism could still be present in the formation process of the CDW fluctuations in 1T-TiSe$_2$, but more likely as a higher-order correcting effect. This is further corroborated by the fact that the equivalent electron-related collective soft mode (e.g., plasmon or exciton), which could produce the above features in scattering with Ti 3d or Se 4p states, is not observed in the normal phase above $T_{\rm CDW}$, neither in experiments\,\cite{lin2022dramatic,kogar2017sign} nor theoretical studies\,\cite{chao2019hampers}.

In conclusion, our study provides direct evidence of charge density wave (CDW) fluctuations in 1T-TiSe$_2$ at room temperature using time-resolved extreme ultraviolet (XUV) momentum microscopy. We observe the persistence of the backfolded Se 4p band well above $T_{\rm CDW}$, demonstrating that CDW fluctuations remain detectable even at elevated temperatures. The ultrafast melting and recovery of these fluctuations, featuring the presence of coherent CDW amplitude phonon modes, highlight the importance of electron-phonon interactions in stabilizing the CDW phase at elevated temperatures.

Our experimental results, combined with density functional theory (DFT) calculations incorporating electron self-energy corrections, show that the electron-phonon interaction remains the dominant contributor to the CDW mechanism at room temperature. The observation of coherent amplitude phonon oscillations further supports this assertion, challenging the notion that the CDW fluctuating phase is purely driven by excitonic correlations. Instead, our findings suggest that while electron-hole coupling could also play a role, electron-phonon interactions are essential for the persistence of CDW fluctuations above $T_{\rm CDW}$ with strong implications for the low-temperature CDW phase formation.

These results not only refine the understanding of CDW physics in 1T-TiSe$_2$ but also have broader consequences for the study of fluctuating phases in other transition metal dichalcogenides and strongly correlated materials. Given the emerging connection between CDW fluctuations and unconventional superconductivity, our findings may provide valuable insights into the interplay between these two phenomena. Future studies leveraging advanced ultrafast techniques could further elucidate the microscopic mechanisms governing the fluctuating CDW phase, potentially unveiling new pathways for manipulating correlated electron systems in quantum materials.

\section*{Methods}

\textbf{Experimental Methods.} The experimental apparatus is centered around a custom-designed, polarization-tunable ultrafast XUV beamline~\cite{Comby22}, interfaced with a Momentum Microscope end-station~\cite{Medjanik17}. This beamline is driven by a commercial high-repetition-rate Yb fiber laser (1030 nm, 135 fs FWHM, 50 W) from Amplitude Laser Group, operating at 166 kHz. In the probe branch, a segment of the laser output is frequency-doubled using a BBO crystal, yielding 515 nm pulses (5 W) that are tightly focused into a thin, high-density argon gas jet configured in an annular geometry to facilitate high-order harmonic generation (HHG). Spatial filtering through pinholes effectively separates the intense 515 nm fundamental from the emerging XUV beamlet. To extract the 9th harmonic (21.6 eV), a combination of SiC/Mg multilayer mirrors (NTTAT) and a 200 nm Sn metallic filter is used for spectral selection. On the pump side, a small portion of the original 1030 nm laser pulse (135 fs FWHM) is directed as the IR excitation. The IR pump and XUV probe pulses are recombined collinearly via a drilled mirror and focused onto the target. Bulk iodine vapor transport-grown
$\mathrm{1T}$-$\mathrm{TiSe_2}$ single crystals~\cite{watson2019orbital} are freshly cleaved and positioned on a motorized 6-axis hexapod for high-precision alignment within the main vacuum chamber, which maintains an ultra-high base pressure of 2$\times$10$^{-10}$ mbar. Photoemission data acquisition is carried out with a custom-designed time-of-flight momentum microscope featuring an advanced front lens capable of multiple operating modes (GST mbH)~\cite{tkach2024multimode}. This detection system enables simultaneous mapping of the entire surface Brillouin zone over a broad binding energy range without necessitating sample orientation adjustments~\cite{Medjanik17}. Measurements in this study were conducted at room temperature, where the total energy resolution reaches \(113~\text{meV}\). This resolution, \(\Delta E_{\text{tot}}\), results from the convolution of the Fermi-Dirac distribution width at a given temperature \(T\), the bandwidth of the ionizing radiation \(\Delta E_{h\nu}\), and the time-of-flight detector resolution \(\Delta E_{\text{ToF}}\). The thermal broadening of the Fermi-Dirac distribution is estimated as \(\Delta E_{\text{therm}} \approx 4kT\), corresponding to approximately \(100~\text{meV}\) at room temperature.  For data post-processing, an open-source workflow~\cite{Xian20, Xian19_2} is employed to efficiently convert raw single-event-based datasets into calibrated, binned data hypervolumes, incorporating appropriate axis calibration. Further details about the experimental setup are provided in Ref.~\cite{Fragkos2025-ob}.

For STM, the surface was cleaved \textit{in situ} and directly measured at room temperature in ultrahigh vacuum with a base pressure below $1 \times 10^{-10}$~mbar. STM measurements were performed with an Omicron LT-STM instrument and performed in constant current mode by applying a bias voltage to the sample.

The native defects observed with STM on the sample perfectly match those observed in \cite{arpaia2019dyn}, with an estimated defect density of $\sim 0.35 \pm 0.1~\%$. Only very few defects of type \textit{D} associated to intercalated Ti atoms could be observed (density $< 0.01\%$), validating the low Ti doping of the sample. No signs of $2 \times 2$ CDW modulation have been observed with STM even close to $E_F$ [Supplementary Figures 5(a) and 5(b) in SI]. Indeed, the FFT-amplitude plot [Supplementary Figure 5(c)] only shows spots corresponding to the $1 \times 1$ unit cell charge modulation.

\textbf{Computational Methods.}  Ground state calculations were performed within the framework of the DFT using the \textsc{Quantum Espresso} package\,\cite{giannozzi2017qe}. We use optimized norm-conserving Vanderbilt pseudopotentials\,\cite{hamann13} with the PBE exchange-correlation functional\,\cite{pbe} from the PseudoDojo library\,\cite{pseudodojo}.
The plane wave energy cutoff was fixed to $80$\,Ry. To reduce the interaction between the periodic images we modeled the 2D layers in periodic cells with $20$\,\AA~vacuum in the direction normal to the layer.
The convergence criterion for energy is set to 10$^{-8}$\,eV and the atomic positions are relaxed until the Hellmann-Feynman forces are less than 10$^{-4}$\,eV/\AA. A lattice parameter of $a=3.537\,\text{\AA}$ is obtained with the PBE functional. The relaxed atomic positions in fractional (crystal) coordinates are Ti (0.0000000000, 0.0000000000, 0.5000000000), Se (0.3333333333, 0.6666666667, 0.4163666236), and Se (0.6666666667, 0.3333333333, 0.5836333764).
To obtain the ground state charge density the 2D Brillouin zone is sampled with $48 \times 48 \times 1$ k-point mesh. In order to determine the $T_{\rm CDW}$, i.e., the temperature at which the relevant acoustic phonon mode becomes unstable, we use the Fermi-Dirac smearing functions with the electronic temperatures $T_{\rm el}$ ranging from 100\,K to 1600\,K, while for the calculations of the electron self-energy we range $T_{\rm el}$ from 1100\,K to 1600\,K.

For the electron-phonon-coupling calculation, we use maximally localized Wannier functions~\cite{Marzari2012}, with 11 initial projections of d orbitals on the Ti sites and p orbitals on the Se atom sites, which have been implemented in the EPW code\,\cite{lee2023epw}. The electronic structure, dynamical matrices, and electron-phonon matrix elements were obtained from DFT and density functional perturbation theory (DFPT)\,\cite{baroni2001dfpt} calculations, and they were used as the initial data for electron-phonon coupling. Initial DFPT phonon calculations were done on corse grids of $\mathbf{k}=24\times24\times 1$ and $\mathbf{q}=6\times6\times 1$ for the undistorted original unit cell, as well as for various $T_{\rm el}$ mentioned above. The soft acoustic phonon mode goes to zero at $T_{\rm el}=1105$\,K. In order to get a more accurate transition temperature $T_{\rm CDW}$ one would need to include the anharmonic corrections\,\cite{zhou2020anharmonicity}.

To account for the dynamical renormalization of the electronic structure due to electron-phonon interactions we calculate the Fan-Migdal electron self-energy\,\cite{giustino2017epc}
\begin{align}
\Sigma_{n\mathbf{k}}(\omega)= \sum_{m\nu \mathbf q,s=\pm 1}\left|g^\nu_{nm} (\mathbf{k},\mathbf{q})\right|^2 \frac{n(\omega_{\nu\mathbf{q}})+f(s\varepsilon_{m\mathbf{k+q}})}{\omega + i\eta - \varepsilon_{m\mathbf{k+q}} + s\omega_{\nu\mathbf{q}}},
\label{eq:FM}
\end{align}
where $g^\nu_{nm} (\mathbf{k},\mathbf{q})$ are the electron-phonon matrix elements, $\varepsilon_{m\mathbf{k+q}}$ are the DFT electron energies, $\omega_{\nu\mathbf{q}}$ are the phonon frequencies, while $n$ and $f$ are the Bose-Einstein and Fermi-Dirac distribution functions. The fine $\mathbf{q} $ momentum grid of $400\times 400 \times 1$ was used for the summation in Eq.\,\eqref{eq:FM}, while $\eta$ was set to 30\,meV. The electron spectral function is calculated as
\begin{align}
A_{\mathbf{k}}(\omega)= \frac{1}{\pi}\sum_{n} \frac{-\mathrm{Im}\,\Sigma_{n\mathbf{k}}(\omega)}{[\omega-\varepsilon_{n\mathbf{k}}-\mathrm{Re}\,\Sigma_{n\mathbf{k}}(\omega)]^2+\mathrm{Im}\,\Sigma_{n\mathbf{k}}(\omega)^2}.
\label{eq:specfun}
\end{align}
In the CDW systems, the input parameters entering Eqs.\,\eqref{eq:FM} and \eqref{eq:specfun}, such as $g^\nu_{nm} (\mathbf{k},\mathbf{q})$ and $\omega_{\nu\mathbf{q}}$ depend on the temperature (e.g., the soft M phonon mode softens as one approaches $T_{\rm CDW}$), therefore, $A_{\mathbf{k}}(\omega)$ is calculated for several temperatures close and far from $T_{\rm CDW}$.

The dynamical phonon spectral function is calculated as
\begin{align}
        B_{\nu\mathbf{q}}(\omega)=\frac{1}{\pi}\mathrm{Im}\left[ \frac{-2\omega_{\nu\mathbf{q}}}{\omega^2-\omega_{\nu\mathbf{q}}^2-2\omega_{\nu\mathbf{q}}\pi_{\nu\mathbf{q}}(\omega)} \right],
\label{eq:phspec}
\end{align}
where $\pi_{\nu\mathbf{q}}(\omega)$ is the phonon self-energy accounting for the dynamical renormalizations of the phonons due to coupling with electrons. The real part of the phonon self-energy shifts the DFPT frequencies $\omega_{\nu\mathbf{q}}^{\rm dyn} \approx \omega_{\nu\mathbf{q}} + \text{Re}[\pi_{\nu\mathbf{q}}(\omega_{\nu\mathbf{q}}^{\rm dyn})] -  \text{Re}[\pi_{\nu\mathbf{q}}(0)]$, while the imaginary part introduces the broadening $\gamma_{\nu\mathbf{q}}= -2\text{Im}[\pi_{\nu\mathbf{q}}(\omega_{\nu\mathbf{q}}^{\rm dyn})]$. The phonon self-energy due to electron-phonon coupling is calculated as
\begin{equation}
\pi_{\nu\mathbf{q}}(\omega)=\sum_{\mathbf{k}nm}\left| g^{\nu}_{nm}(\mathbf{k},\mathbf{q}) \right|^2\frac{f(\varepsilon_{\text{n}\mathbf{k}})-f(\varepsilon_{\text{m}\mathbf{k+q}})}{\omega+i\eta+\varepsilon_{n\mathbf{k}}-\varepsilon_{m\mathbf{k+q}}}.
\label{eq:phself}
\end{equation}
The summation of the latter equation was done on $\mathbf{k}=600\times600\times 1$ momentum grid.

\section*{Data Availability}

The ARPES and DFT data that support the findings of this study are available in Zenodo under the identifier https://doi.org/10.5281/zenodo.17813602. Additional data can be obtained from the authors upon reasonable request.

\begin{acknowledgments}
We thank Claude Monney for useful discussions. We thank Nikita Fedorov, Romain Delos, Pierre Hericourt, Rodrigue Bouillaud, Laurent Merzeau, and Frank Blais for technical assistance and Aymen Mahmoudi, Joel Morf, and Frederic Chassot for their assistance during STM measurements.
J.S., S.S., and J.M. would like to thank the QM4ST project with Reg. No. \texttt{CZ.02.01.01/00/22\_008/0004572}, cofunded by the ERDF as part of the M\v{S}MT. The research leading to these results has received funding from LASERLAB-EUROPE (grant agreement no. 871124, European Union’s Horizon 2020 research and innovation programme). We acknowledge the financial support of the IdEx University of Bordeaux / Grand Research Program "GPR LIGHT". We acknowledge support from ERC Starting Grant ERC-2022-STG No.101076639, Quantum Matter Bordeaux, AAP CNRS Tremplin, and AAP SMR from Université de Bordeaux. This work is part of the ULTRAFAST and TORNADO projects of PEPR LUMA and was supported by the French National Research Agency, as a part of the France 2030 program, under grants ANR-23-EXLU-0002 and ANR-23-EXLU-0004. SF acknowledges funding from the European Union’s Horizon Europe research and innovation programme under the Marie Skłodowska-Curie 2024 Postdoctoral Fellowship No 101198277 (TopQMat). JS acknowledges funding from the European Union’s Horizon Europe research and innovation programme under the Marie Skłodowska‑Curie grant agreement No 101209345—ART.QM funded by the European Union. Views and opinions expressed are however those of the author(s) only and do not necessarily reflect those of the European Union or European Research Executive Agency. Neither the European Union nor the granting authority can be held responsible for them. This work was funded by the Würzburg-Dresden Cluster of Excellence on Complexity, Topology and Dynamics in Quantum Matter – ctd.qmat (EXC 2147, project-id 390858490) and by the DFG through SFB1170 “Tocotronics” and directly via RE 1469/13-2. N.G.E., and D.N. acknowledge financial support from the project "Podizanje znanstvene izvrsnosti Centra za napredne laserske tehnike (CALTboost)" financed by the European Union through the National Recovery and Resilience Plan 2021-2026 (NRPP), European Regional Development Fund for the project ‘Materials for clean energy, advanced sensors and quantum technologies’ (Grant No. PK.1.1.10.0002), and Croatian Science Foundation (Grant no. IP-2025-02-5926). QC also acknowledges support from the TERAQUANTUM project of the Région Nouvelle-Aquitaine. This work was supported by the TWISTnSHINE , funded as project No. LL2314 by Programme ERC CZ.

\end{acknowledgments}

\section*{Author contributions}
S.F., H.O., S.B., and J.S. performed the experiments, supported by S.S., A.J., Q.C., Y.M., and M.P.T.M.. S.F., H.O., S.B., and J.S. analyzed the experimental data.  D.D. and S.P. participated in maintaining the laser system. N.G.E. and D.N. performed the theoretical calculations. F.D. and K.R. synthesized and characterized the samples. B.H. and J.S. performed the STM measurements. S.F., H.O., N.G.E., A.J., S.S., Q.C., M.P.T.M., M.Ü., F.D., B.H., D.D., S.P., F.B., J.M., Y.M., F.R., K.R., D.N., S.B., J.S. contributed to the interpretation and discussion of the results. J.S., S.B., D.N., H.O., and S.F. wrote the manuscript with inputs from coauthors.
J.S., S.B., and D.N. conceived and planned the project.

\section*{Competing interests}

The authors declare no competing interests.


\bibliography{TiSe2_CDW} 

\begin{thebibliography}{82}%
\makeatletter
\providecommand \@ifxundefined [1]{%
 \@ifx{#1\undefined}
}%
\providecommand \@ifnum [1]{%
 \ifnum #1\expandafter \@firstoftwo
 \else \expandafter \@secondoftwo
 \fi
}%
\providecommand \@ifx [1]{%
 \ifx #1\expandafter \@firstoftwo
 \else \expandafter \@secondoftwo
 \fi
}%
\providecommand \natexlab [1]{#1}%
\providecommand \enquote  [1]{``#1''}%
\providecommand \bibnamefont  [1]{#1}%
\providecommand \bibfnamefont [1]{#1}%
\providecommand \citenamefont [1]{#1}%
\providecommand \href@noop [0]{\@secondoftwo}%
\providecommand \href [0]{\begingroup \@sanitize@url \@href}%
\providecommand \@href[1]{\@@startlink{#1}\@@href}%
\providecommand \@@href[1]{\endgroup#1\@@endlink}%
\providecommand \@sanitize@url [0]{\catcode `\\12\catcode `\$12\catcode `\&12\catcode `\#12\catcode `\^12\catcode `\_12\catcode `\%12\relax}%
\providecommand \@@startlink[1]{}%
\providecommand \@@endlink[0]{}%
\providecommand \url  [0]{\begingroup\@sanitize@url \@url }%
\providecommand \@url [1]{\endgroup\@href {#1}{\urlprefix }}%
\providecommand \urlprefix  [0]{URL }%
\providecommand \Eprint [0]{\href }%
\providecommand \doibase [0]{http://dx.doi.org/}%
\providecommand \selectlanguage [0]{\@gobble}%
\providecommand \bibinfo  [0]{\@secondoftwo}%
\providecommand \bibfield  [0]{\@secondoftwo}%
\providecommand \translation [1]{[#1]}%
\providecommand \BibitemOpen [0]{}%
\providecommand \bibitemStop [0]{}%
\providecommand \bibitemNoStop [0]{.\EOS\space}%
\providecommand \EOS [0]{\spacefactor3000\relax}%
\providecommand \BibitemShut  [1]{\csname bibitem#1\endcsname}%
\let\auto@bib@innerbib\@empty
\bibitem [{\citenamefont {Di~Salvo}\ \emph {et~al.}(1976)\citenamefont {Di~Salvo}, \citenamefont {Moncton},\ and\ \citenamefont {Waszczak}}]{di1976electronic}%
  \BibitemOpen
  \bibfield  {author} {\bibinfo {author} {\bibfnamefont {F.~J.}\ \bibnamefont {Di~Salvo}}, \bibinfo {author} {\bibfnamefont {D.~E.}\ \bibnamefont {Moncton}}, \ and\ \bibinfo {author} {\bibfnamefont {J.~V.}\ \bibnamefont {Waszczak}},\ }\bibfield  {title} {\enquote {\bibinfo {title} {Electronic properties and superlattice formation in the semimetal ${\mathrm{tise}}_{2}$},}\ }\href {\doibase 10.1103/PhysRevB.14.4321} {\bibfield  {journal} {\bibinfo  {journal} {Phys. Rev. B}\ }\textbf {\bibinfo {volume} {14}},\ \bibinfo {pages} {4321--4328} (\bibinfo {year} {1976})}\BibitemShut {NoStop}%
\bibitem [{\citenamefont {Woo}\ \emph {et~al.}(1976{\natexlab{a}})\citenamefont {Woo}, \citenamefont {Brown}, \citenamefont {McMillan}, \citenamefont {Miller}, \citenamefont {Schaffman},\ and\ \citenamefont {Sears}}]{woo1076superlattice}%
  \BibitemOpen
  \bibfield  {author} {\bibinfo {author} {\bibfnamefont {K.~C.}\ \bibnamefont {Woo}}, \bibinfo {author} {\bibfnamefont {F.~C.}\ \bibnamefont {Brown}}, \bibinfo {author} {\bibfnamefont {W.~L.}\ \bibnamefont {McMillan}}, \bibinfo {author} {\bibfnamefont {R.~J.}\ \bibnamefont {Miller}}, \bibinfo {author} {\bibfnamefont {M.~J.}\ \bibnamefont {Schaffman}}, \ and\ \bibinfo {author} {\bibfnamefont {M.~P.}\ \bibnamefont {Sears}},\ }\bibfield  {title} {\enquote {\bibinfo {title} {Superlattice formation in titanium diselenide},}\ }\href {\doibase 10.1103/PhysRevB.14.3242} {\bibfield  {journal} {\bibinfo  {journal} {Phys. Rev. B}\ }\textbf {\bibinfo {volume} {14}},\ \bibinfo {pages} {3242--3247} (\bibinfo {year} {1976}{\natexlab{a}})}\BibitemShut {NoStop}%
\bibitem [{\citenamefont {Rossnagel}(2011)}]{Rossnagel_2011}%
  \BibitemOpen
  \bibfield  {author} {\bibinfo {author} {\bibfnamefont {K}~\bibnamefont {Rossnagel}},\ }\bibfield  {title} {\enquote {\bibinfo {title} {On the origin of charge-density waves in select layered transition-metal dichalcogenides},}\ }\href {\doibase 10.1088/0953-8984/23/21/213001} {\bibfield  {journal} {\bibinfo  {journal} {Journal of Physics: Condensed Matter}\ }\textbf {\bibinfo {volume} {23}},\ \bibinfo {pages} {213001} (\bibinfo {year} {2011})}\BibitemShut {NoStop}%
\bibitem [{\citenamefont {J\'erome}\ \emph {et~al.}(1967)\citenamefont {J\'erome}, \citenamefont {Rice},\ and\ \citenamefont {Kohn}}]{jerome1967excitonic}%
  \BibitemOpen
  \bibfield  {author} {\bibinfo {author} {\bibfnamefont {D.}~\bibnamefont {J\'erome}}, \bibinfo {author} {\bibfnamefont {T.~M.}\ \bibnamefont {Rice}}, \ and\ \bibinfo {author} {\bibfnamefont {W.}~\bibnamefont {Kohn}},\ }\bibfield  {title} {\enquote {\bibinfo {title} {Excitonic insulator},}\ }\href {\doibase 10.1103/PhysRev.158.462} {\bibfield  {journal} {\bibinfo  {journal} {Phys. Rev.}\ }\textbf {\bibinfo {volume} {158}},\ \bibinfo {pages} {462} (\bibinfo {year} {1967})}\BibitemShut {NoStop}%
\bibitem [{\citenamefont {Kidd}\ \emph {et~al.}(2002)\citenamefont {Kidd}, \citenamefont {Miller}, \citenamefont {Chou},\ and\ \citenamefont {Chiang}}]{kidd2002electron}%
  \BibitemOpen
  \bibfield  {author} {\bibinfo {author} {\bibfnamefont {T.~E.}\ \bibnamefont {Kidd}}, \bibinfo {author} {\bibfnamefont {T.}~\bibnamefont {Miller}}, \bibinfo {author} {\bibfnamefont {M.~Y.}\ \bibnamefont {Chou}}, \ and\ \bibinfo {author} {\bibfnamefont {T.-C.}\ \bibnamefont {Chiang}},\ }\bibfield  {title} {\enquote {\bibinfo {title} {Electron-hole coupling and the charge density wave transition in ${\mathrm{tise}}_{2}$},}\ }\href {\doibase 10.1103/PhysRevLett.88.226402} {\bibfield  {journal} {\bibinfo  {journal} {Phys. Rev. Lett.}\ }\textbf {\bibinfo {volume} {88}},\ \bibinfo {pages} {226402} (\bibinfo {year} {2002})}\BibitemShut {NoStop}%
\bibitem [{\citenamefont {Cercellier}\ \emph {et~al.}(2007)\citenamefont {Cercellier}, \citenamefont {Monney}, \citenamefont {Clerc}, \citenamefont {Battaglia}, \citenamefont {Despont}, \citenamefont {Garnier}, \citenamefont {Beck}, \citenamefont {Aebi}, \citenamefont {Patthey}, \citenamefont {Berger},\ and\ \citenamefont {Forr\'o}}]{cercellier2007evidence}%
  \BibitemOpen
  \bibfield  {author} {\bibinfo {author} {\bibfnamefont {H.}~\bibnamefont {Cercellier}}, \bibinfo {author} {\bibfnamefont {C.}~\bibnamefont {Monney}}, \bibinfo {author} {\bibfnamefont {F.}~\bibnamefont {Clerc}}, \bibinfo {author} {\bibfnamefont {C.}~\bibnamefont {Battaglia}}, \bibinfo {author} {\bibfnamefont {L.}~\bibnamefont {Despont}}, \bibinfo {author} {\bibfnamefont {M.~G.}\ \bibnamefont {Garnier}}, \bibinfo {author} {\bibfnamefont {H.}~\bibnamefont {Beck}}, \bibinfo {author} {\bibfnamefont {P.}~\bibnamefont {Aebi}}, \bibinfo {author} {\bibfnamefont {L.}~\bibnamefont {Patthey}}, \bibinfo {author} {\bibfnamefont {H.}~\bibnamefont {Berger}}, \ and\ \bibinfo {author} {\bibfnamefont {L.}~\bibnamefont {Forr\'o}},\ }\bibfield  {title} {\enquote {\bibinfo {title} {Evidence for an excitonic insulator phase in $1t\mathrm{\text{\ensuremath{-}}}{\mathrm{tise}}_{2}$},}\ }\href {\doibase 10.1103/PhysRevLett.99.146403} {\bibfield  {journal} {\bibinfo  {journal} {Phys. Rev. Lett.}\ }\textbf {\bibinfo {volume} {99}},\
  \bibinfo {pages} {146403} (\bibinfo {year} {2007})}\BibitemShut {NoStop}%
\bibitem [{\citenamefont {Monney}\ \emph {et~al.}(2009)\citenamefont {Monney}, \citenamefont {Cercellier}, \citenamefont {Clerc}, \citenamefont {Battaglia}, \citenamefont {Schwier}, \citenamefont {Didiot}, \citenamefont {Garnier}, \citenamefont {Beck}, \citenamefont {Aebi}, \citenamefont {Berger}, \citenamefont {Forr\'o},\ and\ \citenamefont {Patthey}}]{monney2009spontaneous}%
  \BibitemOpen
  \bibfield  {author} {\bibinfo {author} {\bibfnamefont {C.}~\bibnamefont {Monney}}, \bibinfo {author} {\bibfnamefont {H.}~\bibnamefont {Cercellier}}, \bibinfo {author} {\bibfnamefont {F.}~\bibnamefont {Clerc}}, \bibinfo {author} {\bibfnamefont {C.}~\bibnamefont {Battaglia}}, \bibinfo {author} {\bibfnamefont {E.~F.}\ \bibnamefont {Schwier}}, \bibinfo {author} {\bibfnamefont {C.}~\bibnamefont {Didiot}}, \bibinfo {author} {\bibfnamefont {M.~G.}\ \bibnamefont {Garnier}}, \bibinfo {author} {\bibfnamefont {H.}~\bibnamefont {Beck}}, \bibinfo {author} {\bibfnamefont {P.}~\bibnamefont {Aebi}}, \bibinfo {author} {\bibfnamefont {H.}~\bibnamefont {Berger}}, \bibinfo {author} {\bibfnamefont {L.}~\bibnamefont {Forr\'o}}, \ and\ \bibinfo {author} {\bibfnamefont {L.}~\bibnamefont {Patthey}},\ }\bibfield  {title} {\enquote {\bibinfo {title} {Spontaneous exciton condensation in $1t{\text{-tise}}_{2}$: Bcs-like approach},}\ }\href {\doibase 10.1103/PhysRevB.79.045116} {\bibfield  {journal} {\bibinfo  {journal} {Phys. Rev. B}\
  }\textbf {\bibinfo {volume} {79}},\ \bibinfo {pages} {045116} (\bibinfo {year} {2009})}\BibitemShut {NoStop}%
\bibitem [{\citenamefont {Monney}\ \emph {et~al.}(2010)\citenamefont {Monney}, \citenamefont {Schwier}, \citenamefont {Garnier}, \citenamefont {Mariotti}, \citenamefont {Didiot}, \citenamefont {Beck}, \citenamefont {Aebi}, \citenamefont {Cercellier}, \citenamefont {Marcus}, \citenamefont {Battaglia}, \citenamefont {Berger},\ and\ \citenamefont {Titov}}]{monney2010temperature}%
  \BibitemOpen
  \bibfield  {author} {\bibinfo {author} {\bibfnamefont {C.}~\bibnamefont {Monney}}, \bibinfo {author} {\bibfnamefont {E.~F.}\ \bibnamefont {Schwier}}, \bibinfo {author} {\bibfnamefont {M.~G.}\ \bibnamefont {Garnier}}, \bibinfo {author} {\bibfnamefont {N.}~\bibnamefont {Mariotti}}, \bibinfo {author} {\bibfnamefont {C.}~\bibnamefont {Didiot}}, \bibinfo {author} {\bibfnamefont {H.}~\bibnamefont {Beck}}, \bibinfo {author} {\bibfnamefont {P.}~\bibnamefont {Aebi}}, \bibinfo {author} {\bibfnamefont {H.}~\bibnamefont {Cercellier}}, \bibinfo {author} {\bibfnamefont {J.}~\bibnamefont {Marcus}}, \bibinfo {author} {\bibfnamefont {C.}~\bibnamefont {Battaglia}}, \bibinfo {author} {\bibfnamefont {H.}~\bibnamefont {Berger}}, \ and\ \bibinfo {author} {\bibfnamefont {A.~N.}\ \bibnamefont {Titov}},\ }\bibfield  {title} {\enquote {\bibinfo {title} {Temperature-dependent photoemission on $1t{\text{-tise}}_{2}$: Interpretation within the exciton condensate phase model},}\ }\href {\doibase 10.1103/PhysRevB.81.155104} {\bibfield
  {journal} {\bibinfo  {journal} {Phys. Rev. B}\ }\textbf {\bibinfo {volume} {81}},\ \bibinfo {pages} {155104} (\bibinfo {year} {2010})}\BibitemShut {NoStop}%
\bibitem [{\citenamefont {Monney}\ \emph {et~al.}(2012)\citenamefont {Monney}, \citenamefont {Monney}, \citenamefont {Aebi},\ and\ \citenamefont {Beck}}]{monney2012electron}%
  \BibitemOpen
  \bibfield  {author} {\bibinfo {author} {\bibfnamefont {C.}~\bibnamefont {Monney}}, \bibinfo {author} {\bibfnamefont {G.}~\bibnamefont {Monney}}, \bibinfo {author} {\bibfnamefont {P.}~\bibnamefont {Aebi}}, \ and\ \bibinfo {author} {\bibfnamefont {H.}~\bibnamefont {Beck}},\ }\bibfield  {title} {\enquote {\bibinfo {title} {Electron-hole fluctuation phase in $1t$-tise${}_{2}$},}\ }\href {\doibase 10.1103/PhysRevB.85.235150} {\bibfield  {journal} {\bibinfo  {journal} {Phys. Rev. B}\ }\textbf {\bibinfo {volume} {85}},\ \bibinfo {pages} {235150} (\bibinfo {year} {2012})}\BibitemShut {NoStop}%
\bibitem [{\citenamefont {Kogar}\ \emph {et~al.}(2017)\citenamefont {Kogar}, \citenamefont {Rak}, \citenamefont {Vig}, \citenamefont {Husain}, \citenamefont {Flicker}, \citenamefont {Joe}, \citenamefont {Venema}, \citenamefont {MacDougall}, \citenamefont {Chiang}, \citenamefont {Fradkin}, \citenamefont {van Wezel},\ and\ \citenamefont {Abbamonte}}]{kogar2017sign}%
  \BibitemOpen
  \bibfield  {author} {\bibinfo {author} {\bibfnamefont {Anshul}\ \bibnamefont {Kogar}}, \bibinfo {author} {\bibfnamefont {Melinda~S.}\ \bibnamefont {Rak}}, \bibinfo {author} {\bibfnamefont {Sean}\ \bibnamefont {Vig}}, \bibinfo {author} {\bibfnamefont {Ali~A.}\ \bibnamefont {Husain}}, \bibinfo {author} {\bibfnamefont {Felix}\ \bibnamefont {Flicker}}, \bibinfo {author} {\bibfnamefont {Young~Il}\ \bibnamefont {Joe}}, \bibinfo {author} {\bibfnamefont {Luc}\ \bibnamefont {Venema}}, \bibinfo {author} {\bibfnamefont {Greg~J.}\ \bibnamefont {MacDougall}}, \bibinfo {author} {\bibfnamefont {Tai~C.}\ \bibnamefont {Chiang}}, \bibinfo {author} {\bibfnamefont {Eduardo}\ \bibnamefont {Fradkin}}, \bibinfo {author} {\bibfnamefont {Jasper}\ \bibnamefont {van Wezel}}, \ and\ \bibinfo {author} {\bibfnamefont {Peter}\ \bibnamefont {Abbamonte}},\ }\bibfield  {title} {\enquote {\bibinfo {title} {Signatures of exciton condensation in a transition metal dichalcogenide},}\ }\href {\doibase 10.1126/science.aam6432} {\bibfield
  {journal} {\bibinfo  {journal} {Science}\ }\textbf {\bibinfo {volume} {358}},\ \bibinfo {pages} {1314} (\bibinfo {year} {2017})}\BibitemShut {NoStop}%
\bibitem [{\citenamefont {Hughes}(1977)}]{hughes1977structural}%
  \BibitemOpen
  \bibfield  {author} {\bibinfo {author} {\bibfnamefont {H~P}\ \bibnamefont {Hughes}},\ }\bibfield  {title} {\enquote {\bibinfo {title} {Structural distortion in tise2 and related materials-a possible jahn-teller effect?}}\ }\href {\doibase 10.1088/0022-3719/10/11/009} {\bibfield  {journal} {\bibinfo  {journal} {Journal of Physics C: Solid State Physics}\ }\textbf {\bibinfo {volume} {10}},\ \bibinfo {pages} {L319} (\bibinfo {year} {1977})}\BibitemShut {NoStop}%
\bibitem [{\citenamefont {Yoshida}\ and\ \citenamefont {Motizuki}(1980)}]{yoshida1980epc}%
  \BibitemOpen
  \bibfield  {author} {\bibinfo {author} {\bibfnamefont {Yukimasa}\ \bibnamefont {Yoshida}}\ and\ \bibinfo {author} {\bibfnamefont {Kazuko}\ \bibnamefont {Motizuki}},\ }\bibfield  {title} {\enquote {\bibinfo {title} {Electron lattice interaction and lattice instability of 1t-tise$_2$},}\ }\href {\doibase 10.1143/JPSJ.49.898} {\bibfield  {journal} {\bibinfo  {journal} {Journal of the Physical Society of Japan}\ }\textbf {\bibinfo {volume} {49}},\ \bibinfo {pages} {898} (\bibinfo {year} {1980})}\BibitemShut {NoStop}%
\bibitem [{\citenamefont {Whangbo}\ and\ \citenamefont {Canadell}(1992)}]{whangbo1992analogies}%
  \BibitemOpen
  \bibfield  {author} {\bibinfo {author} {\bibfnamefont {Myung~Hwan}\ \bibnamefont {Whangbo}}\ and\ \bibinfo {author} {\bibfnamefont {Enric}\ \bibnamefont {Canadell}},\ }\bibfield  {title} {\enquote {\bibinfo {title} {Analogies between the concepts of molecular chemistry and solid-state physics concerning structural instabilities. electronic origin of the structural modulations in layered transition metal dichalcogenides},}\ }\href {\doibase 10.1021/ja00050a044} {\bibfield  {journal} {\bibinfo  {journal} {Journal of the American Chemical Society}\ }\textbf {\bibinfo {volume} {114}},\ \bibinfo {pages} {9587--9600} (\bibinfo {year} {1992})}\BibitemShut {NoStop}%
\bibitem [{\citenamefont {Rossnagel}\ \emph {et~al.}(2002)\citenamefont {Rossnagel}, \citenamefont {Kipp},\ and\ \citenamefont {Skibowski}}]{rossnagel2002charge}%
  \BibitemOpen
  \bibfield  {author} {\bibinfo {author} {\bibfnamefont {K.}~\bibnamefont {Rossnagel}}, \bibinfo {author} {\bibfnamefont {L.}~\bibnamefont {Kipp}}, \ and\ \bibinfo {author} {\bibfnamefont {M.}~\bibnamefont {Skibowski}},\ }\bibfield  {title} {\enquote {\bibinfo {title} {Charge-density-wave phase transition in $1t\ensuremath{-}{\mathrm{tise}}_{2}:$ excitonic insulator versus band-type jahn-teller mechanism},}\ }\href {\doibase 10.1103/PhysRevB.65.235101} {\bibfield  {journal} {\bibinfo  {journal} {Phys. Rev. B}\ }\textbf {\bibinfo {volume} {65}},\ \bibinfo {pages} {235101} (\bibinfo {year} {2002})}\BibitemShut {NoStop}%
\bibitem [{\citenamefont {Weber}\ \emph {et~al.}(2011)\citenamefont {Weber}, \citenamefont {Rosenkranz}, \citenamefont {Castellan}, \citenamefont {Osborn}, \citenamefont {Karapetrov}, \citenamefont {Hott}, \citenamefont {Heid}, \citenamefont {Bohnen},\ and\ \citenamefont {Alatas}}]{weber2011epc}%
  \BibitemOpen
  \bibfield  {author} {\bibinfo {author} {\bibfnamefont {F.}~\bibnamefont {Weber}}, \bibinfo {author} {\bibfnamefont {S.}~\bibnamefont {Rosenkranz}}, \bibinfo {author} {\bibfnamefont {J.-P.}\ \bibnamefont {Castellan}}, \bibinfo {author} {\bibfnamefont {R.}~\bibnamefont {Osborn}}, \bibinfo {author} {\bibfnamefont {G.}~\bibnamefont {Karapetrov}}, \bibinfo {author} {\bibfnamefont {R.}~\bibnamefont {Hott}}, \bibinfo {author} {\bibfnamefont {R.}~\bibnamefont {Heid}}, \bibinfo {author} {\bibfnamefont {K.-P.}\ \bibnamefont {Bohnen}}, \ and\ \bibinfo {author} {\bibfnamefont {A.}~\bibnamefont {Alatas}},\ }\bibfield  {title} {\enquote {\bibinfo {title} {Electron-phonon coupling and the soft phonon mode in ${\mathrm{tise}}_{2}$},}\ }\href {\doibase 10.1103/PhysRevLett.107.266401} {\bibfield  {journal} {\bibinfo  {journal} {Phys. Rev. Lett.}\ }\textbf {\bibinfo {volume} {107}},\ \bibinfo {pages} {266401} (\bibinfo {year} {2011})}\BibitemShut {NoStop}%
\bibitem [{\citenamefont {Porer}\ \emph {et~al.}(2014)\citenamefont {Porer}, \citenamefont {Leierseder}, \citenamefont {M{\'{e}}nard}, \citenamefont {Dachraoui}, \citenamefont {Mouchliadis}, \citenamefont {Perakis}, \citenamefont {Heinzmann}, \citenamefont {Demsar}, \citenamefont {Rossnagel},\ and\ \citenamefont {Huber}}]{porer2014nonthermal}%
  \BibitemOpen
  \bibfield  {author} {\bibinfo {author} {\bibfnamefont {M.}~\bibnamefont {Porer}}, \bibinfo {author} {\bibfnamefont {U.}~\bibnamefont {Leierseder}}, \bibinfo {author} {\bibfnamefont {J.-M.}\ \bibnamefont {M{\'{e}}nard}}, \bibinfo {author} {\bibfnamefont {H.}~\bibnamefont {Dachraoui}}, \bibinfo {author} {\bibfnamefont {L.}~\bibnamefont {Mouchliadis}}, \bibinfo {author} {\bibfnamefont {I.~E.}\ \bibnamefont {Perakis}}, \bibinfo {author} {\bibfnamefont {U.}~\bibnamefont {Heinzmann}}, \bibinfo {author} {\bibfnamefont {J.}~\bibnamefont {Demsar}}, \bibinfo {author} {\bibfnamefont {K.}~\bibnamefont {Rossnagel}}, \ and\ \bibinfo {author} {\bibfnamefont {R.}~\bibnamefont {Huber}},\ }\bibfield  {title} {\enquote {\bibinfo {title} {Non-thermal separation of electronic and structural orders in a persisting charge density wave},}\ }\href {\doibase 10.1038/nmat4042} {\bibfield  {journal} {\bibinfo  {journal} {Nature Materials}\ }\textbf {\bibinfo {volume} {13}},\ \bibinfo {pages} {857--861} (\bibinfo {year}
  {2014})}\BibitemShut {NoStop}%
\bibitem [{\citenamefont {Monney}\ \emph {et~al.}(2016)\citenamefont {Monney}, \citenamefont {Puppin}, \citenamefont {Nicholson}, \citenamefont {Hoesch}, \citenamefont {Chapman}, \citenamefont {Springate}, \citenamefont {Berger}, \citenamefont {Magrez}, \citenamefont {Cacho}, \citenamefont {Ernstorfer},\ and\ \citenamefont {Wolf}}]{monney2016revealing}%
  \BibitemOpen
  \bibfield  {author} {\bibinfo {author} {\bibfnamefont {C.}~\bibnamefont {Monney}}, \bibinfo {author} {\bibfnamefont {M.}~\bibnamefont {Puppin}}, \bibinfo {author} {\bibfnamefont {C.~W.}\ \bibnamefont {Nicholson}}, \bibinfo {author} {\bibfnamefont {M.}~\bibnamefont {Hoesch}}, \bibinfo {author} {\bibfnamefont {R.~T.}\ \bibnamefont {Chapman}}, \bibinfo {author} {\bibfnamefont {E.}~\bibnamefont {Springate}}, \bibinfo {author} {\bibfnamefont {H.}~\bibnamefont {Berger}}, \bibinfo {author} {\bibfnamefont {A.}~\bibnamefont {Magrez}}, \bibinfo {author} {\bibfnamefont {C.}~\bibnamefont {Cacho}}, \bibinfo {author} {\bibfnamefont {R.}~\bibnamefont {Ernstorfer}}, \ and\ \bibinfo {author} {\bibfnamefont {M.}~\bibnamefont {Wolf}},\ }\bibfield  {title} {\enquote {\bibinfo {title} {Revealing the role of electrons and phonons in the ultrafast recovery of charge density wave correlations in $\mathbf{1}t\text{\ensuremath{-}}{\mathbf{tise}}_{\mathbf{2}}$},}\ }\href {\doibase 10.1103/PhysRevB.94.165165} {\bibfield  {journal}
  {\bibinfo  {journal} {Phys. Rev. B}\ }\textbf {\bibinfo {volume} {94}},\ \bibinfo {pages} {165165} (\bibinfo {year} {2016})}\BibitemShut {NoStop}%
\bibitem [{\citenamefont {Karam}\ \emph {et~al.}(2018)\citenamefont {Karam}, \citenamefont {Hu},\ and\ \citenamefont {Blake}}]{karam2018strong}%
  \BibitemOpen
  \bibfield  {author} {\bibinfo {author} {\bibfnamefont {Tony~E.}\ \bibnamefont {Karam}}, \bibinfo {author} {\bibfnamefont {Jianbo}\ \bibnamefont {Hu}}, \ and\ \bibinfo {author} {\bibfnamefont {Geoffrey~A.}\ \bibnamefont {Blake}},\ }\bibfield  {title} {\enquote {\bibinfo {title} {Strongly coupled electron–phonon dynamics in few-layer tise2 exfoliates},}\ }\href {\doibase 10.1021/acsphotonics.7b00878} {\bibfield  {journal} {\bibinfo  {journal} {ACS Photonics}\ }\textbf {\bibinfo {volume} {5}},\ \bibinfo {pages} {1228} (\bibinfo {year} {2018})}\BibitemShut {NoStop}%
\bibitem [{\citenamefont {Hedayat}\ \emph {et~al.}(2019)\citenamefont {Hedayat}, \citenamefont {Sayers}, \citenamefont {Bugini}, \citenamefont {Dallera}, \citenamefont {Wolverson}, \citenamefont {Batten}, \citenamefont {Karbassi}, \citenamefont {Friedemann}, \citenamefont {Cerullo}, \citenamefont {van Wezel}, \citenamefont {Clark}, \citenamefont {Carpene},\ and\ \citenamefont {Da~Como}}]{hedayat2019exc}%
  \BibitemOpen
  \bibfield  {author} {\bibinfo {author} {\bibfnamefont {H.}~\bibnamefont {Hedayat}}, \bibinfo {author} {\bibfnamefont {C.~J.}\ \bibnamefont {Sayers}}, \bibinfo {author} {\bibfnamefont {D.}~\bibnamefont {Bugini}}, \bibinfo {author} {\bibfnamefont {C.}~\bibnamefont {Dallera}}, \bibinfo {author} {\bibfnamefont {D.}~\bibnamefont {Wolverson}}, \bibinfo {author} {\bibfnamefont {T.}~\bibnamefont {Batten}}, \bibinfo {author} {\bibfnamefont {S.}~\bibnamefont {Karbassi}}, \bibinfo {author} {\bibfnamefont {S.}~\bibnamefont {Friedemann}}, \bibinfo {author} {\bibfnamefont {G.}~\bibnamefont {Cerullo}}, \bibinfo {author} {\bibfnamefont {J.}~\bibnamefont {van Wezel}}, \bibinfo {author} {\bibfnamefont {S.~R.}\ \bibnamefont {Clark}}, \bibinfo {author} {\bibfnamefont {E.}~\bibnamefont {Carpene}}, \ and\ \bibinfo {author} {\bibfnamefont {E.}~\bibnamefont {Da~Como}},\ }\bibfield  {title} {\enquote {\bibinfo {title} {Excitonic and lattice contributions to the charge density wave in $1t\ensuremath{-}\mathrm{TiS}{\mathrm{e}}_{2}$
  revealed by a phonon bottleneck},}\ }\href {\doibase 10.1103/PhysRevResearch.1.023029} {\bibfield  {journal} {\bibinfo  {journal} {Phys. Rev. Research}\ }\textbf {\bibinfo {volume} {1}},\ \bibinfo {pages} {023029} (\bibinfo {year} {2019})}\BibitemShut {NoStop}%
\bibitem [{\citenamefont {Lian}\ \emph {et~al.}(2020)\citenamefont {Lian}, \citenamefont {Zhang}, \citenamefont {Hu}, \citenamefont {Guan},\ and\ \citenamefont {Meng}}]{Lian2020}%
  \BibitemOpen
  \bibfield  {author} {\bibinfo {author} {\bibfnamefont {Chao}\ \bibnamefont {Lian}}, \bibinfo {author} {\bibfnamefont {Sheng~Jie}\ \bibnamefont {Zhang}}, \bibinfo {author} {\bibfnamefont {Shi~Qi}\ \bibnamefont {Hu}}, \bibinfo {author} {\bibfnamefont {Meng~Xue}\ \bibnamefont {Guan}}, \ and\ \bibinfo {author} {\bibfnamefont {Sheng}\ \bibnamefont {Meng}},\ }\bibfield  {title} {\enquote {\bibinfo {title} {Ultrafast charge ordering by self-amplified exciton–phonon dynamics in tise2},}\ }\href {\doibase 10.1038/s41467-019-13672-7} {\bibfield  {journal} {\bibinfo  {journal} {Nature Communications}\ }\textbf {\bibinfo {volume} {11}},\ \bibinfo {pages} {43} (\bibinfo {year} {2020})}\BibitemShut {NoStop}%
\bibitem [{\citenamefont {Otto}\ \emph {et~al.}(2021)\citenamefont {Otto}, \citenamefont {Pöhls}, \citenamefont {de~Cotret}, \citenamefont {Stern}, \citenamefont {Sutton},\ and\ \citenamefont {Siwick}}]{otto2021epc}%
  \BibitemOpen
  \bibfield  {author} {\bibinfo {author} {\bibfnamefont {Martin~R.}\ \bibnamefont {Otto}}, \bibinfo {author} {\bibfnamefont {Jan-Hendrik}\ \bibnamefont {Pöhls}}, \bibinfo {author} {\bibfnamefont {Laurent P.~René}\ \bibnamefont {de~Cotret}}, \bibinfo {author} {\bibfnamefont {Mark~J.}\ \bibnamefont {Stern}}, \bibinfo {author} {\bibfnamefont {Mark}\ \bibnamefont {Sutton}}, \ and\ \bibinfo {author} {\bibfnamefont {Bradley~J.}\ \bibnamefont {Siwick}},\ }\bibfield  {title} {\enquote {\bibinfo {title} {Mechanisms of electron-phonon coupling unraveled in momentum and time: The case of soft phonons in tise<sub>2</sub>},}\ }\href {\doibase 10.1126/sciadv.abf2810} {\bibfield  {journal} {\bibinfo  {journal} {Science Advances}\ }\textbf {\bibinfo {volume} {7}},\ \bibinfo {pages} {eabf2810} (\bibinfo {year} {2021})}\BibitemShut {NoStop}%
\bibitem [{\citenamefont {Cheng}\ \emph {et~al.}(2022)\citenamefont {Cheng}, \citenamefont {Zong}, \citenamefont {Li}, \citenamefont {Xia}, \citenamefont {Duan}, \citenamefont {Zhao}, \citenamefont {Li}, \citenamefont {Qi}, \citenamefont {Wu}, \citenamefont {Zhao} \emph {et~al.}}]{cheng2022light}%
  \BibitemOpen
  \bibfield  {author} {\bibinfo {author} {\bibfnamefont {Yun}\ \bibnamefont {Cheng}}, \bibinfo {author} {\bibfnamefont {Alfred}\ \bibnamefont {Zong}}, \bibinfo {author} {\bibfnamefont {Jun}\ \bibnamefont {Li}}, \bibinfo {author} {\bibfnamefont {Wei}\ \bibnamefont {Xia}}, \bibinfo {author} {\bibfnamefont {Shaofeng}\ \bibnamefont {Duan}}, \bibinfo {author} {\bibfnamefont {Wenxuan}\ \bibnamefont {Zhao}}, \bibinfo {author} {\bibfnamefont {Yidian}\ \bibnamefont {Li}}, \bibinfo {author} {\bibfnamefont {Fengfeng}\ \bibnamefont {Qi}}, \bibinfo {author} {\bibfnamefont {Jun}\ \bibnamefont {Wu}}, \bibinfo {author} {\bibfnamefont {Lingrong}\ \bibnamefont {Zhao}},  \emph {et~al.},\ }\bibfield  {title} {\enquote {\bibinfo {title} {Light-induced dimension crossover dictated by excitonic correlations},}\ }\href {\doibase https://doi.org/10.1038/s41467-022-28309-5} {\bibfield  {journal} {\bibinfo  {journal} {Nature communications}\ }\textbf {\bibinfo {volume} {13}},\ \bibinfo {pages} {963} (\bibinfo {year}
  {2022})}\BibitemShut {NoStop}%
\bibitem [{\citenamefont {Heinrich}\ \emph {et~al.}(2023)\citenamefont {Heinrich}, \citenamefont {Chang}, \citenamefont {Zayko}, \citenamefont {Rossnagel}, \citenamefont {Sivis},\ and\ \citenamefont {Ropers}}]{heinrich2023elec}%
  \BibitemOpen
  \bibfield  {author} {\bibinfo {author} {\bibfnamefont {Tobias}\ \bibnamefont {Heinrich}}, \bibinfo {author} {\bibfnamefont {Hung-Tzu}\ \bibnamefont {Chang}}, \bibinfo {author} {\bibfnamefont {Sergey}\ \bibnamefont {Zayko}}, \bibinfo {author} {\bibfnamefont {Kai}\ \bibnamefont {Rossnagel}}, \bibinfo {author} {\bibfnamefont {Murat}\ \bibnamefont {Sivis}}, \ and\ \bibinfo {author} {\bibfnamefont {Claus}\ \bibnamefont {Ropers}},\ }\bibfield  {title} {\enquote {\bibinfo {title} {Electronic and structural fingerprints of charge-density-wave excitations in extreme ultraviolet transient absorption spectroscopy},}\ }\href {\doibase 10.1103/PhysRevX.13.021033} {\bibfield  {journal} {\bibinfo  {journal} {Phys. Rev. X}\ }\textbf {\bibinfo {volume} {13}},\ \bibinfo {pages} {021033} (\bibinfo {year} {2023})}\BibitemShut {NoStop}%
\bibitem [{\citenamefont {van Wezel}\ \emph {et~al.}(2010)\citenamefont {van Wezel}, \citenamefont {Nahai-Williamson},\ and\ \citenamefont {Saxena}}]{van2010exciton}%
  \BibitemOpen
  \bibfield  {author} {\bibinfo {author} {\bibfnamefont {Jasper}\ \bibnamefont {van Wezel}}, \bibinfo {author} {\bibfnamefont {Paul}\ \bibnamefont {Nahai-Williamson}}, \ and\ \bibinfo {author} {\bibfnamefont {Siddarth~S.}\ \bibnamefont {Saxena}},\ }\bibfield  {title} {\enquote {\bibinfo {title} {Exciton-phonon-driven charge density wave in ${\text{tise}}_{2}$},}\ }\href {\doibase 10.1103/PhysRevB.81.165109} {\bibfield  {journal} {\bibinfo  {journal} {Phys. Rev. B}\ }\textbf {\bibinfo {volume} {81}},\ \bibinfo {pages} {165109} (\bibinfo {year} {2010})}\BibitemShut {NoStop}%
\bibitem [{\citenamefont {Rohwer}\ \emph {et~al.}(2011)\citenamefont {Rohwer}, \citenamefont {Hellmann}, \citenamefont {Wiesenmayer}, \citenamefont {Sohrt}, \citenamefont {Stange}, \citenamefont {Slomski}, \citenamefont {Carr}, \citenamefont {Liu}, \citenamefont {Miaja-Avila}, \citenamefont {Kallaene}, \citenamefont {Mathias}, \citenamefont {Kipp}, \citenamefont {Rossnagel},\ and\ \citenamefont {Bauer}}]{rohwer2011collapse}%
  \BibitemOpen
  \bibfield  {author} {\bibinfo {author} {\bibfnamefont {Timm}\ \bibnamefont {Rohwer}}, \bibinfo {author} {\bibfnamefont {Stefan}\ \bibnamefont {Hellmann}}, \bibinfo {author} {\bibfnamefont {Martin}\ \bibnamefont {Wiesenmayer}}, \bibinfo {author} {\bibfnamefont {Christian}\ \bibnamefont {Sohrt}}, \bibinfo {author} {\bibfnamefont {Ankatrin}\ \bibnamefont {Stange}}, \bibinfo {author} {\bibfnamefont {Bartosz}\ \bibnamefont {Slomski}}, \bibinfo {author} {\bibfnamefont {Adra}\ \bibnamefont {Carr}}, \bibinfo {author} {\bibfnamefont {Yanwei}\ \bibnamefont {Liu}}, \bibinfo {author} {\bibfnamefont {Luis}\ \bibnamefont {Miaja-Avila}}, \bibinfo {author} {\bibfnamefont {Matthias}\ \bibnamefont {Kallaene}}, \bibinfo {author} {\bibfnamefont {Stefan}\ \bibnamefont {Mathias}}, \bibinfo {author} {\bibfnamefont {Lutz}\ \bibnamefont {Kipp}}, \bibinfo {author} {\bibfnamefont {Kai}\ \bibnamefont {Rossnagel}}, \ and\ \bibinfo {author} {\bibfnamefont {Michael}\ \bibnamefont {Bauer}},\ }\bibfield  {title} {\enquote {\bibinfo {title}
  {Collapse of long-range charge order tracked by time-resolved photoemission at high momenta},}\ }\href {\doibase 10.1038/nature09829} {\bibfield  {journal} {\bibinfo  {journal} {Nature}\ }\textbf {\bibinfo {volume} {471}},\ \bibinfo {pages} {490} (\bibinfo {year} {2011})}\BibitemShut {NoStop}%
\bibitem [{\citenamefont {Hellmann}\ \emph {et~al.}(2012)\citenamefont {Hellmann}, \citenamefont {Rohwer}, \citenamefont {Kall{\"a}ne}, \citenamefont {Hanff}, \citenamefont {Sohrt}, \citenamefont {Stange}, \citenamefont {Carr}, \citenamefont {Murnane}, \citenamefont {Kapteyn}, \citenamefont {Kipp} \emph {et~al.}}]{hellmann2012time}%
  \BibitemOpen
  \bibfield  {author} {\bibinfo {author} {\bibfnamefont {S}~\bibnamefont {Hellmann}}, \bibinfo {author} {\bibfnamefont {T}~\bibnamefont {Rohwer}}, \bibinfo {author} {\bibfnamefont {M}~\bibnamefont {Kall{\"a}ne}}, \bibinfo {author} {\bibfnamefont {K}~\bibnamefont {Hanff}}, \bibinfo {author} {\bibfnamefont {C}~\bibnamefont {Sohrt}}, \bibinfo {author} {\bibfnamefont {A}~\bibnamefont {Stange}}, \bibinfo {author} {\bibfnamefont {A}~\bibnamefont {Carr}}, \bibinfo {author} {\bibfnamefont {MM}~\bibnamefont {Murnane}}, \bibinfo {author} {\bibfnamefont {HC}~\bibnamefont {Kapteyn}}, \bibinfo {author} {\bibfnamefont {L}~\bibnamefont {Kipp}},  \emph {et~al.},\ }\bibfield  {title} {\enquote {\bibinfo {title} {Time-domain classification of charge-density-wave insulators},}\ }\href {\doibase https://doi.org/10.1038/ncomms2078} {\bibfield  {journal} {\bibinfo  {journal} {Nature communications}\ }\textbf {\bibinfo {volume} {3}},\ \bibinfo {pages} {1069} (\bibinfo {year} {2012})}\BibitemShut {NoStop}%
\bibitem [{\citenamefont {Duan}\ \emph {et~al.}(2021)\citenamefont {Duan}, \citenamefont {Cheng}, \citenamefont {Xia}, \citenamefont {Yang}, \citenamefont {Xu}, \citenamefont {Qi}, \citenamefont {Huang}, \citenamefont {Tang}, \citenamefont {Guo}, \citenamefont {Luo}, \citenamefont {Qian}, \citenamefont {Xiang}, \citenamefont {Zhang},\ and\ \citenamefont {Zhang}}]{duan2021optical}%
  \BibitemOpen
  \bibfield  {author} {\bibinfo {author} {\bibfnamefont {Shaofeng}\ \bibnamefont {Duan}}, \bibinfo {author} {\bibfnamefont {Yun}\ \bibnamefont {Cheng}}, \bibinfo {author} {\bibfnamefont {Wei}\ \bibnamefont {Xia}}, \bibinfo {author} {\bibfnamefont {Yuanyuan}\ \bibnamefont {Yang}}, \bibinfo {author} {\bibfnamefont {Chengyang}\ \bibnamefont {Xu}}, \bibinfo {author} {\bibfnamefont {Fengfeng}\ \bibnamefont {Qi}}, \bibinfo {author} {\bibfnamefont {Chaozhi}\ \bibnamefont {Huang}}, \bibinfo {author} {\bibfnamefont {Tianwei}\ \bibnamefont {Tang}}, \bibinfo {author} {\bibfnamefont {Yanfeng}\ \bibnamefont {Guo}}, \bibinfo {author} {\bibfnamefont {Weidong}\ \bibnamefont {Luo}}, \bibinfo {author} {\bibfnamefont {Dong}\ \bibnamefont {Qian}}, \bibinfo {author} {\bibfnamefont {Dao}\ \bibnamefont {Xiang}}, \bibinfo {author} {\bibfnamefont {Jie}\ \bibnamefont {Zhang}}, \ and\ \bibinfo {author} {\bibfnamefont {Wentao}\ \bibnamefont {Zhang}},\ }\bibfield  {title} {\enquote {\bibinfo {title} {Optical manipulation of electronic
  dimensionality in a quantum material},}\ }\href {\doibase 10.1038/s41586-021-03643-8} {\bibfield  {journal} {\bibinfo  {journal} {Nature}\ }\textbf {\bibinfo {volume} {595}},\ \bibinfo {pages} {239} (\bibinfo {year} {2021})}\BibitemShut {NoStop}%
\bibitem [{\citenamefont {Woo}\ \emph {et~al.}(1976{\natexlab{b}})\citenamefont {Woo}, \citenamefont {Brown}, \citenamefont {McMillan}, \citenamefont {Miller}, \citenamefont {Schaffman},\ and\ \citenamefont {Sears}}]{woo1976lattice}%
  \BibitemOpen
  \bibfield  {author} {\bibinfo {author} {\bibfnamefont {K.~C.}\ \bibnamefont {Woo}}, \bibinfo {author} {\bibfnamefont {F.~C.}\ \bibnamefont {Brown}}, \bibinfo {author} {\bibfnamefont {W.~L.}\ \bibnamefont {McMillan}}, \bibinfo {author} {\bibfnamefont {R.~J.}\ \bibnamefont {Miller}}, \bibinfo {author} {\bibfnamefont {M.~J.}\ \bibnamefont {Schaffman}}, \ and\ \bibinfo {author} {\bibfnamefont {M.~P.}\ \bibnamefont {Sears}},\ }\bibfield  {title} {\enquote {\bibinfo {title} {Superlattice formation in titanium diselenide},}\ }\href {\doibase 10.1103/PhysRevB.14.3242} {\bibfield  {journal} {\bibinfo  {journal} {Phys. Rev. B}\ }\textbf {\bibinfo {volume} {14}},\ \bibinfo {pages} {3242--3247} (\bibinfo {year} {1976}{\natexlab{b}})}\BibitemShut {NoStop}%
\bibitem [{\citenamefont {Miyahara}\ \emph {et~al.}(1995)\citenamefont {Miyahara}, \citenamefont {Bando},\ and\ \citenamefont {Ozaki}}]{miyahara1995sts}%
  \BibitemOpen
  \bibfield  {author} {\bibinfo {author} {\bibfnamefont {Y}~\bibnamefont {Miyahara}}, \bibinfo {author} {\bibfnamefont {H}~\bibnamefont {Bando}}, \ and\ \bibinfo {author} {\bibfnamefont {H}~\bibnamefont {Ozaki}},\ }\bibfield  {title} {\enquote {\bibinfo {title} {Tunnelling spectroscopic study of the cdw energy gap in tise2},}\ }\href {\doibase 10.1088/0953-8984/7/13/006} {\bibfield  {journal} {\bibinfo  {journal} {Journal of Physics: Condensed Matter}\ }\textbf {\bibinfo {volume} {7}},\ \bibinfo {pages} {2553} (\bibinfo {year} {1995})}\BibitemShut {NoStop}%
\bibitem [{\citenamefont {Holt}\ \emph {et~al.}(2001)\citenamefont {Holt}, \citenamefont {Zschack}, \citenamefont {Hong}, \citenamefont {Chou},\ and\ \citenamefont {Chiang}}]{holt2001x}%
  \BibitemOpen
  \bibfield  {author} {\bibinfo {author} {\bibfnamefont {M.}~\bibnamefont {Holt}}, \bibinfo {author} {\bibfnamefont {P.}~\bibnamefont {Zschack}}, \bibinfo {author} {\bibfnamefont {Hawoong}\ \bibnamefont {Hong}}, \bibinfo {author} {\bibfnamefont {M.~Y.}\ \bibnamefont {Chou}}, \ and\ \bibinfo {author} {\bibfnamefont {T.-C.}\ \bibnamefont {Chiang}},\ }\bibfield  {title} {\enquote {\bibinfo {title} {X-ray studies of phonon softening in ${\mathrm{tise}}_{2}$},}\ }\href {\doibase 10.1103/PhysRevLett.86.3799} {\bibfield  {journal} {\bibinfo  {journal} {Phys. Rev. Lett.}\ }\textbf {\bibinfo {volume} {86}},\ \bibinfo {pages} {3799--3802} (\bibinfo {year} {2001})}\BibitemShut {NoStop}%
\bibitem [{\citenamefont {Chen}\ \emph {et~al.}(2016)\citenamefont {Chen}, \citenamefont {Chan}, \citenamefont {Fang}, \citenamefont {Mo}, \citenamefont {Hussain}, \citenamefont {Fedorov}, \citenamefont {Chou},\ and\ \citenamefont {Chiang}}]{chen2016hidden}%
  \BibitemOpen
  \bibfield  {author} {\bibinfo {author} {\bibfnamefont {P}~\bibnamefont {Chen}}, \bibinfo {author} {\bibfnamefont {Y-H}\ \bibnamefont {Chan}}, \bibinfo {author} {\bibfnamefont {X-Y}\ \bibnamefont {Fang}}, \bibinfo {author} {\bibfnamefont {S-K}\ \bibnamefont {Mo}}, \bibinfo {author} {\bibfnamefont {Z}~\bibnamefont {Hussain}}, \bibinfo {author} {\bibfnamefont {A-V}\ \bibnamefont {Fedorov}}, \bibinfo {author} {\bibfnamefont {MY}~\bibnamefont {Chou}}, \ and\ \bibinfo {author} {\bibfnamefont {T-C}\ \bibnamefont {Chiang}},\ }\bibfield  {title} {\enquote {\bibinfo {title} {Hidden order and dimensional crossover of the charge density waves in tise2},}\ }\href {\doibase https://doi.org/10.1038/srep37910} {\bibfield  {journal} {\bibinfo  {journal} {Scientific reports}\ }\textbf {\bibinfo {volume} {6}},\ \bibinfo {pages} {37910} (\bibinfo {year} {2016})}\BibitemShut {NoStop}%
\bibitem [{\citenamefont {Mizukoshi}\ \emph {et~al.}(2023)\citenamefont {Mizukoshi}, \citenamefont {Fukuda}, \citenamefont {Komori}, \citenamefont {Ishikawa}, \citenamefont {Ueno},\ and\ \citenamefont {Hase}}]{mizukoshi2023ultrafast}%
  \BibitemOpen
  \bibfield  {author} {\bibinfo {author} {\bibfnamefont {Yu}~\bibnamefont {Mizukoshi}}, \bibinfo {author} {\bibfnamefont {Takumi}\ \bibnamefont {Fukuda}}, \bibinfo {author} {\bibfnamefont {Yuta}\ \bibnamefont {Komori}}, \bibinfo {author} {\bibfnamefont {Ryo}\ \bibnamefont {Ishikawa}}, \bibinfo {author} {\bibfnamefont {Keiji}\ \bibnamefont {Ueno}}, \ and\ \bibinfo {author} {\bibfnamefont {Muneaki}\ \bibnamefont {Hase}},\ }\bibfield  {title} {\enquote {\bibinfo {title} {Ultrafast melting of charge-density wave fluctuations at room temperature in 1t-tise2 monitored under non-equilibrium conditions},}\ }\href {\doibase 10.1063/5.0153161} {\bibfield  {journal} {\bibinfo  {journal} {Applied Physics Letters}\ }\textbf {\bibinfo {volume} {122}},\ \bibinfo {pages} {243101} (\bibinfo {year} {2023})}\BibitemShut {NoStop}%
\bibitem [{\citenamefont {Zhang}\ \emph {et~al.}(2022)\citenamefont {Zhang}, \citenamefont {Ruan}, \citenamefont {Yu}, \citenamefont {Gao}, \citenamefont {Berger}, \citenamefont {Forr\'o}, \citenamefont {Watanabe}, \citenamefont {Taniguchi}, \citenamefont {Ranjbar}, \citenamefont {Belosludov}, \citenamefont {K\"uhne}, \citenamefont {Bahramy},\ and\ \citenamefont {Xi}}]{zhang2022second}%
  \BibitemOpen
  \bibfield  {author} {\bibinfo {author} {\bibfnamefont {Ruiming}\ \bibnamefont {Zhang}}, \bibinfo {author} {\bibfnamefont {Wei}\ \bibnamefont {Ruan}}, \bibinfo {author} {\bibfnamefont {Junyao}\ \bibnamefont {Yu}}, \bibinfo {author} {\bibfnamefont {Libo}\ \bibnamefont {Gao}}, \bibinfo {author} {\bibfnamefont {Helmuth}\ \bibnamefont {Berger}}, \bibinfo {author} {\bibfnamefont {L\'aszl\'o}\ \bibnamefont {Forr\'o}}, \bibinfo {author} {\bibfnamefont {Kenji}\ \bibnamefont {Watanabe}}, \bibinfo {author} {\bibfnamefont {Takashi}\ \bibnamefont {Taniguchi}}, \bibinfo {author} {\bibfnamefont {Ahmad}\ \bibnamefont {Ranjbar}}, \bibinfo {author} {\bibfnamefont {Rodion~V.}\ \bibnamefont {Belosludov}}, \bibinfo {author} {\bibfnamefont {Thomas~D.}\ \bibnamefont {K\"uhne}}, \bibinfo {author} {\bibfnamefont {Mohammad~Saeed}\ \bibnamefont {Bahramy}}, \ and\ \bibinfo {author} {\bibfnamefont {Xiaoxiang}\ \bibnamefont {Xi}},\ }\bibfield  {title} {\enquote {\bibinfo {title} {Second-harmonic generation in atomically thin
  $1t\text{\ensuremath{-}}\mathrm{Ti}{\mathrm{se}}_{2}$ and its possible origin from charge density wave transitions},}\ }\href {\doibase 10.1103/PhysRevB.105.085409} {\bibfield  {journal} {\bibinfo  {journal} {Phys. Rev. B}\ }\textbf {\bibinfo {volume} {105}},\ \bibinfo {pages} {085409} (\bibinfo {year} {2022})}\BibitemShut {NoStop}%
\bibitem [{\citenamefont {Guo}\ \emph {et~al.}(2025)\citenamefont {Guo}, \citenamefont {Kogar}, \citenamefont {Henke}, \citenamefont {Flicker}, \citenamefont {de~Juan}, \citenamefont {Sun}, \citenamefont {Khayr}, \citenamefont {Peng}, \citenamefont {Lee}, \citenamefont {Krogstad}, \citenamefont {Rosenkranz}, \citenamefont {Osborn}, \citenamefont {Ruff}, \citenamefont {Lioi}, \citenamefont {Karapetrov}, \citenamefont {Campbell}, \citenamefont {Paglione}, \citenamefont {van Wezel}, \citenamefont {Chiang},\ and\ \citenamefont {Abbamonte}}]{guo2025xrd}%
  \BibitemOpen
  \bibfield  {author} {\bibinfo {author} {\bibfnamefont {Xuefei}\ \bibnamefont {Guo}}, \bibinfo {author} {\bibfnamefont {Anshul}\ \bibnamefont {Kogar}}, \bibinfo {author} {\bibfnamefont {Jans}\ \bibnamefont {Henke}}, \bibinfo {author} {\bibfnamefont {Felix}\ \bibnamefont {Flicker}}, \bibinfo {author} {\bibfnamefont {Fernando}\ \bibnamefont {de~Juan}}, \bibinfo {author} {\bibfnamefont {Stella X.~L.}\ \bibnamefont {Sun}}, \bibinfo {author} {\bibfnamefont {Issam}\ \bibnamefont {Khayr}}, \bibinfo {author} {\bibfnamefont {Yingying}\ \bibnamefont {Peng}}, \bibinfo {author} {\bibfnamefont {Sangjun}\ \bibnamefont {Lee}}, \bibinfo {author} {\bibfnamefont {Matthew~J.}\ \bibnamefont {Krogstad}}, \bibinfo {author} {\bibfnamefont {Stephan}\ \bibnamefont {Rosenkranz}}, \bibinfo {author} {\bibfnamefont {Raymond}\ \bibnamefont {Osborn}}, \bibinfo {author} {\bibfnamefont {Jacob P.~C.}\ \bibnamefont {Ruff}}, \bibinfo {author} {\bibfnamefont {David~B.}\ \bibnamefont {Lioi}}, \bibinfo {author} {\bibfnamefont {Goran}\
  \bibnamefont {Karapetrov}}, \bibinfo {author} {\bibfnamefont {Daniel~J.}\ \bibnamefont {Campbell}}, \bibinfo {author} {\bibfnamefont {Johnpierre}\ \bibnamefont {Paglione}}, \bibinfo {author} {\bibfnamefont {Jasper}\ \bibnamefont {van Wezel}}, \bibinfo {author} {\bibfnamefont {Tai~C.}\ \bibnamefont {Chiang}}, \ and\ \bibinfo {author} {\bibfnamefont {Peter}\ \bibnamefont {Abbamonte}},\ }\bibfield  {title} {\enquote {\bibinfo {title} {In-plane anisotropy of charge density wave fluctuations in 1$t$-tise$_2$},}\ }\href {https://arxiv.org/abs/2501.09968} {\bibfield  {journal} {\bibinfo  {journal} {arXiv}\ ,\ \bibinfo {pages} {2501.09968}} (\bibinfo {year} {2025})}\BibitemShut {NoStop}%
\bibitem [{\citenamefont {Jaouen}\ \emph {et~al.}(2019)\citenamefont {Jaouen}, \citenamefont {Hildebrand}, \citenamefont {Mottas}, \citenamefont {Di~Giovannantonio}, \citenamefont {Ruffieux}, \citenamefont {Rumo}, \citenamefont {Nicholson}, \citenamefont {Razzoli}, \citenamefont {Barreteau}, \citenamefont {Ubaldini}, \citenamefont {Giannini}, \citenamefont {Vanini}, \citenamefont {Beck}, \citenamefont {Monney},\ and\ \citenamefont {Aebi}}]{Jaouen2019}%
  \BibitemOpen
  \bibfield  {author} {\bibinfo {author} {\bibfnamefont {T.}~\bibnamefont {Jaouen}}, \bibinfo {author} {\bibfnamefont {B.}~\bibnamefont {Hildebrand}}, \bibinfo {author} {\bibfnamefont {M.-L.}\ \bibnamefont {Mottas}}, \bibinfo {author} {\bibfnamefont {M.}~\bibnamefont {Di~Giovannantonio}}, \bibinfo {author} {\bibfnamefont {P.}~\bibnamefont {Ruffieux}}, \bibinfo {author} {\bibfnamefont {M.}~\bibnamefont {Rumo}}, \bibinfo {author} {\bibfnamefont {C.~W.}\ \bibnamefont {Nicholson}}, \bibinfo {author} {\bibfnamefont {E.}~\bibnamefont {Razzoli}}, \bibinfo {author} {\bibfnamefont {C.}~\bibnamefont {Barreteau}}, \bibinfo {author} {\bibfnamefont {A.}~\bibnamefont {Ubaldini}}, \bibinfo {author} {\bibfnamefont {E.}~\bibnamefont {Giannini}}, \bibinfo {author} {\bibfnamefont {F.}~\bibnamefont {Vanini}}, \bibinfo {author} {\bibfnamefont {H.}~\bibnamefont {Beck}}, \bibinfo {author} {\bibfnamefont {C.}~\bibnamefont {Monney}}, \ and\ \bibinfo {author} {\bibfnamefont {P.}~\bibnamefont {Aebi}},\ }\bibfield  {title} {\enquote
  {\bibinfo {title} {Phase separation in the vicinity of fermi surface hot spots},}\ }\href {\doibase 10.1103/PhysRevB.100.075152} {\bibfield  {journal} {\bibinfo  {journal} {Phys. Rev. B}\ }\textbf {\bibinfo {volume} {100}},\ \bibinfo {pages} {075152} (\bibinfo {year} {2019})}\BibitemShut {NoStop}%
\bibitem [{\citenamefont {Lian}\ \emph {et~al.}(2019)\citenamefont {Lian}, \citenamefont {Ali},\ and\ \citenamefont {Wong}}]{chao2019hampers}%
  \BibitemOpen
  \bibfield  {author} {\bibinfo {author} {\bibfnamefont {Chao}\ \bibnamefont {Lian}}, \bibinfo {author} {\bibfnamefont {Zulfikhar~A.}\ \bibnamefont {Ali}}, \ and\ \bibinfo {author} {\bibfnamefont {Bryan~M.}\ \bibnamefont {Wong}},\ }\bibfield  {title} {\enquote {\bibinfo {title} {Charge density wave hampers exciton condensation in $1t\text{\ensuremath{-}}{\mathrm{tise}}_{2}$},}\ }\href {\doibase 10.1103/PhysRevB.100.205423} {\bibfield  {journal} {\bibinfo  {journal} {Phys. Rev. B}\ }\textbf {\bibinfo {volume} {100}},\ \bibinfo {pages} {205423} (\bibinfo {year} {2019})}\BibitemShut {NoStop}%
\bibitem [{\citenamefont {Joe}\ \emph {et~al.}(2014)\citenamefont {Joe}, \citenamefont {Chen}, \citenamefont {Ghaemi}, \citenamefont {Finkelstein}, \citenamefont {de~la Pe{\~{n}}a}, \citenamefont {Gan}, \citenamefont {Lee}, \citenamefont {Yuan}, \citenamefont {Geck}, \citenamefont {MacDougall}, \citenamefont {Chiang}, \citenamefont {Cooper}, \citenamefont {Fradkin},\ and\ \citenamefont {Abbamonte}}]{joe2014walls}%
  \BibitemOpen
  \bibfield  {author} {\bibinfo {author} {\bibfnamefont {Y.~I.}\ \bibnamefont {Joe}}, \bibinfo {author} {\bibfnamefont {X.~M.}\ \bibnamefont {Chen}}, \bibinfo {author} {\bibfnamefont {P.}~\bibnamefont {Ghaemi}}, \bibinfo {author} {\bibfnamefont {K.~D.}\ \bibnamefont {Finkelstein}}, \bibinfo {author} {\bibfnamefont {G.~A.}\ \bibnamefont {de~la Pe{\~{n}}a}}, \bibinfo {author} {\bibfnamefont {Y.}~\bibnamefont {Gan}}, \bibinfo {author} {\bibfnamefont {J.~C.~T.}\ \bibnamefont {Lee}}, \bibinfo {author} {\bibfnamefont {S.}~\bibnamefont {Yuan}}, \bibinfo {author} {\bibfnamefont {J.}~\bibnamefont {Geck}}, \bibinfo {author} {\bibfnamefont {G.~J.}\ \bibnamefont {MacDougall}}, \bibinfo {author} {\bibfnamefont {T.~C.}\ \bibnamefont {Chiang}}, \bibinfo {author} {\bibfnamefont {S.~L.}\ \bibnamefont {Cooper}}, \bibinfo {author} {\bibfnamefont {E.}~\bibnamefont {Fradkin}}, \ and\ \bibinfo {author} {\bibfnamefont {P.}~\bibnamefont {Abbamonte}},\ }\bibfield  {title} {\enquote {\bibinfo {title} {Emergence of charge density wave
  domain walls above the superconducting dome in 1t-tise2},}\ }\href {\doibase 10.1038/nphys2935} {\bibfield  {journal} {\bibinfo  {journal} {Nature Physics}\ }\textbf {\bibinfo {volume} {10}},\ \bibinfo {pages} {421} (\bibinfo {year} {2014})}\BibitemShut {NoStop}%
\bibitem [{\citenamefont {Hinlopen}\ \emph {et~al.}(2024)\citenamefont {Hinlopen}, \citenamefont {Moulding}, \citenamefont {Broad}, \citenamefont {Buhot}, \citenamefont {Bangma}, \citenamefont {McCollam}, \citenamefont {Ayres}, \citenamefont {Sayers}, \citenamefont {Como}, \citenamefont {Flicker}, \citenamefont {van Wezel},\ and\ \citenamefont {Friedemann}}]{roemer2024lifshitz}%
  \BibitemOpen
  \bibfield  {author} {\bibinfo {author} {\bibfnamefont {Roemer D.~H.}\ \bibnamefont {Hinlopen}}, \bibinfo {author} {\bibfnamefont {Owen~N.}\ \bibnamefont {Moulding}}, \bibinfo {author} {\bibfnamefont {William~R.}\ \bibnamefont {Broad}}, \bibinfo {author} {\bibfnamefont {Jonathan}\ \bibnamefont {Buhot}}, \bibinfo {author} {\bibfnamefont {Femke}\ \bibnamefont {Bangma}}, \bibinfo {author} {\bibfnamefont {Alix}\ \bibnamefont {McCollam}}, \bibinfo {author} {\bibfnamefont {Jake}\ \bibnamefont {Ayres}}, \bibinfo {author} {\bibfnamefont {Charles~J.}\ \bibnamefont {Sayers}}, \bibinfo {author} {\bibfnamefont {Enrico~Da}\ \bibnamefont {Como}}, \bibinfo {author} {\bibfnamefont {Felix}\ \bibnamefont {Flicker}}, \bibinfo {author} {\bibfnamefont {Jasper}\ \bibnamefont {van Wezel}}, \ and\ \bibinfo {author} {\bibfnamefont {Sven}\ \bibnamefont {Friedemann}},\ }\bibfield  {title} {\enquote {\bibinfo {title} {Lifshitz transition enabling superconducting dome around a charge-order critical point},}\ }\href {\doibase
  10.1126/sciadv.adl3921} {\bibfield  {journal} {\bibinfo  {journal} {Science Advances}\ }\textbf {\bibinfo {volume} {10}},\ \bibinfo {pages} {eadl3921} (\bibinfo {year} {2024})}\BibitemShut {NoStop}%
\bibitem [{\citenamefont {Borisenko}\ \emph {et~al.}(2008)\citenamefont {Borisenko}, \citenamefont {Kordyuk}, \citenamefont {Yaresko}, \citenamefont {Zabolotnyy}, \citenamefont {Inosov}, \citenamefont {Schuster}, \citenamefont {B\"uchner}, \citenamefont {Weber}, \citenamefont {Follath}, \citenamefont {Patthey},\ and\ \citenamefont {Berger}}]{borisenko2008pg}%
  \BibitemOpen
  \bibfield  {author} {\bibinfo {author} {\bibfnamefont {S.~V.}\ \bibnamefont {Borisenko}}, \bibinfo {author} {\bibfnamefont {A.~A.}\ \bibnamefont {Kordyuk}}, \bibinfo {author} {\bibfnamefont {A.~N.}\ \bibnamefont {Yaresko}}, \bibinfo {author} {\bibfnamefont {V.~B.}\ \bibnamefont {Zabolotnyy}}, \bibinfo {author} {\bibfnamefont {D.~S.}\ \bibnamefont {Inosov}}, \bibinfo {author} {\bibfnamefont {R.}~\bibnamefont {Schuster}}, \bibinfo {author} {\bibfnamefont {B.}~\bibnamefont {B\"uchner}}, \bibinfo {author} {\bibfnamefont {R.}~\bibnamefont {Weber}}, \bibinfo {author} {\bibfnamefont {R.}~\bibnamefont {Follath}}, \bibinfo {author} {\bibfnamefont {L.}~\bibnamefont {Patthey}}, \ and\ \bibinfo {author} {\bibfnamefont {H.}~\bibnamefont {Berger}},\ }\bibfield  {title} {\enquote {\bibinfo {title} {Pseudogap and charge density waves in two dimensions},}\ }\href {\doibase 10.1103/PhysRevLett.100.196402} {\bibfield  {journal} {\bibinfo  {journal} {Phys. Rev. Lett.}\ }\textbf {\bibinfo {volume} {100}},\ \bibinfo {pages}
  {196402} (\bibinfo {year} {2008})}\BibitemShut {NoStop}%
\bibitem [{\citenamefont {Arpaia}\ \emph {et~al.}(2019)\citenamefont {Arpaia}, \citenamefont {Caprara}, \citenamefont {Fumagalli}, \citenamefont {Vecchi}, \citenamefont {Peng}, \citenamefont {Andersson}, \citenamefont {Betto}, \citenamefont {Luca}, \citenamefont {Brookes}, \citenamefont {Lombardi}, \citenamefont {Salluzzo}, \citenamefont {Braicovich}, \citenamefont {Castro}, \citenamefont {Grilli},\ and\ \citenamefont {Ghiringhelli}}]{arpaia2019dyn}%
  \BibitemOpen
  \bibfield  {author} {\bibinfo {author} {\bibfnamefont {R.}~\bibnamefont {Arpaia}}, \bibinfo {author} {\bibfnamefont {S.}~\bibnamefont {Caprara}}, \bibinfo {author} {\bibfnamefont {R.}~\bibnamefont {Fumagalli}}, \bibinfo {author} {\bibfnamefont {G.~De}\ \bibnamefont {Vecchi}}, \bibinfo {author} {\bibfnamefont {Y.~Y.}\ \bibnamefont {Peng}}, \bibinfo {author} {\bibfnamefont {E.}~\bibnamefont {Andersson}}, \bibinfo {author} {\bibfnamefont {D.}~\bibnamefont {Betto}}, \bibinfo {author} {\bibfnamefont {G.~M.~De}\ \bibnamefont {Luca}}, \bibinfo {author} {\bibfnamefont {N.~B.}\ \bibnamefont {Brookes}}, \bibinfo {author} {\bibfnamefont {F.}~\bibnamefont {Lombardi}}, \bibinfo {author} {\bibfnamefont {M.}~\bibnamefont {Salluzzo}}, \bibinfo {author} {\bibfnamefont {L.}~\bibnamefont {Braicovich}}, \bibinfo {author} {\bibfnamefont {C.~Di}\ \bibnamefont {Castro}}, \bibinfo {author} {\bibfnamefont {M.}~\bibnamefont {Grilli}}, \ and\ \bibinfo {author} {\bibfnamefont {G.}~\bibnamefont {Ghiringhelli}},\ }\bibfield  {title}
  {\enquote {\bibinfo {title} {Dynamical charge density fluctuations pervading the phase diagram of a cu-based high-<i>t</i><sub>c</sub> superconductor},}\ }\href {\doibase 10.1126/science.aav1315} {\bibfield  {journal} {\bibinfo  {journal} {Science}\ }\textbf {\bibinfo {volume} {365}},\ \bibinfo {pages} {906} (\bibinfo {year} {2019})}\BibitemShut {NoStop}%
\bibitem [{\citenamefont {Zhong}\ \emph {et~al.}(2024)\citenamefont {Zhong}, \citenamefont {Suzuki}, \citenamefont {Liu}, \citenamefont {Liu}, \citenamefont {Nie}, \citenamefont {Shi}, \citenamefont {Meng}, \citenamefont {Lv}, \citenamefont {Ding}, \citenamefont {Kanai}, \citenamefont {Itatani}, \citenamefont {Shin},\ and\ \citenamefont {Okazaki}}]{zhong2024}%
  \BibitemOpen
  \bibfield  {author} {\bibinfo {author} {\bibfnamefont {Yigui}\ \bibnamefont {Zhong}}, \bibinfo {author} {\bibfnamefont {Takeshi}\ \bibnamefont {Suzuki}}, \bibinfo {author} {\bibfnamefont {Hongxiong}\ \bibnamefont {Liu}}, \bibinfo {author} {\bibfnamefont {Kecheng}\ \bibnamefont {Liu}}, \bibinfo {author} {\bibfnamefont {Zhengwei}\ \bibnamefont {Nie}}, \bibinfo {author} {\bibfnamefont {Youguo}\ \bibnamefont {Shi}}, \bibinfo {author} {\bibfnamefont {Sheng}\ \bibnamefont {Meng}}, \bibinfo {author} {\bibfnamefont {Baiqing}\ \bibnamefont {Lv}}, \bibinfo {author} {\bibfnamefont {Hong}\ \bibnamefont {Ding}}, \bibinfo {author} {\bibfnamefont {Teruto}\ \bibnamefont {Kanai}}, \bibinfo {author} {\bibfnamefont {Jiro}\ \bibnamefont {Itatani}}, \bibinfo {author} {\bibfnamefont {Shik}\ \bibnamefont {Shin}}, \ and\ \bibinfo {author} {\bibfnamefont {Kozo}\ \bibnamefont {Okazaki}},\ }\bibfield  {title} {\enquote {\bibinfo {title} {Unveiling van hove singularity modulation and fluctuated charge order in kagome superconductor
  $\mathrm{Cs}{\mathrm{v}}_{3}\mathrm{S}{\mathrm{b}}_{5}$ via time-resolved arpes},}\ }\href {\doibase 10.1103/PhysRevResearch.6.043328} {\bibfield  {journal} {\bibinfo  {journal} {Phys. Rev. Res.}\ }\textbf {\bibinfo {volume} {6}},\ \bibinfo {pages} {043328} (\bibinfo {year} {2024})}\BibitemShut {NoStop}%
\bibitem [{\citenamefont {Liu}\ \emph {et~al.}(2025)\citenamefont {Liu}, \citenamefont {Duan}, \citenamefont {Liu}, \citenamefont {Liu}, \citenamefont {Wang}, \citenamefont {Gu}, \citenamefont {Huang}, \citenamefont {Yang}, \citenamefont {Liu}, \citenamefont {Qian}, \citenamefont {Guo},\ and\ \citenamefont {Zhang}}]{liu2025fluct}%
  \BibitemOpen
  \bibfield  {author} {\bibinfo {author} {\bibfnamefont {Haoran}\ \bibnamefont {Liu}}, \bibinfo {author} {\bibfnamefont {Shaofeng}\ \bibnamefont {Duan}}, \bibinfo {author} {\bibfnamefont {Xiangqi}\ \bibnamefont {Liu}}, \bibinfo {author} {\bibfnamefont {Zhihua}\ \bibnamefont {Liu}}, \bibinfo {author} {\bibfnamefont {Shichong}\ \bibnamefont {Wang}}, \bibinfo {author} {\bibfnamefont {Lingxiao}\ \bibnamefont {Gu}}, \bibinfo {author} {\bibfnamefont {Jiongyu}\ \bibnamefont {Huang}}, \bibinfo {author} {\bibfnamefont {Wenxuan}\ \bibnamefont {Yang}}, \bibinfo {author} {\bibfnamefont {Jianzhe}\ \bibnamefont {Liu}}, \bibinfo {author} {\bibfnamefont {Dong}\ \bibnamefont {Qian}}, \bibinfo {author} {\bibfnamefont {Yanfeng}\ \bibnamefont {Guo}}, \ and\ \bibinfo {author} {\bibfnamefont {Wentao}\ \bibnamefont {Zhang}},\ }\bibfield  {title} {\enquote {\bibinfo {title} {Fluctuated lattice-driven charge density wave far above the condensation temperature in kagome superconductor kv3sb5},}\ }\href {\doibase
  https://doi.org/10.1016/j.scib.2025.02.018} {\bibfield  {journal} {\bibinfo  {journal} {Science Bulletin}\ }\textbf {\bibinfo {volume} {70}},\ \bibinfo {pages} {1211--1214} (\bibinfo {year} {2025})}\BibitemShut {NoStop}%
\bibitem [{\citenamefont {Duan}\ \emph {et~al.}(2023)\citenamefont {Duan}, \citenamefont {Xia}, \citenamefont {Huang}, \citenamefont {Wang}, \citenamefont {Gu}, \citenamefont {Liu}, \citenamefont {Xiang}, \citenamefont {Qian}, \citenamefont {Guo},\ and\ \citenamefont {Zhang}}]{duan2023}%
  \BibitemOpen
  \bibfield  {author} {\bibinfo {author} {\bibfnamefont {Shaofeng}\ \bibnamefont {Duan}}, \bibinfo {author} {\bibfnamefont {Wei}\ \bibnamefont {Xia}}, \bibinfo {author} {\bibfnamefont {Chaozhi}\ \bibnamefont {Huang}}, \bibinfo {author} {\bibfnamefont {Shichong}\ \bibnamefont {Wang}}, \bibinfo {author} {\bibfnamefont {Lingxiao}\ \bibnamefont {Gu}}, \bibinfo {author} {\bibfnamefont {Haoran}\ \bibnamefont {Liu}}, \bibinfo {author} {\bibfnamefont {Dao}\ \bibnamefont {Xiang}}, \bibinfo {author} {\bibfnamefont {Dong}\ \bibnamefont {Qian}}, \bibinfo {author} {\bibfnamefont {Yanfeng}\ \bibnamefont {Guo}}, \ and\ \bibinfo {author} {\bibfnamefont {Wentao}\ \bibnamefont {Zhang}},\ }\bibfield  {title} {\enquote {\bibinfo {title} {Ultrafast switching from the charge density wave phase to a metastable metallic state in $1t\text{\ensuremath{-}}{\mathrm{tise}}_{2}$},}\ }\href {\doibase 10.1103/PhysRevLett.130.226501} {\bibfield  {journal} {\bibinfo  {journal} {Phys. Rev. Lett.}\ }\textbf {\bibinfo {volume} {130}},\ \bibinfo
  {pages} {226501} (\bibinfo {year} {2023})}\BibitemShut {NoStop}%
\bibitem [{\citenamefont {Huber}\ \emph {et~al.}(2024)\citenamefont {Huber}, \citenamefont {Lin}, \citenamefont {Marini}, \citenamefont {Moreschini}, \citenamefont {Jozwiak}, \citenamefont {Bostwick}, \citenamefont {Calandra},\ and\ \citenamefont {Lanzara}}]{huber2024ultrafast}%
  \BibitemOpen
  \bibfield  {author} {\bibinfo {author} {\bibfnamefont {Maximilian}\ \bibnamefont {Huber}}, \bibinfo {author} {\bibfnamefont {Yi}~\bibnamefont {Lin}}, \bibinfo {author} {\bibfnamefont {Giovanni}\ \bibnamefont {Marini}}, \bibinfo {author} {\bibfnamefont {Luca}\ \bibnamefont {Moreschini}}, \bibinfo {author} {\bibfnamefont {Chris}\ \bibnamefont {Jozwiak}}, \bibinfo {author} {\bibfnamefont {Aaron}\ \bibnamefont {Bostwick}}, \bibinfo {author} {\bibfnamefont {Matteo}\ \bibnamefont {Calandra}}, \ and\ \bibinfo {author} {\bibfnamefont {Alessandra}\ \bibnamefont {Lanzara}},\ }\bibfield  {title} {\enquote {\bibinfo {title} {Ultrafast creation of a light-induced semimetallic state in strongly excited 1t-tise<sub>2</sub>},}\ }\href {\doibase 10.1126/sciadv.adl4481} {\bibfield  {journal} {\bibinfo  {journal} {Science Advances}\ }\textbf {\bibinfo {volume} {10}},\ \bibinfo {pages} {eadl4481} (\bibinfo {year} {2024})},\ \Eprint {http://arxiv.org/abs/https://www.science.org/doi/pdf/10.1126/sciadv.adl4481}
  {https://www.science.org/doi/pdf/10.1126/sciadv.adl4481} \BibitemShut {NoStop}%
\bibitem [{\citenamefont {Ishioka}\ \emph {et~al.}(2010)\citenamefont {Ishioka}, \citenamefont {Liu}, \citenamefont {Shimatake}, \citenamefont {Kurosawa}, \citenamefont {Ichimura}, \citenamefont {Toda}, \citenamefont {Oda},\ and\ \citenamefont {Tanda}}]{ishioka2010chiral}%
  \BibitemOpen
  \bibfield  {author} {\bibinfo {author} {\bibfnamefont {J.}~\bibnamefont {Ishioka}}, \bibinfo {author} {\bibfnamefont {Y.~H.}\ \bibnamefont {Liu}}, \bibinfo {author} {\bibfnamefont {K.}~\bibnamefont {Shimatake}}, \bibinfo {author} {\bibfnamefont {T.}~\bibnamefont {Kurosawa}}, \bibinfo {author} {\bibfnamefont {K.}~\bibnamefont {Ichimura}}, \bibinfo {author} {\bibfnamefont {Y.}~\bibnamefont {Toda}}, \bibinfo {author} {\bibfnamefont {M.}~\bibnamefont {Oda}}, \ and\ \bibinfo {author} {\bibfnamefont {S.}~\bibnamefont {Tanda}},\ }\bibfield  {title} {\enquote {\bibinfo {title} {Chiral charge-density waves},}\ }\href {\doibase 10.1103/PhysRevLett.105.176401} {\bibfield  {journal} {\bibinfo  {journal} {Phys. Rev. Lett.}\ }\textbf {\bibinfo {volume} {105}},\ \bibinfo {pages} {176401} (\bibinfo {year} {2010})}\BibitemShut {NoStop}%
\bibitem [{\citenamefont {Xu}\ \emph {et~al.}(2020)\citenamefont {Xu}, \citenamefont {Ma}, \citenamefont {Gao}, \citenamefont {Kogar}, \citenamefont {Zong}, \citenamefont {Mier~Valdivia}, \citenamefont {Dinh}, \citenamefont {Huang}, \citenamefont {Singh}, \citenamefont {Hsu}, \citenamefont {Chang}, \citenamefont {Ruff}, \citenamefont {Watanabe}, \citenamefont {Taniguchi}, \citenamefont {Lin}, \citenamefont {Karapetrov}, \citenamefont {Xiao}, \citenamefont {Jarillo-Herrero},\ and\ \citenamefont {Gedik}}]{xu2020gyro}%
  \BibitemOpen
  \bibfield  {author} {\bibinfo {author} {\bibfnamefont {Su-Yang}\ \bibnamefont {Xu}}, \bibinfo {author} {\bibfnamefont {Qiong}\ \bibnamefont {Ma}}, \bibinfo {author} {\bibfnamefont {Yang}\ \bibnamefont {Gao}}, \bibinfo {author} {\bibfnamefont {Anshul}\ \bibnamefont {Kogar}}, \bibinfo {author} {\bibfnamefont {Alfred}\ \bibnamefont {Zong}}, \bibinfo {author} {\bibfnamefont {Andrés~M.}\ \bibnamefont {Mier~Valdivia}}, \bibinfo {author} {\bibfnamefont {Thao~H.}\ \bibnamefont {Dinh}}, \bibinfo {author} {\bibfnamefont {Shin-Ming}\ \bibnamefont {Huang}}, \bibinfo {author} {\bibfnamefont {Bahadur}\ \bibnamefont {Singh}}, \bibinfo {author} {\bibfnamefont {Chuang-Han}\ \bibnamefont {Hsu}}, \bibinfo {author} {\bibfnamefont {Tay-Rong}\ \bibnamefont {Chang}}, \bibinfo {author} {\bibfnamefont {Jacob P.~C.}\ \bibnamefont {Ruff}}, \bibinfo {author} {\bibfnamefont {Kenji}\ \bibnamefont {Watanabe}}, \bibinfo {author} {\bibfnamefont {Takashi}\ \bibnamefont {Taniguchi}}, \bibinfo {author} {\bibfnamefont {Hsin}\ \bibnamefont
  {Lin}}, \bibinfo {author} {\bibfnamefont {Goran}\ \bibnamefont {Karapetrov}}, \bibinfo {author} {\bibfnamefont {Di}~\bibnamefont {Xiao}}, \bibinfo {author} {\bibfnamefont {Pablo}\ \bibnamefont {Jarillo-Herrero}}, \ and\ \bibinfo {author} {\bibfnamefont {Nuh}\ \bibnamefont {Gedik}},\ }\bibfield  {title} {\enquote {\bibinfo {title} {Spontaneous gyrotropic electronic order in a transition-metal dichalcogenide},}\ }\href {\doibase 10.1038/s41586-020-2011-8} {\bibfield  {journal} {\bibinfo  {journal} {Nature}\ }\textbf {\bibinfo {volume} {578}},\ \bibinfo {pages} {545–549} (\bibinfo {year} {2020})}\BibitemShut {NoStop}%
\bibitem [{\citenamefont {Morosan}\ \emph {et~al.}(2006)\citenamefont {Morosan}, \citenamefont {Zandbergen}, \citenamefont {Dennis}, \citenamefont {Bos}, \citenamefont {Onose}, \citenamefont {Klimczuk}, \citenamefont {Ramirez}, \citenamefont {Ong},\ and\ \citenamefont {Cava}}]{morosan2006superconductivity}%
  \BibitemOpen
  \bibfield  {author} {\bibinfo {author} {\bibfnamefont {E.}~\bibnamefont {Morosan}}, \bibinfo {author} {\bibfnamefont {H.~W.}\ \bibnamefont {Zandbergen}}, \bibinfo {author} {\bibfnamefont {B.~S.}\ \bibnamefont {Dennis}}, \bibinfo {author} {\bibfnamefont {J.~W.~G.}\ \bibnamefont {Bos}}, \bibinfo {author} {\bibfnamefont {Y.}~\bibnamefont {Onose}}, \bibinfo {author} {\bibfnamefont {T.}~\bibnamefont {Klimczuk}}, \bibinfo {author} {\bibfnamefont {A.~P.}\ \bibnamefont {Ramirez}}, \bibinfo {author} {\bibfnamefont {N.~P.}\ \bibnamefont {Ong}}, \ and\ \bibinfo {author} {\bibfnamefont {R.~J.}\ \bibnamefont {Cava}},\ }\bibfield  {title} {\enquote {\bibinfo {title} {Superconductivity in cu$_x$tise$_2$},}\ }\href {\doibase 10.1038/nphys360} {\bibfield  {journal} {\bibinfo  {journal} {Nature Physics}\ }\textbf {\bibinfo {volume} {2}},\ \bibinfo {pages} {544} (\bibinfo {year} {2006})}\BibitemShut {NoStop}%
\bibitem [{\citenamefont {Fragkos}\ \emph {et~al.}(2025)\citenamefont {Fragkos}, \citenamefont {Courtade}, \citenamefont {Tkach}, \citenamefont {Gaudin}, \citenamefont {Descamps}, \citenamefont {Barrette}, \citenamefont {Petit}, \citenamefont {Sch{\"o}nhense}, \citenamefont {Mairesse},\ and\ \citenamefont {Beaulieu}}]{Fragkos2025-ob}%
  \BibitemOpen
  \bibfield  {author} {\bibinfo {author} {\bibfnamefont {Sotirios}\ \bibnamefont {Fragkos}}, \bibinfo {author} {\bibfnamefont {Quentin}\ \bibnamefont {Courtade}}, \bibinfo {author} {\bibfnamefont {Olena}\ \bibnamefont {Tkach}}, \bibinfo {author} {\bibfnamefont {J{\'e}r{\^o}me}\ \bibnamefont {Gaudin}}, \bibinfo {author} {\bibfnamefont {Dominique}\ \bibnamefont {Descamps}}, \bibinfo {author} {\bibfnamefont {Guillaume}\ \bibnamefont {Barrette}}, \bibinfo {author} {\bibfnamefont {St{\'e}phane}\ \bibnamefont {Petit}}, \bibinfo {author} {\bibfnamefont {Gerd}\ \bibnamefont {Sch{\"o}nhense}}, \bibinfo {author} {\bibfnamefont {Yann}\ \bibnamefont {Mairesse}}, \ and\ \bibinfo {author} {\bibfnamefont {Samuel}\ \bibnamefont {Beaulieu}},\ }\bibfield  {title} {\enquote {\bibinfo {title} {Time- and polarization-resolved extreme ultraviolet momentum microscopy},}\ }\href@noop {} {\bibfield  {journal} {\bibinfo  {journal} {Rev. Sci. Instrum.}\ }\textbf {\bibinfo {volume} {96}},\ \bibinfo {pages} {115201} (\bibinfo {year}
  {2025})}\BibitemShut {NoStop}%
\bibitem [{\citenamefont {Tkach}\ and\ \citenamefont {Schönhense}(2025)}]{tkach24}%
  \BibitemOpen
  \bibfield  {author} {\bibinfo {author} {\bibfnamefont {Olena}\ \bibnamefont {Tkach}}\ and\ \bibinfo {author} {\bibfnamefont {Gerd}\ \bibnamefont {Schönhense}},\ }\bibfield  {title} {\enquote {\bibinfo {title} {Multimode objective lens for momentum microscopy and xpeem: Theory},}\ }\href {\doibase https://doi.org/10.1016/j.ultramic.2025.114167} {\bibfield  {journal} {\bibinfo  {journal} {Ultramicroscopy}\ }\textbf {\bibinfo {volume} {276}},\ \bibinfo {pages} {114167} (\bibinfo {year} {2025})}\BibitemShut {NoStop}%
\bibitem [{\citenamefont {Tkach}\ \emph {et~al.}(2024{\natexlab{a}})\citenamefont {Tkach}, \citenamefont {Fragkos}, \citenamefont {Nguyen}, \citenamefont {Chernov}, \citenamefont {Scholz}, \citenamefont {Wind}, \citenamefont {Babenkov}, \citenamefont {Fedchenko}, \citenamefont {Lytvynenko}, \citenamefont {Zimmer}, \citenamefont {Hloskovskii}, \citenamefont {Kutnyakhov}, \citenamefont {Pressacco}, \citenamefont {Dilling}, \citenamefont {Bruckmeier}, \citenamefont {Heber}, \citenamefont {Scholz}, \citenamefont {Sobota}, \citenamefont {Koralek}, \citenamefont {Sirica}, \citenamefont {Kallmayer}, \citenamefont {Hoesch}, \citenamefont {Schlueter}, \citenamefont {Odnodvorets}, \citenamefont {Mairesse}, \citenamefont {Rossnagel}, \citenamefont {Elmers}, \citenamefont {Beaulieu},\ and\ \citenamefont {Schoenhense}}]{tkach24-2}%
  \BibitemOpen
  \bibfield  {author} {\bibinfo {author} {\bibfnamefont {O.}~\bibnamefont {Tkach}}, \bibinfo {author} {\bibfnamefont {S.}~\bibnamefont {Fragkos}}, \bibinfo {author} {\bibfnamefont {Q.}~\bibnamefont {Nguyen}}, \bibinfo {author} {\bibfnamefont {S.}~\bibnamefont {Chernov}}, \bibinfo {author} {\bibfnamefont {M.}~\bibnamefont {Scholz}}, \bibinfo {author} {\bibfnamefont {N.}~\bibnamefont {Wind}}, \bibinfo {author} {\bibfnamefont {S.}~\bibnamefont {Babenkov}}, \bibinfo {author} {\bibfnamefont {O.}~\bibnamefont {Fedchenko}}, \bibinfo {author} {\bibfnamefont {Y.}~\bibnamefont {Lytvynenko}}, \bibinfo {author} {\bibfnamefont {D.}~\bibnamefont {Zimmer}}, \bibinfo {author} {\bibfnamefont {A.}~\bibnamefont {Hloskovskii}}, \bibinfo {author} {\bibfnamefont {D.}~\bibnamefont {Kutnyakhov}}, \bibinfo {author} {\bibfnamefont {F.}~\bibnamefont {Pressacco}}, \bibinfo {author} {\bibfnamefont {J.}~\bibnamefont {Dilling}}, \bibinfo {author} {\bibfnamefont {L.}~\bibnamefont {Bruckmeier}}, \bibinfo {author} {\bibfnamefont
  {M.}~\bibnamefont {Heber}}, \bibinfo {author} {\bibfnamefont {F.}~\bibnamefont {Scholz}}, \bibinfo {author} {\bibfnamefont {J.}~\bibnamefont {Sobota}}, \bibinfo {author} {\bibfnamefont {J.}~\bibnamefont {Koralek}}, \bibinfo {author} {\bibfnamefont {N.}~\bibnamefont {Sirica}}, \bibinfo {author} {\bibfnamefont {M.}~\bibnamefont {Kallmayer}}, \bibinfo {author} {\bibfnamefont {M.}~\bibnamefont {Hoesch}}, \bibinfo {author} {\bibfnamefont {C.}~\bibnamefont {Schlueter}}, \bibinfo {author} {\bibfnamefont {L.~V.}\ \bibnamefont {Odnodvorets}}, \bibinfo {author} {\bibfnamefont {Y.}~\bibnamefont {Mairesse}}, \bibinfo {author} {\bibfnamefont {K.}~\bibnamefont {Rossnagel}}, \bibinfo {author} {\bibfnamefont {H.~J.}\ \bibnamefont {Elmers}}, \bibinfo {author} {\bibfnamefont {S.}~\bibnamefont {Beaulieu}}, \ and\ \bibinfo {author} {\bibfnamefont {G.}~\bibnamefont {Schoenhense}},\ }\href {https://arxiv.org/abs/2401.10084} {\enquote {\bibinfo {title} {Multi-mode front lens for momentum microscopy: Part {II} experiments},}\ }
  (\bibinfo {year} {2024}{\natexlab{a}}),\ \Eprint {http://arxiv.org/abs/2401.10084} {arXiv:2401.10084 [cond-mat.mtrl-sci]} \BibitemShut {NoStop}%
\bibitem [{\citenamefont {Moser}(2017)}]{Moser17}%
  \BibitemOpen
  \bibfield  {author} {\bibinfo {author} {\bibfnamefont {Simon}\ \bibnamefont {Moser}},\ }\bibfield  {title} {\enquote {\bibinfo {title} {An experimentalist's guide to the matrix element in angle resolved photoemission},}\ }\href {\doibase https://doi.org/10.1016/j.elspec.2016.11.007} {\bibfield  {journal} {\bibinfo  {journal} {Journal of Electron Spectroscopy and Related Phenomena}\ }\textbf {\bibinfo {volume} {214}},\ \bibinfo {pages} {29--52} (\bibinfo {year} {2017})}\BibitemShut {NoStop}%
\bibitem [{\citenamefont {Schusser}\ \emph {et~al.}(2024)\citenamefont {Schusser}, \citenamefont {Orio}, \citenamefont {Ünzelmann}, \citenamefont {Heßdörfer}, \citenamefont {Masilamani}, \citenamefont {Diekmann}, \citenamefont {Rossnagel},\ and\ \citenamefont {Reinert}}]{Schusser2024}%
  \BibitemOpen
  \bibfield  {author} {\bibinfo {author} {\bibfnamefont {J}~\bibnamefont {Schusser}}, \bibinfo {author} {\bibfnamefont {H}~\bibnamefont {Orio}}, \bibinfo {author} {\bibfnamefont {M}~\bibnamefont {Ünzelmann}}, \bibinfo {author} {\bibfnamefont {J}~\bibnamefont {Heßdörfer}}, \bibinfo {author} {\bibfnamefont {M~P~T}\ \bibnamefont {Masilamani}}, \bibinfo {author} {\bibfnamefont {F}~\bibnamefont {Diekmann}}, \bibinfo {author} {\bibfnamefont {K}~\bibnamefont {Rossnagel}}, \ and\ \bibinfo {author} {\bibfnamefont {F}~\bibnamefont {Reinert}},\ }\bibfield  {title} {\enquote {\bibinfo {title} {Towards robust dichroism in angle-resolved photoemission},}\ }\href {\doibase 10.1038/s42005-024-01762-y} {\bibfield  {journal} {\bibinfo  {journal} {Communications Physics}\ }\textbf {\bibinfo {volume} {7}},\ \bibinfo {pages} {270} (\bibinfo {year} {2024})}\BibitemShut {NoStop}%
\bibitem [{\citenamefont {Schusser}\ \emph {et~al.}(2022)\citenamefont {Schusser}, \citenamefont {Bentmann}, \citenamefont {\"Unzelmann}, \citenamefont {Figgemeier}, \citenamefont {Min}, \citenamefont {Moser}, \citenamefont {Neu}, \citenamefont {Siegrist},\ and\ \citenamefont {Reinert}}]{PhysRevLett.129.246404}%
  \BibitemOpen
  \bibfield  {author} {\bibinfo {author} {\bibfnamefont {J.}~\bibnamefont {Schusser}}, \bibinfo {author} {\bibfnamefont {H.}~\bibnamefont {Bentmann}}, \bibinfo {author} {\bibfnamefont {M.}~\bibnamefont {\"Unzelmann}}, \bibinfo {author} {\bibfnamefont {T.}~\bibnamefont {Figgemeier}}, \bibinfo {author} {\bibfnamefont {C.-H.}\ \bibnamefont {Min}}, \bibinfo {author} {\bibfnamefont {S.}~\bibnamefont {Moser}}, \bibinfo {author} {\bibfnamefont {J.~N.}\ \bibnamefont {Neu}}, \bibinfo {author} {\bibfnamefont {T.}~\bibnamefont {Siegrist}}, \ and\ \bibinfo {author} {\bibfnamefont {F.}~\bibnamefont {Reinert}},\ }\bibfield  {title} {\enquote {\bibinfo {title} {Assessing nontrivial topology in weyl semimetals by dichroic photoemission},}\ }\href {\doibase 10.1103/PhysRevLett.129.246404} {\bibfield  {journal} {\bibinfo  {journal} {Phys. Rev. Lett.}\ }\textbf {\bibinfo {volume} {129}},\ \bibinfo {pages} {246404} (\bibinfo {year} {2022})}\BibitemShut {NoStop}%
\bibitem [{\citenamefont {Borisenko}\ \emph {et~al.}(2009)\citenamefont {Borisenko}, \citenamefont {Kordyuk}, \citenamefont {Zabolotnyy}, \citenamefont {Inosov}, \citenamefont {Evtushinsky}, \citenamefont {B\"uchner}, \citenamefont {Yaresko}, \citenamefont {Varykhalov}, \citenamefont {Follath}, \citenamefont {Eberhardt}, \citenamefont {Patthey},\ and\ \citenamefont {Berger}}]{borisenko2009pg}%
  \BibitemOpen
  \bibfield  {author} {\bibinfo {author} {\bibfnamefont {S.~V.}\ \bibnamefont {Borisenko}}, \bibinfo {author} {\bibfnamefont {A.~A.}\ \bibnamefont {Kordyuk}}, \bibinfo {author} {\bibfnamefont {V.~B.}\ \bibnamefont {Zabolotnyy}}, \bibinfo {author} {\bibfnamefont {D.~S.}\ \bibnamefont {Inosov}}, \bibinfo {author} {\bibfnamefont {D.}~\bibnamefont {Evtushinsky}}, \bibinfo {author} {\bibfnamefont {B.}~\bibnamefont {B\"uchner}}, \bibinfo {author} {\bibfnamefont {A.~N.}\ \bibnamefont {Yaresko}}, \bibinfo {author} {\bibfnamefont {A.}~\bibnamefont {Varykhalov}}, \bibinfo {author} {\bibfnamefont {R.}~\bibnamefont {Follath}}, \bibinfo {author} {\bibfnamefont {W.}~\bibnamefont {Eberhardt}}, \bibinfo {author} {\bibfnamefont {L.}~\bibnamefont {Patthey}}, \ and\ \bibinfo {author} {\bibfnamefont {H.}~\bibnamefont {Berger}},\ }\bibfield  {title} {\enquote {\bibinfo {title} {Two energy gaps and fermi-surface ``arcs'' in ${\mathrm{nbse}}_{2}$},}\ }\href {\doibase 10.1103/PhysRevLett.102.166402} {\bibfield  {journal} {\bibinfo
  {journal} {Phys. Rev. Lett.}\ }\textbf {\bibinfo {volume} {102}},\ \bibinfo {pages} {166402} (\bibinfo {year} {2009})}\BibitemShut {NoStop}%
\bibitem [{\citenamefont {Huber}\ \emph {et~al.}(2022)\citenamefont {Huber}, \citenamefont {Lin}, \citenamefont {Dale}, \citenamefont {Sailus}, \citenamefont {Tongay}, \citenamefont {Kaindl},\ and\ \citenamefont {Lanzara}}]{huber2022fluct}%
  \BibitemOpen
  \bibfield  {author} {\bibinfo {author} {\bibfnamefont {Maximilian}\ \bibnamefont {Huber}}, \bibinfo {author} {\bibfnamefont {Yi}~\bibnamefont {Lin}}, \bibinfo {author} {\bibfnamefont {Nicholas}\ \bibnamefont {Dale}}, \bibinfo {author} {\bibfnamefont {Renee}\ \bibnamefont {Sailus}}, \bibinfo {author} {\bibfnamefont {Sefaattin}\ \bibnamefont {Tongay}}, \bibinfo {author} {\bibfnamefont {Robert~A.}\ \bibnamefont {Kaindl}}, \ and\ \bibinfo {author} {\bibfnamefont {Alessandra}\ \bibnamefont {Lanzara}},\ }\bibfield  {title} {\enquote {\bibinfo {title} {Revealing the order parameter dynamics of 1t-tise$_2$ following optical excitation},}\ }\href {\doibase 10.1038/s41598-022-19319-w} {\bibfield  {journal} {\bibinfo  {journal} {Scientific Reports}\ }\textbf {\bibinfo {volume} {12}},\ \bibinfo {pages} {15860} (\bibinfo {year} {2022})}\BibitemShut {NoStop}%
\bibitem [{\citenamefont {Boschini}\ \emph {et~al.}(2024)\citenamefont {Boschini}, \citenamefont {Zonno},\ and\ \citenamefont {Damascelli}}]{boschini2024}%
  \BibitemOpen
  \bibfield  {author} {\bibinfo {author} {\bibfnamefont {Fabio}\ \bibnamefont {Boschini}}, \bibinfo {author} {\bibfnamefont {Marta}\ \bibnamefont {Zonno}}, \ and\ \bibinfo {author} {\bibfnamefont {Andrea}\ \bibnamefont {Damascelli}},\ }\bibfield  {title} {\enquote {\bibinfo {title} {Time-resolved arpes studies of quantum materials},}\ }\href {\doibase 10.1103/RevModPhys.96.015003} {\bibfield  {journal} {\bibinfo  {journal} {Rev. Mod. Phys.}\ }\textbf {\bibinfo {volume} {96}},\ \bibinfo {pages} {015003} (\bibinfo {year} {2024})}\BibitemShut {NoStop}%
\bibitem [{\citenamefont {Buchberger}\ \emph {et~al.}(2025)\citenamefont {Buchberger}, \citenamefont {in~'t Veld}, \citenamefont {Rajan}, \citenamefont {Murgatroyd}, \citenamefont {Edwards}, \citenamefont {Saika}, \citenamefont {Kushwaha}, \citenamefont {Visscher}, \citenamefont {Berges}, \citenamefont {Carbone}, \citenamefont {Osiecki}, \citenamefont {Polley}, \citenamefont {Wehling},\ and\ \citenamefont {King}}]{buchberger2025}%
  \BibitemOpen
  \bibfield  {author} {\bibinfo {author} {\bibfnamefont {Sebastian}\ \bibnamefont {Buchberger}}, \bibinfo {author} {\bibfnamefont {Yann}\ \bibnamefont {in~'t Veld}}, \bibinfo {author} {\bibfnamefont {Akhil}\ \bibnamefont {Rajan}}, \bibinfo {author} {\bibfnamefont {Philip A.~E.}\ \bibnamefont {Murgatroyd}}, \bibinfo {author} {\bibfnamefont {Brendan}\ \bibnamefont {Edwards}}, \bibinfo {author} {\bibfnamefont {Bruno~K.}\ \bibnamefont {Saika}}, \bibinfo {author} {\bibfnamefont {Naina}\ \bibnamefont {Kushwaha}}, \bibinfo {author} {\bibfnamefont {Maria~H.}\ \bibnamefont {Visscher}}, \bibinfo {author} {\bibfnamefont {Jan}\ \bibnamefont {Berges}}, \bibinfo {author} {\bibfnamefont {Dina}\ \bibnamefont {Carbone}}, \bibinfo {author} {\bibfnamefont {Jacek}\ \bibnamefont {Osiecki}}, \bibinfo {author} {\bibfnamefont {Craig}\ \bibnamefont {Polley}}, \bibinfo {author} {\bibfnamefont {Tim}\ \bibnamefont {Wehling}}, \ and\ \bibinfo {author} {\bibfnamefont {Phil D.~C.}\ \bibnamefont {King}},\ }\href
  {https://arxiv.org/abs/2506.01470} {\enquote {\bibinfo {title} {Persistence of charge ordering instability to coulomb engineering in the excitonic insulator candidate tise$_2$},}\ } (\bibinfo {year} {2025}),\ \Eprint {http://arxiv.org/abs/2506.01470} {arXiv:2506.01470 [cond-mat.mtrl-sci]} \BibitemShut {NoStop}%
\bibitem [{\citenamefont {Kurtz}\ \emph {et~al.}(2024)\citenamefont {Kurtz}, \citenamefont {Dauwe}, \citenamefont {Yalunin}, \citenamefont {Storeck}, \citenamefont {Horstmann}, \citenamefont {Böckmann},\ and\ \citenamefont {Ropers}}]{kurtz2024nontherm}%
  \BibitemOpen
  \bibfield  {author} {\bibinfo {author} {\bibfnamefont {Felix}\ \bibnamefont {Kurtz}}, \bibinfo {author} {\bibfnamefont {Tim~N.}\ \bibnamefont {Dauwe}}, \bibinfo {author} {\bibfnamefont {Sergey~V.}\ \bibnamefont {Yalunin}}, \bibinfo {author} {\bibfnamefont {Gero}\ \bibnamefont {Storeck}}, \bibinfo {author} {\bibfnamefont {Jan~Gerrit}\ \bibnamefont {Horstmann}}, \bibinfo {author} {\bibfnamefont {Hannes}\ \bibnamefont {Böckmann}}, \ and\ \bibinfo {author} {\bibfnamefont {Claus}\ \bibnamefont {Ropers}},\ }\bibfield  {title} {\enquote {\bibinfo {title} {Non-thermal phonon dynamics and a quenched exciton condensate probed by surface-sensitive electron diffraction},}\ }\href {\doibase 10.1038/s41563-024-01880-6} {\bibfield  {journal} {\bibinfo  {journal} {Nature Materials}\ }\textbf {\bibinfo {volume} {23}},\ \bibinfo {pages} {890} (\bibinfo {year} {2024})}\BibitemShut {NoStop}%
\bibitem [{\citenamefont {Giustino}(2017)}]{giustino2017epc}%
  \BibitemOpen
  \bibfield  {author} {\bibinfo {author} {\bibfnamefont {Feliciano}\ \bibnamefont {Giustino}},\ }\bibfield  {title} {\enquote {\bibinfo {title} {Electron-phonon interactions from first principles},}\ }\href {\doibase 10.1103/RevModPhys.89.015003} {\bibfield  {journal} {\bibinfo  {journal} {Rev. Mod. Phys.}\ }\textbf {\bibinfo {volume} {89}},\ \bibinfo {pages} {015003} (\bibinfo {year} {2017})}\BibitemShut {NoStop}%
\bibitem [{\citenamefont {Lee}\ \emph {et~al.}(2023)\citenamefont {Lee}, \citenamefont {Poncé}, \citenamefont {Bushick}, \citenamefont {Hajinazar}, \citenamefont {Lafuente-Bartolome}, \citenamefont {Leveillee}, \citenamefont {Lian}, \citenamefont {Lihm}, \citenamefont {Macheda}, \citenamefont {Mori}, \citenamefont {Paudyal}, \citenamefont {Sio}, \citenamefont {Tiwari}, \citenamefont {Zacharias}, \citenamefont {Zhang}, \citenamefont {Bonini}, \citenamefont {Kioupakis}, \citenamefont {Margine},\ and\ \citenamefont {Giustino}}]{lee2023epw}%
  \BibitemOpen
  \bibfield  {author} {\bibinfo {author} {\bibfnamefont {Hyungjun}\ \bibnamefont {Lee}}, \bibinfo {author} {\bibfnamefont {Samuel}\ \bibnamefont {Poncé}}, \bibinfo {author} {\bibfnamefont {Kyle}\ \bibnamefont {Bushick}}, \bibinfo {author} {\bibfnamefont {Samad}\ \bibnamefont {Hajinazar}}, \bibinfo {author} {\bibfnamefont {Jon}\ \bibnamefont {Lafuente-Bartolome}}, \bibinfo {author} {\bibfnamefont {Joshua}\ \bibnamefont {Leveillee}}, \bibinfo {author} {\bibfnamefont {Chao}\ \bibnamefont {Lian}}, \bibinfo {author} {\bibfnamefont {Jae-Mo}\ \bibnamefont {Lihm}}, \bibinfo {author} {\bibfnamefont {Francesco}\ \bibnamefont {Macheda}}, \bibinfo {author} {\bibfnamefont {Hitoshi}\ \bibnamefont {Mori}}, \bibinfo {author} {\bibfnamefont {Hari}\ \bibnamefont {Paudyal}}, \bibinfo {author} {\bibfnamefont {Weng~Hong}\ \bibnamefont {Sio}}, \bibinfo {author} {\bibfnamefont {Sabyasachi}\ \bibnamefont {Tiwari}}, \bibinfo {author} {\bibfnamefont {Marios}\ \bibnamefont {Zacharias}}, \bibinfo {author} {\bibfnamefont {Xiao}\
  \bibnamefont {Zhang}}, \bibinfo {author} {\bibfnamefont {Nicola}\ \bibnamefont {Bonini}}, \bibinfo {author} {\bibfnamefont {Emmanouil}\ \bibnamefont {Kioupakis}}, \bibinfo {author} {\bibfnamefont {Elena~R.}\ \bibnamefont {Margine}}, \ and\ \bibinfo {author} {\bibfnamefont {Feliciano}\ \bibnamefont {Giustino}},\ }\bibfield  {title} {\enquote {\bibinfo {title} {Electron–phonon physics from first principles using the epw code},}\ }\href {\doibase 10.1038/s41524-023-01107-3} {\bibfield  {journal} {\bibinfo  {journal} {npj Computational Materials}\ }\textbf {\bibinfo {volume} {9}},\ \bibinfo {pages} {156} (\bibinfo {year} {2023})}\BibitemShut {NoStop}%
\bibitem [{\citenamefont {Baroni}\ \emph {et~al.}(2001)\citenamefont {Baroni}, \citenamefont {de~Gironcoli}, \citenamefont {Dal~Corso},\ and\ \citenamefont {Giannozzi}}]{baroni2001dfpt}%
  \BibitemOpen
  \bibfield  {author} {\bibinfo {author} {\bibfnamefont {Stefano}\ \bibnamefont {Baroni}}, \bibinfo {author} {\bibfnamefont {Stefano}\ \bibnamefont {de~Gironcoli}}, \bibinfo {author} {\bibfnamefont {Andrea}\ \bibnamefont {Dal~Corso}}, \ and\ \bibinfo {author} {\bibfnamefont {Paolo}\ \bibnamefont {Giannozzi}},\ }\bibfield  {title} {\enquote {\bibinfo {title} {Phonons and related crystal properties from density-functional perturbation theory},}\ }\href {\doibase 10.1103/RevModPhys.73.515} {\bibfield  {journal} {\bibinfo  {journal} {Rev. Mod. Phys.}\ }\textbf {\bibinfo {volume} {73}},\ \bibinfo {pages} {515--562} (\bibinfo {year} {2001})}\BibitemShut {NoStop}%
\bibitem [{\citenamefont {Giannozzi}\ \emph {et~al.}(2017)\citenamefont {Giannozzi}, \citenamefont {Andreussi}, \citenamefont {Brumme}, \citenamefont {Bunau}, \citenamefont {Buongiorno~Nardelli}, \citenamefont {Calandra}, \citenamefont {Car}, \citenamefont {Cavazzoni}, \citenamefont {Ceresoli}, \citenamefont {Cococcioni}, \citenamefont {Colonna}, \citenamefont {Carnimeo}, \citenamefont {Dal~Corso}, \citenamefont {de~Gironcoli}, \citenamefont {Delugas}, \citenamefont {DiStasio}, \citenamefont {Ferretti}, \citenamefont {Floris}, \citenamefont {Fratesi}, \citenamefont {Fugallo}, \citenamefont {Gebauer}, \citenamefont {Gerstmann}, \citenamefont {Giustino}, \citenamefont {Gorni}, \citenamefont {Jia}, \citenamefont {Kawamura}, \citenamefont {Ko}, \citenamefont {Kokalj}, \citenamefont {Küçükbenli}, \citenamefont {Lazzeri}, \citenamefont {Marsili}, \citenamefont {Marzari}, \citenamefont {Mauri}, \citenamefont {Nguyen}, \citenamefont {Nguyen}, \citenamefont {Otero-de-la Roza}, \citenamefont {Paulatto},
  \citenamefont {Poncé}, \citenamefont {Rocca}, \citenamefont {Sabatini}, \citenamefont {Santra}, \citenamefont {Schlipf}, \citenamefont {Seitsonen}, \citenamefont {Smogunov}, \citenamefont {Timrov}, \citenamefont {Thonhauser}, \citenamefont {Umari}, \citenamefont {Vast}, \citenamefont {Wu},\ and\ \citenamefont {Baroni}}]{giannozzi2017qe}%
  \BibitemOpen
  \bibfield  {author} {\bibinfo {author} {\bibfnamefont {P}~\bibnamefont {Giannozzi}}, \bibinfo {author} {\bibfnamefont {O}~\bibnamefont {Andreussi}}, \bibinfo {author} {\bibfnamefont {T}~\bibnamefont {Brumme}}, \bibinfo {author} {\bibfnamefont {O}~\bibnamefont {Bunau}}, \bibinfo {author} {\bibfnamefont {M}~\bibnamefont {Buongiorno~Nardelli}}, \bibinfo {author} {\bibfnamefont {M}~\bibnamefont {Calandra}}, \bibinfo {author} {\bibfnamefont {R}~\bibnamefont {Car}}, \bibinfo {author} {\bibfnamefont {C}~\bibnamefont {Cavazzoni}}, \bibinfo {author} {\bibfnamefont {D}~\bibnamefont {Ceresoli}}, \bibinfo {author} {\bibfnamefont {M}~\bibnamefont {Cococcioni}}, \bibinfo {author} {\bibfnamefont {N}~\bibnamefont {Colonna}}, \bibinfo {author} {\bibfnamefont {I}~\bibnamefont {Carnimeo}}, \bibinfo {author} {\bibfnamefont {A}~\bibnamefont {Dal~Corso}}, \bibinfo {author} {\bibfnamefont {S}~\bibnamefont {de~Gironcoli}}, \bibinfo {author} {\bibfnamefont {P}~\bibnamefont {Delugas}}, \bibinfo {author} {\bibfnamefont {R~A}\
  \bibnamefont {DiStasio}}, \bibinfo {author} {\bibfnamefont {A}~\bibnamefont {Ferretti}}, \bibinfo {author} {\bibfnamefont {A}~\bibnamefont {Floris}}, \bibinfo {author} {\bibfnamefont {G}~\bibnamefont {Fratesi}}, \bibinfo {author} {\bibfnamefont {G}~\bibnamefont {Fugallo}}, \bibinfo {author} {\bibfnamefont {R}~\bibnamefont {Gebauer}}, \bibinfo {author} {\bibfnamefont {U}~\bibnamefont {Gerstmann}}, \bibinfo {author} {\bibfnamefont {F}~\bibnamefont {Giustino}}, \bibinfo {author} {\bibfnamefont {T}~\bibnamefont {Gorni}}, \bibinfo {author} {\bibfnamefont {J}~\bibnamefont {Jia}}, \bibinfo {author} {\bibfnamefont {M}~\bibnamefont {Kawamura}}, \bibinfo {author} {\bibfnamefont {H-Y}\ \bibnamefont {Ko}}, \bibinfo {author} {\bibfnamefont {A}~\bibnamefont {Kokalj}}, \bibinfo {author} {\bibfnamefont {E}~\bibnamefont {Küçükbenli}}, \bibinfo {author} {\bibfnamefont {M}~\bibnamefont {Lazzeri}}, \bibinfo {author} {\bibfnamefont {M}~\bibnamefont {Marsili}}, \bibinfo {author} {\bibfnamefont {N}~\bibnamefont {Marzari}},
  \bibinfo {author} {\bibfnamefont {F}~\bibnamefont {Mauri}}, \bibinfo {author} {\bibfnamefont {N~L}\ \bibnamefont {Nguyen}}, \bibinfo {author} {\bibfnamefont {H-V}\ \bibnamefont {Nguyen}}, \bibinfo {author} {\bibfnamefont {A}~\bibnamefont {Otero-de-la Roza}}, \bibinfo {author} {\bibfnamefont {L}~\bibnamefont {Paulatto}}, \bibinfo {author} {\bibfnamefont {S}~\bibnamefont {Poncé}}, \bibinfo {author} {\bibfnamefont {D}~\bibnamefont {Rocca}}, \bibinfo {author} {\bibfnamefont {R}~\bibnamefont {Sabatini}}, \bibinfo {author} {\bibfnamefont {B}~\bibnamefont {Santra}}, \bibinfo {author} {\bibfnamefont {M}~\bibnamefont {Schlipf}}, \bibinfo {author} {\bibfnamefont {A~P}\ \bibnamefont {Seitsonen}}, \bibinfo {author} {\bibfnamefont {A}~\bibnamefont {Smogunov}}, \bibinfo {author} {\bibfnamefont {I}~\bibnamefont {Timrov}}, \bibinfo {author} {\bibfnamefont {T}~\bibnamefont {Thonhauser}}, \bibinfo {author} {\bibfnamefont {P}~\bibnamefont {Umari}}, \bibinfo {author} {\bibfnamefont {N}~\bibnamefont {Vast}}, \bibinfo {author}
  {\bibfnamefont {X}~\bibnamefont {Wu}}, \ and\ \bibinfo {author} {\bibfnamefont {S}~\bibnamefont {Baroni}},\ }\bibfield  {title} {\enquote {\bibinfo {title} {Advanced capabilities for materials modelling with quantum espresso},}\ }\href {\doibase 10.1088/1361-648X/aa8f79} {\bibfield  {journal} {\bibinfo  {journal} {Journal of Physics: Condensed Matter}\ }\textbf {\bibinfo {volume} {29}},\ \bibinfo {pages} {465901} (\bibinfo {year} {2017})}\BibitemShut {NoStop}%
\bibitem [{\citenamefont {Sadovskii}(2001)}]{sadovskii2001pg}%
  \BibitemOpen
  \bibfield  {author} {\bibinfo {author} {\bibfnamefont {Mikhail~V}\ \bibnamefont {Sadovskii}},\ }\bibfield  {title} {\enquote {\bibinfo {title} {Pseudogap in high-temperature superconductors},}\ }\href {\doibase 10.1070/PU2001v044n05ABEH000902} {\bibfield  {journal} {\bibinfo  {journal} {Physics-Uspekhi}\ }\textbf {\bibinfo {volume} {44}},\ \bibinfo {pages} {515} (\bibinfo {year} {2001})}\BibitemShut {NoStop}%
\bibitem [{\citenamefont {Lee}\ \emph {et~al.}(1973)\citenamefont {Lee}, \citenamefont {Rice},\ and\ \citenamefont {Anderson}}]{lee1973fluct}%
  \BibitemOpen
  \bibfield  {author} {\bibinfo {author} {\bibfnamefont {P.~A.}\ \bibnamefont {Lee}}, \bibinfo {author} {\bibfnamefont {T.~M.}\ \bibnamefont {Rice}}, \ and\ \bibinfo {author} {\bibfnamefont {P.~W.}\ \bibnamefont {Anderson}},\ }\bibfield  {title} {\enquote {\bibinfo {title} {Fluctuation effects at a peierls transition},}\ }\href {\doibase 10.1103/PhysRevLett.31.462} {\bibfield  {journal} {\bibinfo  {journal} {Phys. Rev. Lett.}\ }\textbf {\bibinfo {volume} {31}},\ \bibinfo {pages} {462--465} (\bibinfo {year} {1973})}\BibitemShut {NoStop}%
\bibitem [{\citenamefont {McKenzie}(1995)}]{mckenzie1995pg}%
  \BibitemOpen
  \bibfield  {author} {\bibinfo {author} {\bibfnamefont {Ross~H.}\ \bibnamefont {McKenzie}},\ }\bibfield  {title} {\enquote {\bibinfo {title} {Microscopic theory of the pseudogap and peierls transition in quasi-one-dimensional materials},}\ }\href {\doibase 10.1103/PhysRevB.52.16428} {\bibfield  {journal} {\bibinfo  {journal} {Phys. Rev. B}\ }\textbf {\bibinfo {volume} {52}},\ \bibinfo {pages} {16428--16442} (\bibinfo {year} {1995})}\BibitemShut {NoStop}%
\bibitem [{\citenamefont {Pines}(1997)}]{pines1997hightc}%
  \BibitemOpen
  \bibfield  {author} {\bibinfo {author} {\bibfnamefont {David}\ \bibnamefont {Pines}},\ }\bibfield  {title} {\enquote {\bibinfo {title} {Understanding high temperature superconductors: Progress and prospects},}\ }\href {\doibase https://doi.org/10.1016/S0921-4534(97)00253-0} {\bibfield  {journal} {\bibinfo  {journal} {Physica C: Superconductivity}\ }\textbf {\bibinfo {volume} {282}},\ \bibinfo {pages} {273--278} (\bibinfo {year} {1997})}\BibitemShut {NoStop}%
\bibitem [{\citenamefont {Kuchinskii}\ \emph {et~al.}(2012)\citenamefont {Kuchinskii}, \citenamefont {Nekrasov},\ and\ \citenamefont {Sadovskii}}]{kuchinskii2012tmds}%
  \BibitemOpen
  \bibfield  {author} {\bibinfo {author} {\bibfnamefont {E.~Z.}\ \bibnamefont {Kuchinskii}}, \bibinfo {author} {\bibfnamefont {I.~A.}\ \bibnamefont {Nekrasov}}, \ and\ \bibinfo {author} {\bibfnamefont {M.~V.}\ \bibnamefont {Sadovskii}},\ }\bibfield  {title} {\enquote {\bibinfo {title} {Electronic structure of two-dimensional hexagonal diselenides: Charge density waves and pseudogap behavior},}\ }\href {\doibase 10.1134/s1063776112020252} {\bibfield  {journal} {\bibinfo  {journal} {Journal of Experimental and Theoretical Physics}\ }\textbf {\bibinfo {volume} {114}},\ \bibinfo {pages} {671} (\bibinfo {year} {2012})}\BibitemShut {NoStop}%
\bibitem [{\citenamefont {Yoshiyama}\ \emph {et~al.}(1986)\citenamefont {Yoshiyama}, \citenamefont {Takaoka}, \citenamefont {Suzuki},\ and\ \citenamefont {Motizuki}}]{yoshiyama1986tise2}%
  \BibitemOpen
  \bibfield  {author} {\bibinfo {author} {\bibfnamefont {H}~\bibnamefont {Yoshiyama}}, \bibinfo {author} {\bibfnamefont {Y}~\bibnamefont {Takaoka}}, \bibinfo {author} {\bibfnamefont {N}~\bibnamefont {Suzuki}}, \ and\ \bibinfo {author} {\bibfnamefont {K}~\bibnamefont {Motizuki}},\ }\bibfield  {title} {\enquote {\bibinfo {title} {Effects on lattice fluctuations on the charge-density-wave transition in transition-metal dichalcogenides},}\ }\href {\doibase 10.1088/0022-3719/19/28/011} {\bibfield  {journal} {\bibinfo  {journal} {Journal of Physics C: Solid State Physics}\ }\textbf {\bibinfo {volume} {19}},\ \bibinfo {pages} {5591} (\bibinfo {year} {1986})}\BibitemShut {NoStop}%
\bibitem [{\citenamefont {Velebit}\ \emph {et~al.}(2016)\citenamefont {Velebit}, \citenamefont {Pop\ifmmode \check{c}\else \v{c}\fi{}evi\ifmmode~\acute{c}\else \'{c}\fi{}}, \citenamefont {Batisti\ifmmode~\acute{c}\else \'{c}\fi{}}, \citenamefont {Eichler}, \citenamefont {Berger}, \citenamefont {Forr\'o}, \citenamefont {Dressel}, \citenamefont {Bari\ifmmode \check{s}\else \v{s}\fi{}i\ifmmode~\acute{c}\else \'{c}\fi{}},\ and\ \citenamefont {Tuti\ifmmode~\check{s}\else \v{s}\fi{}}}]{velebit2016scatt}%
  \BibitemOpen
  \bibfield  {author} {\bibinfo {author} {\bibfnamefont {K.}~\bibnamefont {Velebit}}, \bibinfo {author} {\bibfnamefont {P.}~\bibnamefont {Pop\ifmmode \check{c}\else \v{c}\fi{}evi\ifmmode~\acute{c}\else \'{c}\fi{}}}, \bibinfo {author} {\bibfnamefont {I.}~\bibnamefont {Batisti\ifmmode~\acute{c}\else \'{c}\fi{}}}, \bibinfo {author} {\bibfnamefont {M.}~\bibnamefont {Eichler}}, \bibinfo {author} {\bibfnamefont {H.}~\bibnamefont {Berger}}, \bibinfo {author} {\bibfnamefont {L.}~\bibnamefont {Forr\'o}}, \bibinfo {author} {\bibfnamefont {M.}~\bibnamefont {Dressel}}, \bibinfo {author} {\bibfnamefont {N.}~\bibnamefont {Bari\ifmmode \check{s}\else \v{s}\fi{}i\ifmmode~\acute{c}\else \'{c}\fi{}}}, \ and\ \bibinfo {author} {\bibfnamefont {E.}~\bibnamefont {Tuti\ifmmode~\check{s}\else \v{s}\fi{}}},\ }\bibfield  {title} {\enquote {\bibinfo {title} {Scattering-dominated high-temperature phase of $1t\text{\ensuremath{-}}\mathrm{TiS}{\mathrm{e}}_{2}$: An optical conductivity study},}\ }\href {\doibase 10.1103/PhysRevB.94.075105}
  {\bibfield  {journal} {\bibinfo  {journal} {Phys. Rev. B}\ }\textbf {\bibinfo {volume} {94}},\ \bibinfo {pages} {075105} (\bibinfo {year} {2016})}\BibitemShut {NoStop}%
\bibitem [{\citenamefont {Monney}\ \emph {et~al.}(2015)\citenamefont {Monney}, \citenamefont {Monney}, \citenamefont {Hildebrand}, \citenamefont {Aebi},\ and\ \citenamefont {Beck}}]{monney2015elcorr}%
  \BibitemOpen
  \bibfield  {author} {\bibinfo {author} {\bibfnamefont {G.}~\bibnamefont {Monney}}, \bibinfo {author} {\bibfnamefont {C.}~\bibnamefont {Monney}}, \bibinfo {author} {\bibfnamefont {B.}~\bibnamefont {Hildebrand}}, \bibinfo {author} {\bibfnamefont {P.}~\bibnamefont {Aebi}}, \ and\ \bibinfo {author} {\bibfnamefont {H.}~\bibnamefont {Beck}},\ }\bibfield  {title} {\enquote {\bibinfo {title} {Impact of electron-hole correlations on the $1t\text{\ensuremath{-}}{\mathrm{tise}}_{2}$ electronic structure},}\ }\href {\doibase 10.1103/PhysRevLett.114.086402} {\bibfield  {journal} {\bibinfo  {journal} {Phys. Rev. Lett.}\ }\textbf {\bibinfo {volume} {114}},\ \bibinfo {pages} {086402} (\bibinfo {year} {2015})}\BibitemShut {NoStop}%
\bibitem [{\citenamefont {Lin}\ \emph {et~al.}(2022)\citenamefont {Lin}, \citenamefont {Wang}, \citenamefont {Balassis}, \citenamefont {Echeverry}, \citenamefont {Vasenko}, \citenamefont {Silkin}, \citenamefont {Chulkov}, \citenamefont {Shi}, \citenamefont {Zhang}, \citenamefont {Guo},\ and\ \citenamefont {Zhu}}]{lin2022dramatic}%
  \BibitemOpen
  \bibfield  {author} {\bibinfo {author} {\bibfnamefont {Zijian}\ \bibnamefont {Lin}}, \bibinfo {author} {\bibfnamefont {Cuixiang}\ \bibnamefont {Wang}}, \bibinfo {author} {\bibfnamefont {A.}~\bibnamefont {Balassis}}, \bibinfo {author} {\bibfnamefont {J.~P.}\ \bibnamefont {Echeverry}}, \bibinfo {author} {\bibfnamefont {A.~S.}\ \bibnamefont {Vasenko}}, \bibinfo {author} {\bibfnamefont {V.~M.}\ \bibnamefont {Silkin}}, \bibinfo {author} {\bibfnamefont {E.~V.}\ \bibnamefont {Chulkov}}, \bibinfo {author} {\bibfnamefont {Youguo}\ \bibnamefont {Shi}}, \bibinfo {author} {\bibfnamefont {Jiandi}\ \bibnamefont {Zhang}}, \bibinfo {author} {\bibfnamefont {Jiandong}\ \bibnamefont {Guo}}, \ and\ \bibinfo {author} {\bibfnamefont {Xuetao}\ \bibnamefont {Zhu}},\ }\bibfield  {title} {\enquote {\bibinfo {title} {Dramatic plasmon response to the charge-density-wave gap development in $1t\text{\ensuremath{-}}{\mathrm{tise}}_{2}$},}\ }\href {\doibase 10.1103/PhysRevLett.129.187601} {\bibfield  {journal} {\bibinfo  {journal} {Phys.
  Rev. Lett.}\ }\textbf {\bibinfo {volume} {129}},\ \bibinfo {pages} {187601} (\bibinfo {year} {2022})}\BibitemShut {NoStop}%
\bibitem [{\citenamefont {Comby}\ \emph {et~al.}(2022)\citenamefont {Comby}, \citenamefont {Rajak}, \citenamefont {Descamps}, \citenamefont {Petit}, \citenamefont {Blanchet}, \citenamefont {Mairesse}, \citenamefont {Gaudin},\ and\ \citenamefont {Beaulieu}}]{Comby22}%
  \BibitemOpen
  \bibfield  {author} {\bibinfo {author} {\bibfnamefont {Antoine}\ \bibnamefont {Comby}}, \bibinfo {author} {\bibfnamefont {Debobrata}\ \bibnamefont {Rajak}}, \bibinfo {author} {\bibfnamefont {Dominique}\ \bibnamefont {Descamps}}, \bibinfo {author} {\bibfnamefont {Stéphane}\ \bibnamefont {Petit}}, \bibinfo {author} {\bibfnamefont {Valérie}\ \bibnamefont {Blanchet}}, \bibinfo {author} {\bibfnamefont {Yann}\ \bibnamefont {Mairesse}}, \bibinfo {author} {\bibfnamefont {Jérome}\ \bibnamefont {Gaudin}}, \ and\ \bibinfo {author} {\bibfnamefont {Samuel}\ \bibnamefont {Beaulieu}},\ }\bibfield  {title} {\enquote {\bibinfo {title} {Ultrafast polarization-tunable monochromatic extreme ultraviolet source at high-repetition-rate},}\ }\href {\doibase 10.1088/2040-8986/ac7a49} {\bibfield  {journal} {\bibinfo  {journal} {Journal of Optics}\ }\textbf {\bibinfo {volume} {24}},\ \bibinfo {pages} {084003} (\bibinfo {year} {2022})}\BibitemShut {NoStop}%
\bibitem [{\citenamefont {Medjanik}\ \emph {et~al.}(2017)\citenamefont {Medjanik}, \citenamefont {Fedchenko}, \citenamefont {Chernov}, \citenamefont {Kutnyakhov}, \citenamefont {Ellguth}, \citenamefont {Oelsner}, \citenamefont {Sch{\"o}nhense}, \citenamefont {Peixoto}, \citenamefont {Lutz}, \citenamefont {Min}, \citenamefont {Reinert}, \citenamefont {D{\"a}ster}, \citenamefont {Acremann}, \citenamefont {Viefhaus}, \citenamefont {Wurth}, \citenamefont {Elmers},\ and\ \citenamefont {Sch{\"o}nhense}}]{Medjanik17}%
  \BibitemOpen
  \bibfield  {author} {\bibinfo {author} {\bibfnamefont {K.}~\bibnamefont {Medjanik}}, \bibinfo {author} {\bibfnamefont {O.}~\bibnamefont {Fedchenko}}, \bibinfo {author} {\bibfnamefont {S.}~\bibnamefont {Chernov}}, \bibinfo {author} {\bibfnamefont {D.}~\bibnamefont {Kutnyakhov}}, \bibinfo {author} {\bibfnamefont {M.}~\bibnamefont {Ellguth}}, \bibinfo {author} {\bibfnamefont {A.}~\bibnamefont {Oelsner}}, \bibinfo {author} {\bibfnamefont {B.}~\bibnamefont {Sch{\"o}nhense}}, \bibinfo {author} {\bibfnamefont {T.~R.~F.}\ \bibnamefont {Peixoto}}, \bibinfo {author} {\bibfnamefont {P.}~\bibnamefont {Lutz}}, \bibinfo {author} {\bibfnamefont {C.-H.}\ \bibnamefont {Min}}, \bibinfo {author} {\bibfnamefont {F.}~\bibnamefont {Reinert}}, \bibinfo {author} {\bibfnamefont {S.}~\bibnamefont {D{\"a}ster}}, \bibinfo {author} {\bibfnamefont {Y.}~\bibnamefont {Acremann}}, \bibinfo {author} {\bibfnamefont {J.}~\bibnamefont {Viefhaus}}, \bibinfo {author} {\bibfnamefont {W.}~\bibnamefont {Wurth}}, \bibinfo {author} {\bibfnamefont
  {H.~J.}\ \bibnamefont {Elmers}}, \ and\ \bibinfo {author} {\bibfnamefont {G.}~\bibnamefont {Sch{\"o}nhense}},\ }\bibfield  {title} {\enquote {\bibinfo {title} {Direct 3d mapping of the fermi surface and fermi velocity},}\ }\href {\doibase 10.1038/nmat4875} {\bibfield  {journal} {\bibinfo  {journal} {Nature Materials}\ }\textbf {\bibinfo {volume} {16}},\ \bibinfo {pages} {615--621} (\bibinfo {year} {2017})}\BibitemShut {NoStop}%
\bibitem [{\citenamefont {Watson}\ \emph {et~al.}(2019)\citenamefont {Watson}, \citenamefont {Clark}, \citenamefont {Mazzola}, \citenamefont {Markovi\ifmmode~\acute{c}\else \'{c}\fi{}}, \citenamefont {Sunko}, \citenamefont {Kim}, \citenamefont {Rossnagel},\ and\ \citenamefont {King}}]{watson2019orbital}%
  \BibitemOpen
  \bibfield  {author} {\bibinfo {author} {\bibfnamefont {Matthew~D.}\ \bibnamefont {Watson}}, \bibinfo {author} {\bibfnamefont {Oliver~J.}\ \bibnamefont {Clark}}, \bibinfo {author} {\bibfnamefont {Federico}\ \bibnamefont {Mazzola}}, \bibinfo {author} {\bibfnamefont {Igor}\ \bibnamefont {Markovi\ifmmode~\acute{c}\else \'{c}\fi{}}}, \bibinfo {author} {\bibfnamefont {Veronika}\ \bibnamefont {Sunko}}, \bibinfo {author} {\bibfnamefont {Timur~K.}\ \bibnamefont {Kim}}, \bibinfo {author} {\bibfnamefont {Kai}\ \bibnamefont {Rossnagel}}, \ and\ \bibinfo {author} {\bibfnamefont {Philip D.~C.}\ \bibnamefont {King}},\ }\bibfield  {title} {\enquote {\bibinfo {title} {Orbital- and ${k}_{z}$-selective hybridization of se $4p$ and ti $3d$ states in the charge density wave phase of ${\mathrm{tise}}_{2}$},}\ }\href {\doibase 10.1103/PhysRevLett.122.076404} {\bibfield  {journal} {\bibinfo  {journal} {Phys. Rev. Lett.}\ }\textbf {\bibinfo {volume} {122}},\ \bibinfo {pages} {076404} (\bibinfo {year} {2019})}\BibitemShut {NoStop}%
\bibitem [{\citenamefont {Tkach}\ \emph {et~al.}(2024{\natexlab{b}})\citenamefont {Tkach}, \citenamefont {Fragkos}, \citenamefont {Nguyen}, \citenamefont {Chernov}, \citenamefont {Scholz}, \citenamefont {Wind}, \citenamefont {Babenkov}, \citenamefont {Fedchenko}, \citenamefont {Lytvynenko}, \citenamefont {Zimmer}, \citenamefont {Hloskovskii}, \citenamefont {Kutnyakhov}, \citenamefont {Pressacco}, \citenamefont {Dilling}, \citenamefont {Bruckmeier}, \citenamefont {Heber}, \citenamefont {Scholz}, \citenamefont {Sobota}, \citenamefont {Koralek}, \citenamefont {Sirica}, \citenamefont {Kallmayer}, \citenamefont {Hoesch}, \citenamefont {Schlueter}, \citenamefont {Odnodvorets}, \citenamefont {Mairesse}, \citenamefont {Rossnagel}, \citenamefont {Elmers}, \citenamefont {Beaulieu},\ and\ \citenamefont {Schoenhense}}]{tkach2024multimode}%
  \BibitemOpen
  \bibfield  {author} {\bibinfo {author} {\bibfnamefont {O.}~\bibnamefont {Tkach}}, \bibinfo {author} {\bibfnamefont {S.}~\bibnamefont {Fragkos}}, \bibinfo {author} {\bibfnamefont {Q.}~\bibnamefont {Nguyen}}, \bibinfo {author} {\bibfnamefont {S.}~\bibnamefont {Chernov}}, \bibinfo {author} {\bibfnamefont {M.}~\bibnamefont {Scholz}}, \bibinfo {author} {\bibfnamefont {N.}~\bibnamefont {Wind}}, \bibinfo {author} {\bibfnamefont {S.}~\bibnamefont {Babenkov}}, \bibinfo {author} {\bibfnamefont {O.}~\bibnamefont {Fedchenko}}, \bibinfo {author} {\bibfnamefont {Y.}~\bibnamefont {Lytvynenko}}, \bibinfo {author} {\bibfnamefont {D.}~\bibnamefont {Zimmer}}, \bibinfo {author} {\bibfnamefont {A.}~\bibnamefont {Hloskovskii}}, \bibinfo {author} {\bibfnamefont {D.}~\bibnamefont {Kutnyakhov}}, \bibinfo {author} {\bibfnamefont {F.}~\bibnamefont {Pressacco}}, \bibinfo {author} {\bibfnamefont {J.}~\bibnamefont {Dilling}}, \bibinfo {author} {\bibfnamefont {L.}~\bibnamefont {Bruckmeier}}, \bibinfo {author} {\bibfnamefont
  {M.}~\bibnamefont {Heber}}, \bibinfo {author} {\bibfnamefont {F.}~\bibnamefont {Scholz}}, \bibinfo {author} {\bibfnamefont {J.}~\bibnamefont {Sobota}}, \bibinfo {author} {\bibfnamefont {J.}~\bibnamefont {Koralek}}, \bibinfo {author} {\bibfnamefont {N.}~\bibnamefont {Sirica}}, \bibinfo {author} {\bibfnamefont {M.}~\bibnamefont {Kallmayer}}, \bibinfo {author} {\bibfnamefont {M.}~\bibnamefont {Hoesch}}, \bibinfo {author} {\bibfnamefont {C.}~\bibnamefont {Schlueter}}, \bibinfo {author} {\bibfnamefont {L.~V.}\ \bibnamefont {Odnodvorets}}, \bibinfo {author} {\bibfnamefont {Y.}~\bibnamefont {Mairesse}}, \bibinfo {author} {\bibfnamefont {K.}~\bibnamefont {Rossnagel}}, \bibinfo {author} {\bibfnamefont {H.~J.}\ \bibnamefont {Elmers}}, \bibinfo {author} {\bibfnamefont {S.}~\bibnamefont {Beaulieu}}, \ and\ \bibinfo {author} {\bibfnamefont {G.}~\bibnamefont {Schoenhense}},\ }\bibfield  {title} {\enquote {\bibinfo {title} {Multi-mode front lens for momentum microscopy: Part i theory and part ii experiments},}\ }\href
  {https://arxiv.org/abs/2408.10104 and https://arxiv.org/abs/2401.10084} {\bibfield  {journal} {\bibinfo  {journal} {arXiv}\ ,\ \bibinfo {pages} {2408.10104 and 2401.10084}} (\bibinfo {year} {2024}{\natexlab{b}})}\BibitemShut {NoStop}%
\bibitem [{\citenamefont {Xian}\ \emph {et~al.}(2020)\citenamefont {Xian}, \citenamefont {Acremann}, \citenamefont {Agustsson}, \citenamefont {Dendzik}, \citenamefont {B{\"u}hlmann}, \citenamefont {Curcio}, \citenamefont {Kutnyakhov}, \citenamefont {Pressacco}, \citenamefont {Heber}, \citenamefont {Dong}, \citenamefont {Pincelli}, \citenamefont {Demsar}, \citenamefont {Wurth}, \citenamefont {Hofmann}, \citenamefont {Wolf}, \citenamefont {Scheidgen}, \citenamefont {Rettig},\ and\ \citenamefont {Ernstorfer}}]{Xian20}%
  \BibitemOpen
  \bibfield  {author} {\bibinfo {author} {\bibfnamefont {R.~Patrick}\ \bibnamefont {Xian}}, \bibinfo {author} {\bibfnamefont {Yves}\ \bibnamefont {Acremann}}, \bibinfo {author} {\bibfnamefont {Steinn~Y.}\ \bibnamefont {Agustsson}}, \bibinfo {author} {\bibfnamefont {Maciej}\ \bibnamefont {Dendzik}}, \bibinfo {author} {\bibfnamefont {Kevin}\ \bibnamefont {B{\"u}hlmann}}, \bibinfo {author} {\bibfnamefont {Davide}\ \bibnamefont {Curcio}}, \bibinfo {author} {\bibfnamefont {Dmytro}\ \bibnamefont {Kutnyakhov}}, \bibinfo {author} {\bibfnamefont {Federico}\ \bibnamefont {Pressacco}}, \bibinfo {author} {\bibfnamefont {Michael}\ \bibnamefont {Heber}}, \bibinfo {author} {\bibfnamefont {Shuo}\ \bibnamefont {Dong}}, \bibinfo {author} {\bibfnamefont {Tommaso}\ \bibnamefont {Pincelli}}, \bibinfo {author} {\bibfnamefont {Jure}\ \bibnamefont {Demsar}}, \bibinfo {author} {\bibfnamefont {Wilfried}\ \bibnamefont {Wurth}}, \bibinfo {author} {\bibfnamefont {Philip}\ \bibnamefont {Hofmann}}, \bibinfo {author} {\bibfnamefont
  {Martin}\ \bibnamefont {Wolf}}, \bibinfo {author} {\bibfnamefont {Markus}\ \bibnamefont {Scheidgen}}, \bibinfo {author} {\bibfnamefont {Laurenz}\ \bibnamefont {Rettig}}, \ and\ \bibinfo {author} {\bibfnamefont {Ralph}\ \bibnamefont {Ernstorfer}},\ }\bibfield  {title} {\enquote {\bibinfo {title} {An open-source, end-to-end workflow for multidimensional photoemission spectroscopy},}\ }\href {\doibase 10.1038/s41597-020-00769-8} {\bibfield  {journal} {\bibinfo  {journal} {Scientific Data}\ }\textbf {\bibinfo {volume} {7}},\ \bibinfo {pages} {442} (\bibinfo {year} {2020})}\BibitemShut {NoStop}%
\bibitem [{\citenamefont {Xian}\ \emph {et~al.}(2019)\citenamefont {Xian}, \citenamefont {Rettig},\ and\ \citenamefont {Ernstorfer}}]{Xian19_2}%
  \BibitemOpen
  \bibfield  {author} {\bibinfo {author} {\bibfnamefont {R.~Patrick}\ \bibnamefont {Xian}}, \bibinfo {author} {\bibfnamefont {Laurenz}\ \bibnamefont {Rettig}}, \ and\ \bibinfo {author} {\bibfnamefont {Ralph}\ \bibnamefont {Ernstorfer}},\ }\bibfield  {title} {\enquote {\bibinfo {title} {Symmetry-guided nonrigid registration: The case for distortion correction in multidimensional photoemission spectroscopy},}\ }\href {\doibase https://doi.org/10.1016/j.ultramic.2019.04.004} {\bibfield  {journal} {\bibinfo  {journal} {Ultramicroscopy}\ }\textbf {\bibinfo {volume} {202}},\ \bibinfo {pages} {133 -- 139} (\bibinfo {year} {2019})}\BibitemShut {NoStop}%
\bibitem [{\citenamefont {Hamann}(2013)}]{hamann13}%
  \BibitemOpen
  \bibfield  {author} {\bibinfo {author} {\bibfnamefont {D.~R.}\ \bibnamefont {Hamann}},\ }\bibfield  {title} {\enquote {\bibinfo {title} {Optimized norm-conserving vanderbilt pseudopotentials},}\ }\href {\doibase 10.1103/PhysRevB.88.085117} {\bibfield  {journal} {\bibinfo  {journal} {Phys. Rev. B}\ }\textbf {\bibinfo {volume} {88}},\ \bibinfo {pages} {085117} (\bibinfo {year} {2013})}\BibitemShut {NoStop}%
\bibitem [{\citenamefont {Perdew}\ \emph {et~al.}(1996)\citenamefont {Perdew}, \citenamefont {Burke},\ and\ \citenamefont {Ernzerhof}}]{pbe}%
  \BibitemOpen
  \bibfield  {author} {\bibinfo {author} {\bibfnamefont {John~P.}\ \bibnamefont {Perdew}}, \bibinfo {author} {\bibfnamefont {Kieron}\ \bibnamefont {Burke}}, \ and\ \bibinfo {author} {\bibfnamefont {Matthias}\ \bibnamefont {Ernzerhof}},\ }\bibfield  {title} {\enquote {\bibinfo {title} {Generalized gradient approximation made simple},}\ }\href {\doibase 10.1103/PhysRevLett.77.3865} {\bibfield  {journal} {\bibinfo  {journal} {Phys. Rev. Lett.}\ }\textbf {\bibinfo {volume} {77}},\ \bibinfo {pages} {3865--3868} (\bibinfo {year} {1996})}\BibitemShut {NoStop}%
\bibitem [{\citenamefont {{van Setten}}\ \emph {et~al.}(2018)\citenamefont {{van Setten}}, \citenamefont {Giantomassi}, \citenamefont {Bousquet}, \citenamefont {Verstraete}, \citenamefont {Hamann}, \citenamefont {Gonze},\ and\ \citenamefont {Rignanese}}]{pseudodojo}%
  \BibitemOpen
  \bibfield  {author} {\bibinfo {author} {\bibfnamefont {M.J.}\ \bibnamefont {{van Setten}}}, \bibinfo {author} {\bibfnamefont {M.}~\bibnamefont {Giantomassi}}, \bibinfo {author} {\bibfnamefont {E.}~\bibnamefont {Bousquet}}, \bibinfo {author} {\bibfnamefont {M.J.}\ \bibnamefont {Verstraete}}, \bibinfo {author} {\bibfnamefont {D.R.}\ \bibnamefont {Hamann}}, \bibinfo {author} {\bibfnamefont {X.}~\bibnamefont {Gonze}}, \ and\ \bibinfo {author} {\bibfnamefont {G.-M.}\ \bibnamefont {Rignanese}},\ }\bibfield  {title} {\enquote {\bibinfo {title} {The pseudodojo: Training and grading a 85 element optimized norm-conserving pseudopotential table},}\ }\href {\doibase https://doi.org/10.1016/j.cpc.2018.01.012} {\bibfield  {journal} {\bibinfo  {journal} {Computer Physics Communications}\ }\textbf {\bibinfo {volume} {226}},\ \bibinfo {pages} {39} (\bibinfo {year} {2018})}\BibitemShut {NoStop}%
\bibitem [{\citenamefont {Marzari}\ \emph {et~al.}(2012)\citenamefont {Marzari}, \citenamefont {Mostofi}, \citenamefont {Yates}, \citenamefont {Souza},\ and\ \citenamefont {Vanderbilt}}]{Marzari2012}%
  \BibitemOpen
  \bibfield  {author} {\bibinfo {author} {\bibfnamefont {Nicola}\ \bibnamefont {Marzari}}, \bibinfo {author} {\bibfnamefont {Arash~A.}\ \bibnamefont {Mostofi}}, \bibinfo {author} {\bibfnamefont {Jonathan~R.}\ \bibnamefont {Yates}}, \bibinfo {author} {\bibfnamefont {Ivo}\ \bibnamefont {Souza}}, \ and\ \bibinfo {author} {\bibfnamefont {David}\ \bibnamefont {Vanderbilt}},\ }\bibfield  {title} {\enquote {\bibinfo {title} {Maximally localized {Wannier} functions: Theory and applications},}\ }\href {\doibase 10.1103/RevModPhys.84.1419} {\bibfield  {journal} {\bibinfo  {journal} {Rev. Mod. Phys.}\ }\textbf {\bibinfo {volume} {84}},\ \bibinfo {pages} {1419} (\bibinfo {year} {2012})}\BibitemShut {NoStop}%
\bibitem [{\citenamefont {Zhou}\ \emph {et~al.}(2020)\citenamefont {Zhou}, \citenamefont {Monacelli}, \citenamefont {Bianco}, \citenamefont {Errea}, \citenamefont {Mauri},\ and\ \citenamefont {Calandra}}]{zhou2020anharmonicity}%
  \BibitemOpen
  \bibfield  {author} {\bibinfo {author} {\bibfnamefont {Jianqiang~Sky}\ \bibnamefont {Zhou}}, \bibinfo {author} {\bibfnamefont {Lorenzo}\ \bibnamefont {Monacelli}}, \bibinfo {author} {\bibfnamefont {Raffaello}\ \bibnamefont {Bianco}}, \bibinfo {author} {\bibfnamefont {Ion}\ \bibnamefont {Errea}}, \bibinfo {author} {\bibfnamefont {Francesco}\ \bibnamefont {Mauri}}, \ and\ \bibinfo {author} {\bibfnamefont {Matteo}\ \bibnamefont {Calandra}},\ }\bibfield  {title} {\enquote {\bibinfo {title} {Anharmonicity and doping melt the charge density wave in single-layer tise2},}\ }\href {\doibase 10.1021/acs.nanolett.0c00597} {\bibfield  {journal} {\bibinfo  {journal} {Nano Letters}\ }\textbf {\bibinfo {volume} {20}},\ \bibinfo {pages} {4809} (\bibinfo {year} {2020})}\BibitemShut {NoStop}%
\end{thebibliography}%

\end{document}